\documentclass[preprint, 12pt]{elsarticle}

\usepackage{amssymb}
\usepackage{bm}
\usepackage[colorlinks=true, linkcolor=blue, anchorcolor=blue,citecolor=blue]{hyperref}

\usepackage{amsmath}
\usepackage{caption}
\usepackage{subcaption}

\usepackage{cleveref}
\usepackage{multirow}
\usepackage{color,soul}
\usepackage[table]{xcolor}

\usepackage[switch]{}

\newcommand\G{{\bf g}}

\newcommand\N{{\bf n}}

\newcommand\F{{\bf f}}

\newcommand\U{{\bf u}}

\newcommand\X{{\bf x}}

\newcommand\dert{\partial_t}

\newcommand\romandt[1]{\frac{{\rm d} #1}{{\rm d}t}}

\newcommand\be{\begin{equation}}
\newcommand\nd{\end{equation}}
\newcommand\bed{\begin{displaymath}}
\newcommand\ndd{\end{displaymath}}

\newcommand\ba{\begin{array}}
\newcommand\ea{\end{array}}
\newcommand\bea{\begin{eqnarray}}
\newcommand\nda{\end{eqnarray}}

\begin{document}
\sethlcolor{yellow}
\setstcolor{red}
\soulregister\cite7 
\soulregister\ref7

\begin{frontmatter}

\title{An Edge-based Interface Tracking (EBIT) Method for Multiphase-flow Simulation with Surface Tension}

\author[a]{Jieyun Pan\corref{cor1}}
\ead{yi.pan@sorbonne-universite.fr}

\author[a,c]{Tian Long}
\ead{tian.long@dalembert.upmc.fr}

\author[a]{Leonardo Chirco}
\ead{leonardo.chirco@framatome.com}

\author[d]{Ruben Scardovelli}
\ead{ruben.scardovelli@unibo.it}

\author[a]{St\'{e}phane Popinet\footnote{}}
\ead{popinet@basilisk.fr}

\author[a,b]{St\'{e}phane Zaleski}
\ead{stephane.zaleski@sorbonne-universite.fr}

\cortext[cor1]{Corresponding author}
\address[a]{Sorbonne Universit\'{e} and CNRS, Institut Jean Le Rond d'Alembert UMR 7190, F-75005 Paris, France}
\address[b]{Institut Universitaire de France, Paris, France}
\address[c]{School of Aeronautics, Northwestern Polytechnical University, Xi'an, 710072, PR China}
\address[d]{ DIN - Lab. di Montecuccolino, Universit\`a di Bologna, I-40136 Bologna, Italy}

\begin{abstract}

We present a novel Front-Tracking method, the Edge-Based Interface Tracking (EBIT) method for multiphase flow simulations. In the EBIT method, the markers are located on the grid edges and the interface can be reconstructed without storing the connectivity of the markers. This feature makes the process of marker addition or removal easier than in the traditional Front-Tracking method. The EBIT method also allows almost automatic parallelization due to the lack of explicit connectivity. 

In a previous journal article we have presented the kinematic part of the EBIT method, that includes the algorithms for piecewise linear reconstruction and advection of the interface. Here, we complete the presentation of the EBIT method and combine the kinematic algorithm with a Navier--Stokes solver. A circle fit is now implemented to improve the accuracy of mass conservation in the reconstruction phase. Furthermore, to identify the reference phase and to distinguish ambiguous topological configurations, we introduce a new feature: the Color Vertex. For the coupling with the Navier--Stokes equations, we first calculate volume fractions from the position of the markers and the Color Vertex, then viscosity and density fields from the computed volume fractions and finally surface tension stresses with the Height-Function method. In addition, an automatic topology change algorithm is implemented into the EBIT method, making it possible the simulation of more complex flows. The two-dimensional version of the EBIT method has been implemented in the free Basilisk platform, and 
validated with seven standard test cases: stagnation flow, translation with uniform velocity, single vortex, Zalesak's disk, capillary wave, Rayleigh-Taylor instability and rising bubble. The results are compared with those obtained with the Volume-of-Fluid (VOF) method already implemented in Basilisk.
\end{abstract}


\end{frontmatter}


\section{Introduction}
Multiphase flows are ubiquitous in nature and engineering, and their numerical simulation still represents a formidable challenge, especially when a wide range of scales is involved, as in breaking waves on the sea surface, in some industrial processes or in atomizing liquid jets. Scales from meters to microns are typically seen. 
Large Reynolds number turbulence, as well as mass and heat transfers
are the main causes for the introduction of such a wide range of scales.
The CO$_2$ transfer and heat exchange between the oceans and the atmosphere is tightly linked to multiphase flows,
as it takes place through the production and dispersion of small bubbles and droplets.
Similar small structures are observed in technology, for example in the heat exchange and transport 
in nuclear reactors, the atomization of liquids in combustion and other settings, and in most of the synthesis processes in chemical engineering. For many of these natural and engineering problems, numerical modelling is extremely desirable, albeit a monumental challenge. 

However, in the general area of multiphase flow simulation, much progress has recently been achieved in the interrelated issues of multiple scales, discretization, and topology change. These issues are: i) the long-standing one of how to represent numerically (i.e., simulate) a dynamic or moving curve or surface, ii) the more difficult or challenging problem of how to efficiently model and discretize problems at multiple scales, and finally iii) the still open issue of how to take advantage of hierarchical data structures and grids, such as quadtrees and octrees, to address the ``tyranny of small scales'' \cite{or2018tyranny}. 

The first, moving curve issue is divided into two problems: the kinematic problem, where the motion of the curve or surface separating the phases must be described by knowing the fluid velocity field and the rate of phase change, and the dynamic problem, where the momentum and energy conservation equations must be solved for given fluid properties. Methods available for the kinematic problem are sometimes separated into Front-Capturing and Front-Tracking methods. In the first kind, Front-Capturing methods, a tracer or marker function $f(\X,t)$ is integrated in time with the knowledge of an adequate velocity field
$\U(\X,t)$
\be
\dert f + \U \cdot \nabla f = 0
\nd
The tracer function can be a Heaviside function, leading to Volume-of-Fluid (VOF) methods \cite{Hirt_1981_39, Brackbill_1992_100} or a smooth function, leading to Level-Set (LS) methods \cite{Osher_1988_79, Osher_2001_169}.

\begin{figure}
\begin{center}
\begin{tabular}{cc}
\includegraphics[width=0.45\textwidth]{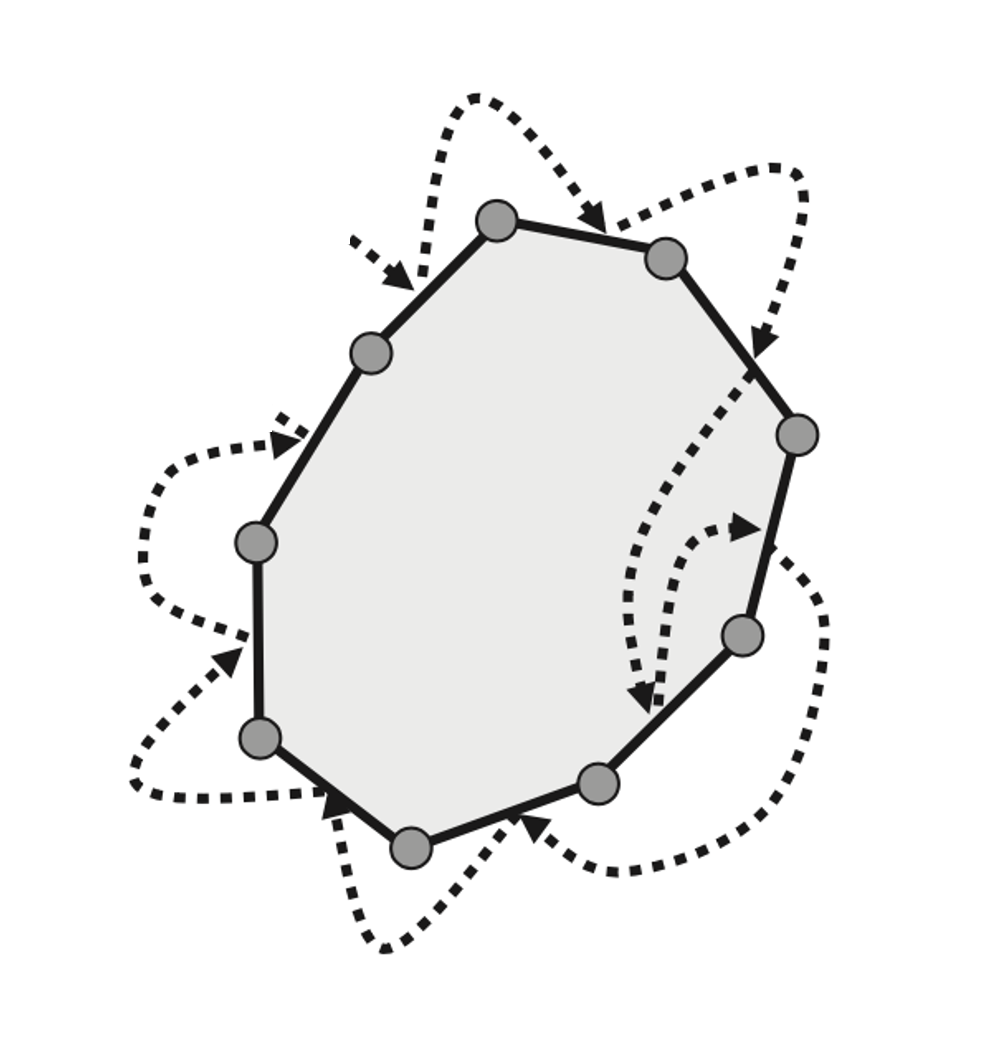} &
\includegraphics[width=0.4\textwidth]{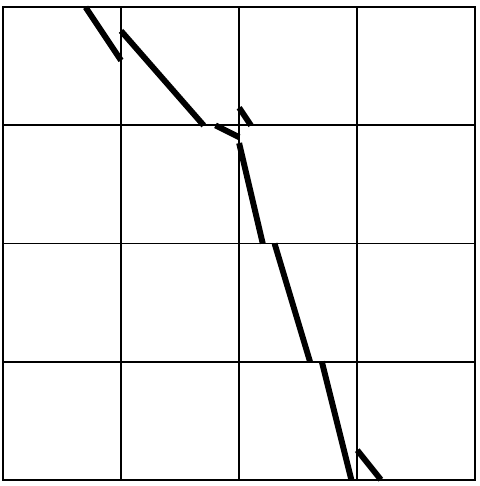}\\
(a) & (b) 
\end{tabular}
\end{center}
\caption{(a) A global data structure with a linked list: the example of the front. Reproduced from \cite{Tryggvason_2011_book}; (b) A local data structure with unlinked VOF linear reconstructions.}
\label{Fig_GT}
\end{figure}

In Front-Tracking methods, the interface or ``front'' is represented by a curve discretization, for example a spline, which evolves with a
prescribed normal velocity $V_S$ , see \cite{Unverdi_1992_100}. Compatibility between the two formulations is achieved if $\U \cdot \N = V_S$. An introduction to the most
popular methods may be found in \cite{Tryggvason_2011_book}. Another point of view on the methods, that is fruitful in connection with
issues ii) and iii) above, can be obtained by investigating how the data
structures are tied to the underlying Eulerian grid. The  data structures are {\em local} when
their components have little or no ``knowledge'' of the
overall connections among pieces of interface or connected regions of a given phase (Fig.~\ref{Fig_GT}b). Thus a discretization of the tracer function $f$ is a local
data structure, tied to the Eulerian grid. On the other hand a {\em global data structure} contains information not only about individual points on the interface, but also about
their connections with the entire interface of a given object. For Front-Tracking methods this is achieved by linked lists, or pointers, that
allow to navigate the data structure along the object (Fig.~\ref{Fig_GT}a). It is then obvious and efficient, using the data structure, to find all the connected pieces of the interface.
The local data structures are not limited to VOF or LS methods. For example, an unconnected marker or particle method, such as Smoothed Particle Hydrodynamics (SPH), or other methods, simply tracks particles of position ${\bf x}_i$ by integrating
\be
\romandt{\X_i} = \U(\X_i,t)
\nd
without any concept of connection between the particles \cite{chen1997surface}. Such an unconnected marker method, despite its similarity with Front-Tracking is in fact a {\em local} method. 
Both the local and the global approaches have advantages. The local methods are easy to parallelize on a grid, are free of constraints that require the solutions of large linear or nonlinear systems (as when constructing interfaces by high-order splines) and are generally computationally efficient. The global methods allow to track and control the topology, so they are also useful for the second issue, the handling of multiscale problems.
Moreover, when dealing with complex multiscale problems, it is often useful to investigate geometrical properties such as skeletons \cite{chen2022characterizing} which are branched manifolds most naturally represented by Front-Tracking. 
Global methods also allow to naturally distinguish between continuous slender objects such as unbroken thin ligaments or threads and strings of small particles or broken ligaments. This latter distinction is of great importance when analyzing statistically highly-fragmented flows \cite{Chirco_2022_467}. 
It is however possible to represent slender objects with local methods in some case. Chiodi \cite{Chiodi_2020_phd} proposed an enhancement of the VOF reconstruction: the reconstruction with 2-planes (R2P) method. The method considers two planes with arbitrary relative orientation, a feature that allows the modelling of sub-grid scale thin sheets, the closure of sheet rims and in general a more accurate sub-grid interface reconstruction. 
The moment-of-fluid (MOF) method \cite{Shashkov_2023_479, Shashkov_2023_494} first proposed by 
Dyadechko and Shashkov \cite{Dyadechko_2008_227}, in which the zeroth, first and second moments of the fragment of material inside a cell are used for interface reconstruction by one half-plane, two half-planes or a circle, is also able to exactly capture corners, filaments and some concave shapes smaller than the grid spacing.
A method that combines some of the properties of global methods, such as Front-Tracking, and of local ones, such as VOF or LS,
seems desirable. There have been indeed some prior attempts at such a combination.

Aulisa et al. \cite{Aulisa_2003_188, Aulisa_2004_197} combined the VOF method with marker points to obtain smooth
interfaces without discontinuity and to improve mass conservation of traditional Front-Tracking methods. L\'{o}pez et al. \cite{Lopez_2005_208} introduced marker points into the VOF method to allow the tracking of fluid structures thinner than the cell size.

The Level Contour Reconstruction Method (LCRM) developed by Shin, Juric and collaborators \cite{Shin_2002_180, Shin_2005_203, Shin_2007_21} combined Front-Tracking and LS methods. It improved the mass conservation problem of traditional LS methods, thanks to the tracking of the interface by Lagrangian elements (instead of advecting the LS function field).
A LS function can then be generated from the Lagrangian elements.
The smoothing of interface elements, as well as topology changes, take place automatically during the reconstruction procedure, thus explicit connectivity information is not needed in their method. Singh and Shyy \cite{Singh_2007_224} used the LCRM to perform topology changes in their traditional Front-Tracking method where connectivity of the Lagrangian elements has to be maintained explicitly.
Shin, Yoon and Juric \cite{Shin_2011_230} later extended the LCRM to obtain a new type of Front-Tracking method, the Local Front Reconstruction Method (LFRM), for both two-dimensional and three-dimensional multiphase flow simulations. The LFRM reconstructs interface elements using the geometrical information directly from the Lagrangian interface elements instead of constructing another LS field.

We suggest a similar method in the present work, based on a purely kinematic approach 
developed by two of us \cite{Chirco_2022_95},
called the Edge-Based Interface-Tracking (EBIT) method.
In that method the position of the interface is tracked by marker points
located on the edges of an Eulerian grid, and 
the connectivity information is implicit.
The basic idea and the split interface advection were discussed in \cite{Chirco_2022_95},
here we improve the mass conservation of the original EBIT method by using a circle fit to reconstruct the interface during advection.
A topology change mechanism is also introduced.
Compared to the LFRM of Shin, Yoon and Juric \cite{Shin_2011_230}, markers in the EBIT method are bound to the
Eulerian grid. These marker are obtained by a reconstruction of
the interface at every time step, thus the Eulerian grid and Lagrangian markers can be distributed to different processors by the same routine
as the one used in the parallelization of the Navier--Stokes solver.
Second, a new feature called Color Vertex,
which amounts to describe the topology of the interface by the color of
markers at the vertices of the grid, is discussed. 
Such a scheme is trivial on a simplex and just slightly more complicated on a square grid. It was
proposed in a similar context by Singh and Shyy \cite{Singh_2007_224}.
Third, we combine the EBIT method with a Navier--Stokes solver for multiphase flow simulations. The coupling is simply
realized by computing volume fractions from the position of the markers and the Color Vertex, then we use the volume fractions as in typical VOF solvers \cite{Tryggvason_2011_book}. 

The paper is organized as follows: the kinematics of the EBIT method is described in Section \ref{Numerical method}.
This includes the interface advection algorithm, the updating rules for the Color Vertices and the automatic topology change algorithm.
Then the coupling algorithm between the EBIT interface description and the multiphase fluid dynamics is presented. 
A two-dimensional flow solver based on a Cartesian or quadtree grid is implemented inside the free platform Basilisk \cite{Popinet_2003_190, Popinet_2009_228}. In Section \ref{Numerical results and discussion}, the EBIT method is validated by the computation of typical examples of
multiphase flow simulations. 
The results obtained by the combined EBIT method and Navier--Stokes solver are presented and compared with those calculated by the PLIC-VOF method implemented inside Basilisk \cite{Popinet_2009_228}.

\section{Numerical method} \label{Numerical method}
\subsection{The EBIT method}

\begin{figure}
\begin{center}
\begin{tabular}{cc}
\includegraphics[width=0.45\textwidth]{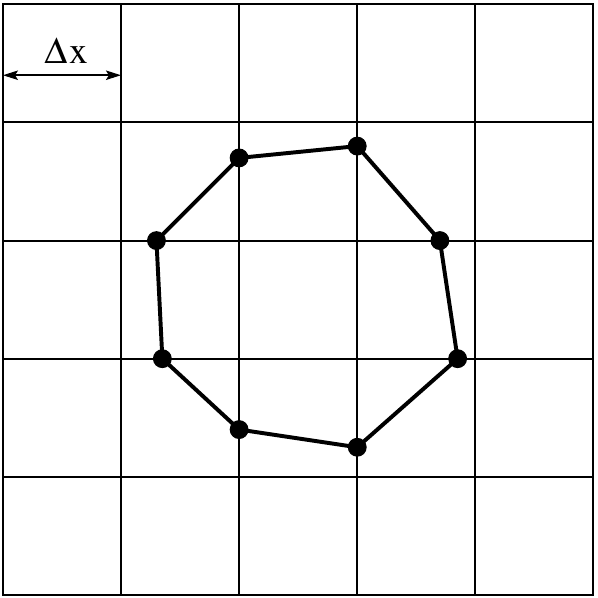} &
\includegraphics[width=0.45\textwidth]{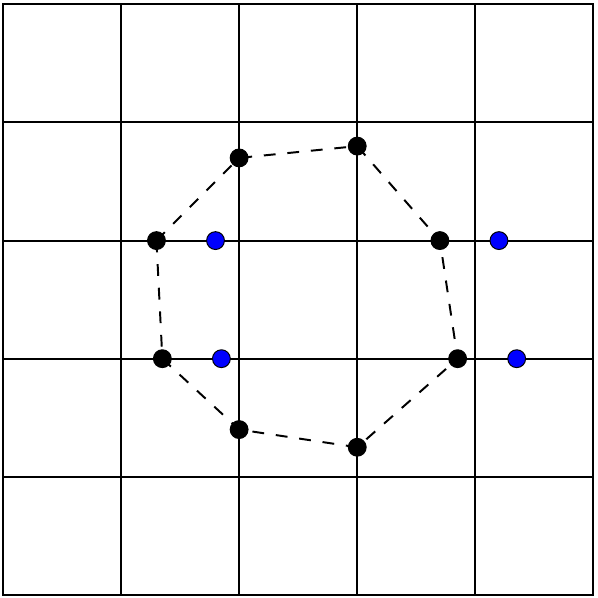}\\
(a) & (b) \\[6pt]
\includegraphics[width=0.45\textwidth]{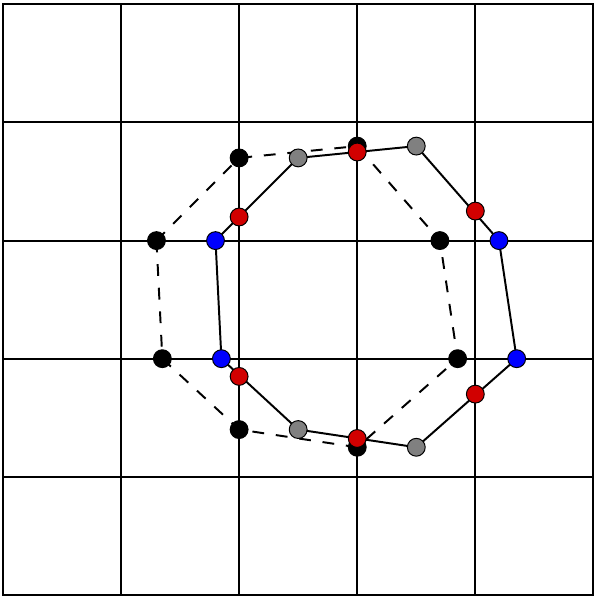} &
\includegraphics[width=0.45\textwidth]{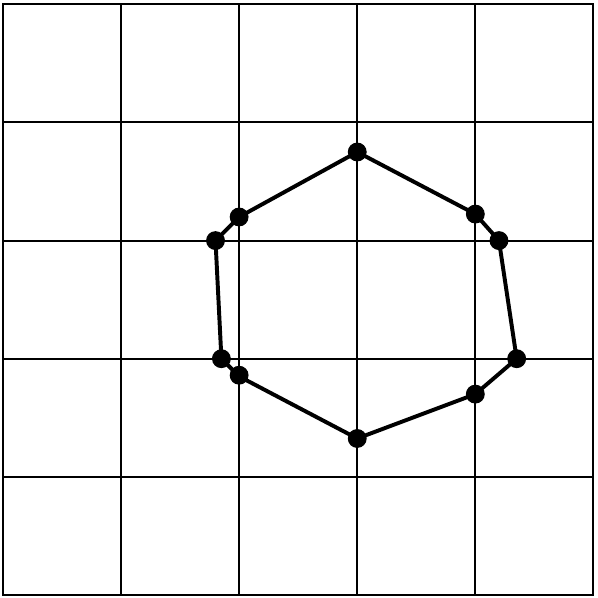}\\
(c) & (d) 
\end{tabular}
\end{center}
\caption{One-dimensional advection of the EBIT method along
the $x$-axis: (a) initial interface line; (b) advection of
the markers on the grid lines aligned with the velocity component
(blue points); (c) advection of the unaligned markers (gray points)
and computation of the intersections with the grid lines (red points);
(d) interface line after the 1D advection.}
\label{Fig_EBIT_advection}
\end{figure}

In the EBIT method, the interface is represented by a set of marker points placed on the grid lines. In order to keep the markers on the grid lines, it is necessary to compute the intersection points between the interface line and the grid lines at the end of each advection step. The equation of motion for a marker point at position $\X_i$ is
\begin{gather}
\frac{d \X_i}{dt} = \U_i
\label{Eq_marker}
\end{gather}
which is discretized by a first-order explicit Euler method
\begin{gather}
\X^{n + 1}_i= \X^{n}_i + \U^{n}_i \Delta t
\label{Eq_marker_dis}
\end{gather}
where the velocity $\U^{n}_i$ at the marker position $\X^{n}_i$ is calculated by a bilinear interpolation.

For a multi-dimensional problem, a split method is used to advect the interface (see Fig.~\ref{Fig_EBIT_advection}), which is similar to that described in \cite{Chirco_2022_95}, but with some improvements. The marker points placed on the grid lines that are aligned with the velocity component of the 1D advection are called \textit{aligned markers}, while the remaining ones are called \textit{unaligned markers}. Starting from the initial position at time step $n$, the new position of the aligned markers is obtained by Eq.~\eqref{Eq_marker_dis} (blue points of Fig.~\ref{Fig_EBIT_advection}b). To compute the new unaligned markers, we first advect them using again Eq.~\eqref{Eq_marker_dis}, obtaining in this way the gray points of Fig.~\ref{Fig_EBIT_advection}c. Finally, the new position of the unaligned markers (red points of Fig.~\ref{Fig_EBIT_advection}d) is obtained by a circle fit method which will be described in the next section.

\begin{figure}
\begin{center}
\begin{tabular}{cc}
\includegraphics[width=0.45\textwidth]{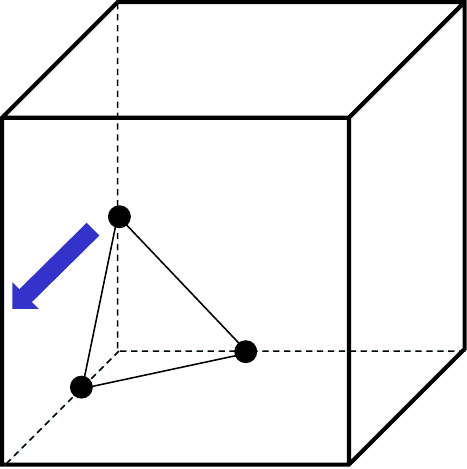} &
\includegraphics[width=0.45\textwidth]{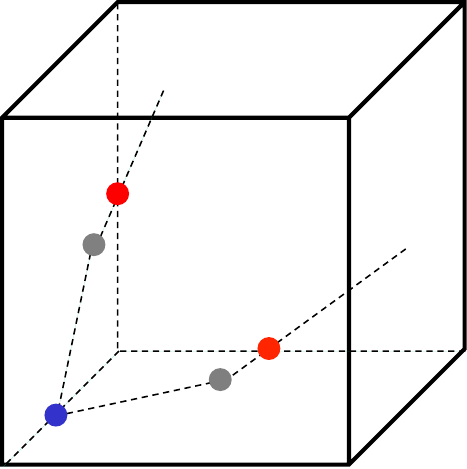}\\
(a) & (b)
\end{tabular}
\end{center}
\caption{One-dimensional advection of the 3D EBIT method:
 (a) initial interface; (b) decomposition of the 3D advection problem.}
\label{Fig_EBIT_3D}
\end{figure}

One important advantage of the split method is that it is relatively straightforward to extend the EBIT method to 3D. For the 1D advection along a coordinate direction of Fig.~\ref{Fig_EBIT_3D}a, the problem can be decomposed into simpler 2D problems on the two cell faces of Fig.~\ref{Fig_EBIT_3D}b, containing the advection direction under consideration and the marker points. In this way we can use most of the algorithms and routines we have developed for the interface advection in 2D. The detailed description of the 3D algorithm and the numerical tests will be presented in a future paper.

\subsection{Circle fit}

\begin{figure}
\centering
\includegraphics[scale=0.8]{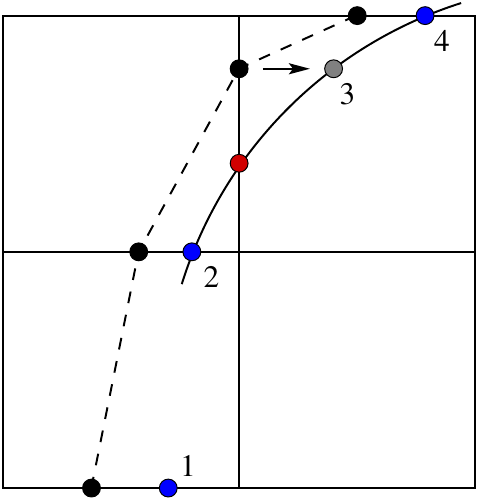}
\caption{Circle fit to compute the position of the unaligned marker (red point).}
\label{Fig_circle_fit}
\end{figure}

In our previous paper about the EBIT method \cite{Chirco_2022_95}, we showed that the computation of the new position of the unaligned markers with a straight line fit was leading to poor accuracy of mass conservation, in particular at low spatial resolution. In this paper, we propose a circle fit as a method to improve the accuracy of mass conservation.
The new position of the unaligned marker (red point of Fig.~\ref{Fig_circle_fit}) is obtained by fitting a circle through the surrounding marker points and by computing its intersection with the corresponding grid line. The gray point is then discarded.
Whenever it is possible, we consider two circles for the fitting, through points 2,3,4 and 1,2,3 of Fig.~\ref{Fig_circle_fit}, respectively. In that case, the final position of the unaligned marker will be the average of the two intersections.

\subsection{Color Vertex}

\begin{figure}
\centering
\includegraphics[width=\textwidth]{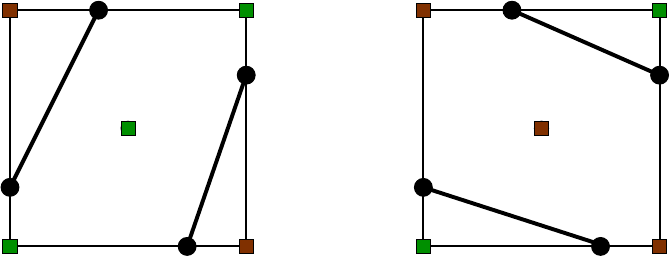}
\caption{Two color vertex configurations (brown and green squares) to select a different connectivity 
in the same set of markers.}
\label{Fig_color_vertex}
\end{figure}

In the EBIT method, the connectivity of the markers is implicit. When there are only two markers on the cell boundary, the interface portion in that cell is given by the segment connecting these two points. However, when there are four markers on the boundary, there are two possible alternative configurations, as shown in Fig.~\ref{Fig_color_vertex}. In order to select one of the two configurations without any ambiguity, we consider a technique called Color Vertex, which was first proposed by Singh and Shyy \cite{Singh_2007_224}, to implement an automatic topology change. The value of the Color Vertex indicates the fluid phase in the corresponding region within the cell, and five color vertices (four in the corners and one in the center of the cell) are enough to select one of the two configurations of Fig.~\ref{Fig_color_vertex}. In other word, we can establish a one-to-one correspondence between the topological configuration and the value of the color vertices within each cell, and then reconstruct the interface segments with no ambiguity.

Furthermore, the direction of the unit normal to the interface can also be determined based on the Color Vertex distribution. The local feature of the Color Vertex makes the EBIT method more suitable for parallelization, when it is compared to the data structure that is used for storing the connectivity in traditional Front-Tracking methods.

\begin{figure}
\centering
\begin{subfigure}[b]{\textwidth}
\centering
\includegraphics[width=\textwidth]{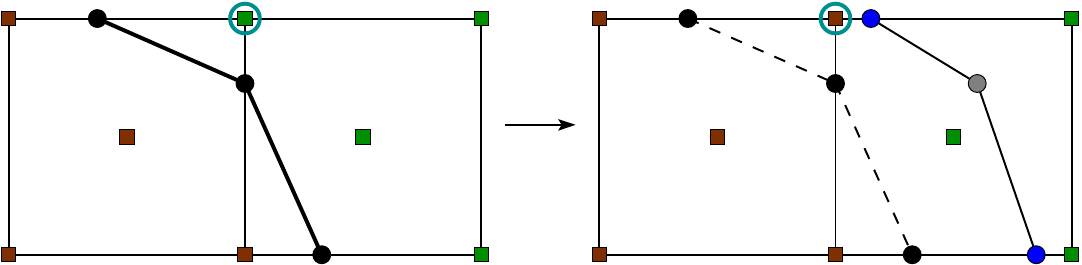}
\end{subfigure}
\caption{Update of the Color Vertex value on a cell corner: its value changes, 
from green to brown, as the marker is advected across the grid lines intersection.}
\label{Fig_color_vertex_update_corner}
\end{figure}

\begin{figure}
\centering
\begin{subfigure}[b]{\textwidth}
\centering
\includegraphics[width=\textwidth]{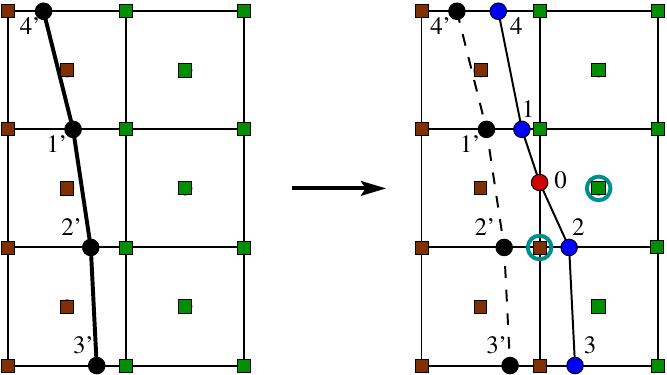}
\caption{}
\label{Fig_color_vertex_update_central_1}
\end{subfigure}
\\[10pt]
\begin{subfigure}[b]{\textwidth}
\centering
\includegraphics[width=\textwidth]{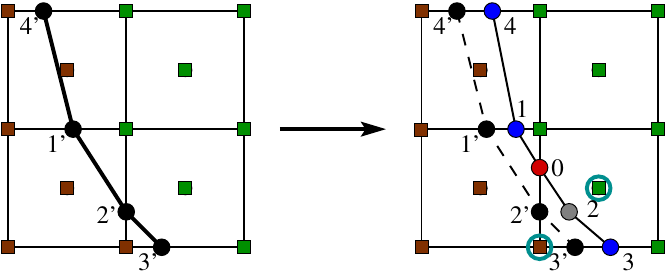}
\caption{}
\label{Fig_color_vertex_update_central_2}
\end{subfigure}
\caption{Update of the Color Vertex value in the cell center with an unaligned marker (red point).}
\label{Fig_color_vertex_update_central}
\end{figure}

As the interface is advected, the value of a Color Vertex should also be updated 
accordingly to ensure that the implicit connectivity information is retained. For a Color Vertex located on a cell corner, we have to change its value if a marker moves across the intersection of the corresponding grid lines, as shown in 
Fig.~\ref{Fig_color_vertex_update_corner}.

In the present implementation of the EBIT algorithm the Color Vertex in the cell center 
is only used to select one of the two configurations shown in Fig.~\ref{Fig_color_vertex}. 
It is important to realize that in such a configuration, regardless of the direction of the advection, there are both aligned markers and unaligned ones.
Therefore, the algorithm for updating the Color Vertex in the cell center proceeds as follows (see Fig.~\ref{Fig_color_vertex_update_central}):

(1) If in the cell under investigation, after the interface advection there is an unaligned marker (red point 0 of Fig.~\ref{Fig_color_vertex_update_central}), we identify the interface segment through points 1 and 2 that brackets the unaligned marker and the corresponding segment through points 1' and 2', before the advection step. These two points are on opposite sides in Fig.~\ref{Fig_color_vertex_update_central_1} and on consecutive
sides in Fig.~\ref{Fig_color_vertex_update_central_2}.

(2) From the connectivity information before the advection step, we identify two more markers, points 3' and 4', that are connected to the segment 1'-2' on opposite sides,
and compute their new positions 3 and 4 after the advection step.

(3) If the four points 1, 2, 3 and 4 are unaligned markers, we do not need to update the value of the Color Vertex in the center, because in this case it is impossible to have an ambiguous configuration.

(4) We check if in the cell under investigation there is an aligned marker (point 2 in Fig.~\ref{Fig_color_vertex_update_central_1} and point 3 in Fig.~\ref{Fig_color_vertex_update_central_2}).

(5) If an aligned marker has been found, we identify the cell corner (bottom left corner of Fig.~\ref{Fig_color_vertex_update_central}) isolated by the segments connecting this marker and the new unaligned marker (point 0 of Fig.~\ref{Fig_color_vertex_update_central}). The value of the Color Vertex in the cell center is set to the opposite value of that of the cell corner.

With these simple rules, the value of the Color Vertex in the cell center is set to the correct value to select one of the two configurations of Fig.~\ref{Fig_color_vertex}.

\subsection{Topology change}

\begin{figure}
\centering
\begin{subfigure}[b]{\textwidth}
\centering
\includegraphics[width=\textwidth]{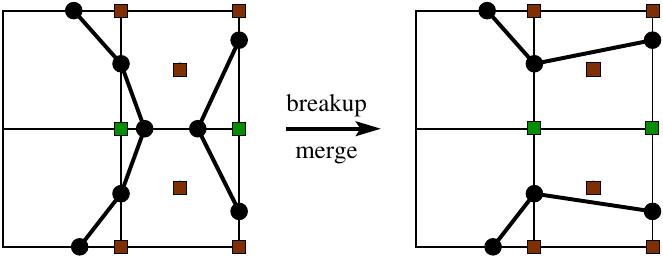}
\label{Fig_topo_change_2}
\end{subfigure}
\caption{Topology change mechanism.}
\label{Fig_topo_change}
\end{figure}

The topology change is controlled by the Color Vertex value distribution in an automatic manner. In the present implementation of the EBIT method, a marker point is present on a cell edge only if the value of the Color Vertex at the two edge endpoints is different. 

Furthermore, only one marker is allowed on a cell edge. When two markers move into the same edge, as shown in Fig.~\ref{Fig_topo_change}, they both will be eliminated automatically, because the value of the Color Vertex at the corresponding endpoints is the same, hence there cannot be any marker on that edge. 
Moreover, the ``surviving'' markers within the cell will be reconnected automatically, see again Fig.~\ref{Fig_topo_change}. This reconnection procedure enables ligament breakup or droplet merging in an automatic way during the interface advection. As a direct consequence of this procedure, the volume occupied by the reference phase can decrease or increase. In particular, it tends to remove droplets or bubbles which are smaller than the grid size. 

This topology change mechanism only affects interfaces that are approaching each other along a direction parallel to the grid lines. Because of the presence of a Color Vertex in the cell center, interfaces approaching along a diagonal direction do not induce any topology change, as long as the four markers remain on a different edge, as shown in Fig.~\ref{Fig_color_vertex}. Compared to the scheme depending on the tetra-marching procedure in Yoon's \cite{Yoon_2010_08} LCRM and Shin's \cite{Shin_2011_230} LFRM, this mechanism results in a more symmetric topology change.

However, in our method approaching interfaces along diagonal directions behave
differently from those approaching along parallel ones. A possible solution to this problem would be removing the restriction on the number of markers per edge, which would also allow us to capture the sub-grid scale interfacial structure and to control the topology change based on a physical mechanism.

\subsection{Governing equations}

The Navier--Stokes equations for incompressible two-phase flow with immiscible fluids written in the one-fluid formulation are
\begin{gather}
\frac{\partial \rho}{\partial t}
+ \U \cdot \nabla \rho= 0 \\
\frac{\partial \rho \U}{\partial t}
+ \nabla \cdot (\rho \U\U) = 
-\nabla p + \nabla \cdot \left[ \mu \left( \nabla \U + \nabla \U^T\right) \right] + \rho \G + \F
\label{Eq:NS}
\\
\nabla \cdot \U = 0
\end{gather}
\noindent where $\rho$ and $\mu$ are density and viscosity, respectively. The gravitational force is taken into account with the $\rho \G$ term. Surface tension is modeled by the term $\F = \sigma \kappa \N \delta_S$, where $\sigma$ is the surface tension coefficient, $\kappa$ the interface curvature, $\N$ the unit normal and $\delta_S$ the surface Dirac delta function. 

The physical properties are calculated as
\begin{gather}
\rho = H \rho_1 + (1 - H) \rho_2, \qquad \mu = H \mu_1 + (1 - H) \mu_2
\label{Eq:physical}
\end{gather}
where $H$ is the Heaviside function, which is equal to 1 inside the reference phase and 0 elsewhere.

Since the markers are located on the grid lines, we consider a simple strategy to couple the EBIT method with the Navier--Stokes equations. From the position of the marker points and the value of the Color Vertices in the cell under investigation, we can easily compute the equation of the straight line connecting the markers and then the volume fraction $C$ \cite{Scardovelli00}. The volume fraction field is used to approximate the Heaviside function in Eq.~\eqref{Eq:physical}, to calculate the curvature by the generalized height function method \cite{Popinet_2009_228} and the Dirac delta function in Eq.~\eqref{Eq:NS}.

The numerical implementation of the EBIT method has been written in the Basilisk C language \cite{Popinet_2003_190, Popinet_2009_228}, which adopts a time-staggered approximate projection method to solve the incompressible Navier--Stokes equations on a Cartesian mesh or a quad/octree mesh. 
The Bell-Colella-Glaz (BCG) \cite{Bell_1989_85} second-order scheme is used to discretize the advection term, and a fully implicit scheme for the diffusion term. A well-balanced Continuous Surface Force (CSF) method is used to calculate the surface tension term \cite{Popinet_2009_228, Popinet_2018_50}.

The geometrical PLIC-VOF scheme generalized for the quad/octree grid \cite{Popinet_2009_228} in Basilisk is used to obtain the results for the comparisons of Sect. \ref{Numerical results and discussion}. The volume fraction $C$ is advanced in time with the following advection equation
\begin{gather}
\frac{\partial C}{\partial t} + \U \cdot \nabla C  = 0
\label{Eq:PLIC-VOF}
\end{gather}
The split scheme proposed by Weymouth and Yue \cite{Weymouth_2010_229} is used to discretize Eq.~\eqref{Eq:PLIC-VOF}. For a two-dimensional problem, the split scheme for the volume fraction advection, here first along the $x$-direction and then along $y$, is
\begin{gather}
\frac{C_{i, j}^ {*} - C_{i, j}^{n-1/2}}{\Delta t} + \frac{\phi_{i + 1/2, j}^{n -1/2} - \phi_{i - 1/2, j}^{n - 1/2}}{\Delta} = C_c \frac{u_{i + 1 / 2, j}^{n} - u_{i - 1 / 2, j}^{n}}{\Delta}
\label{spl1}\\
\frac{C_{i, j}^ {n + 1/2} - C_{i, j}^{*}}{\Delta t} + \frac{\phi_{i, j + 1 / 2}^{*} - \phi_{i, j - 1 / 2}^{*}}{\Delta} = C_c \frac{v_{i, j + 1 / 2}^{n} - v_{i, j - 1 / 2}^{n}}{\Delta}
\label{spl2}\\
C_c = \begin{cases}
      1, & \text{if}\ C_{i, j}^{n - 1 / 2} > 0.5 \\
      0, & \text{if}\ C_{i, j}^{n - 1 / 2} \leq 0.5
\end{cases}
\label{Eq:PLIC-VOF_dis}
\end{gather}
where $\phi_{i + 1/2, j}$ is the volume flux of the reference phase through the right cell face and $\Delta$ the grid spacing. The right-hand-side of Eqs. (\ref{spl1},\ref{spl2}) represents a compression or expansion term, since each component of the velocity field is not divergence-free. The step function $C_c$ is used to guarantee exact mass conservation for a multi-dimensional problem.

\subsection{Adaptive mesh refinement (AMR)}

\begin{figure}
\centering
\begin{subfigure}[b]{0.8\textwidth}
\centering
\includegraphics[width=\textwidth]{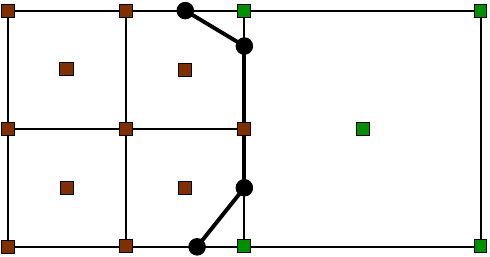}
\end{subfigure}
\caption{The EBIT method with AMR.}
\label{Fig_ebit_amr}
\end{figure}

An efficient adaptive mesh refinement (AMR) technique, which is based on a wavelet decomposition \cite{Popinet_2015_302} of the specified field variables, is implemented in Basilisk, which allows the solution of the flow field at high resolution only in the relevant parts of the domain, reducing in this way the computational cost of the simulation. 

Due to the restriction on the number of markers on each edge, the mesh refinement near the interface should be handled carefully. For the particular situation where the cells on the two sides of the interface are on different levels, as in Fig.~\ref{Fig_ebit_amr}, there are two markers on the same edge of the grid cell on the right. This instance violates the basic restriction of the current implementation of the EBIT method, and it is not allowed.

To avoid this inconsistency, we consider a simple strategy, that is to refine the cells within the $3 \times 3$ stencil of each interfacial cell to the maximum allowable level. With this assumption and the timestep limitation due to the maximum allowable $CFL$ number, which is equal to 1, in the EBIT method, the interface will not be advected between two grid cells at different resolution levels. In principle, this refinement strategy should be less efficient than that based on curvature, however the numerical results of the next section show that the efficiency is still comparable to that based on curvature when other criteria of refinement, such as velocity gradients, are introduced into the AMR strategy.

\section{Numerical results and discussion} \label{Numerical results and discussion}
\subsection{Stagnation flow}
\begin{figure}
\begin{center}
\begin{tabular}{ccc}
\includegraphics[width=0.3\textwidth]{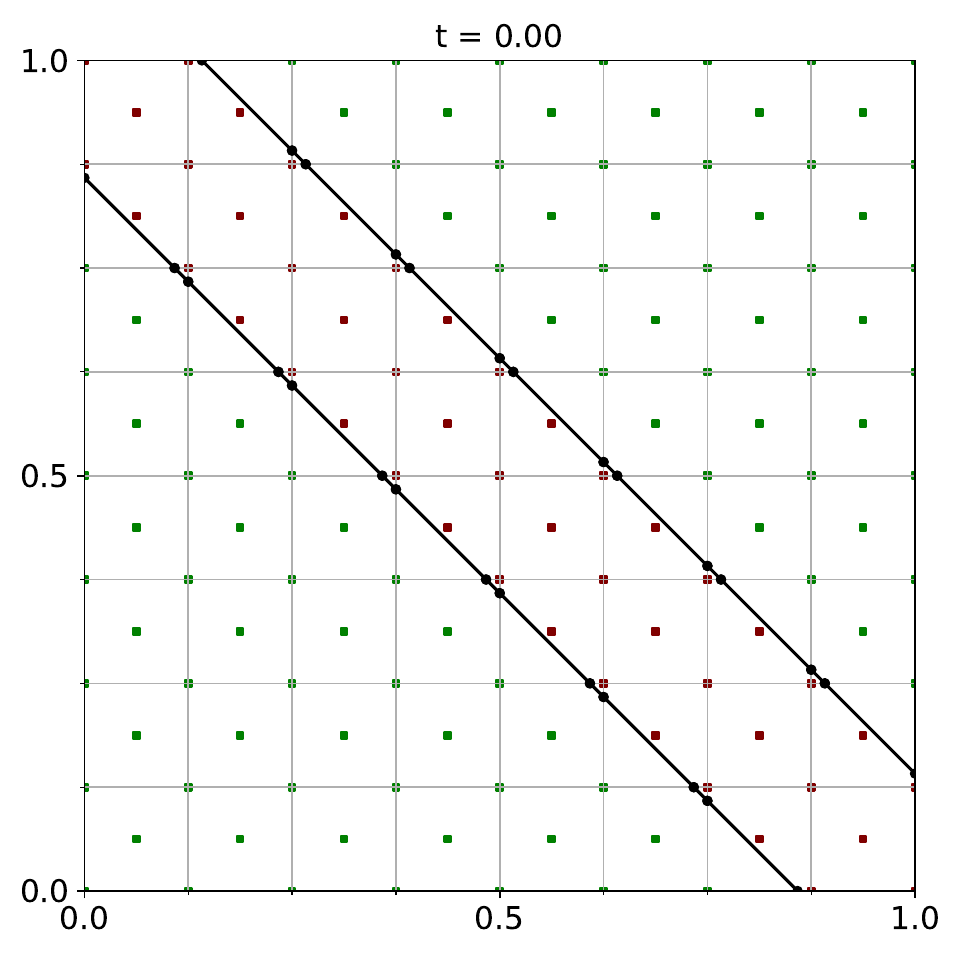} &
\includegraphics[width=0.3\textwidth]{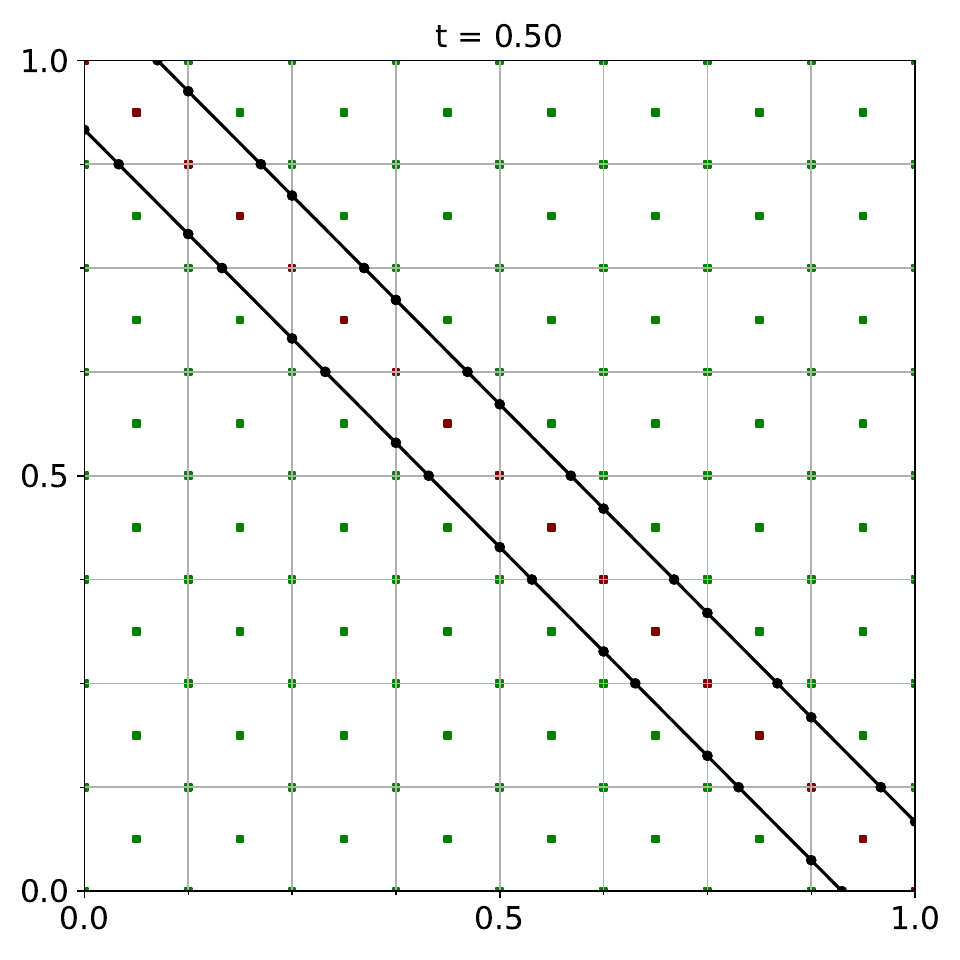} &
\includegraphics[width=0.3\textwidth]{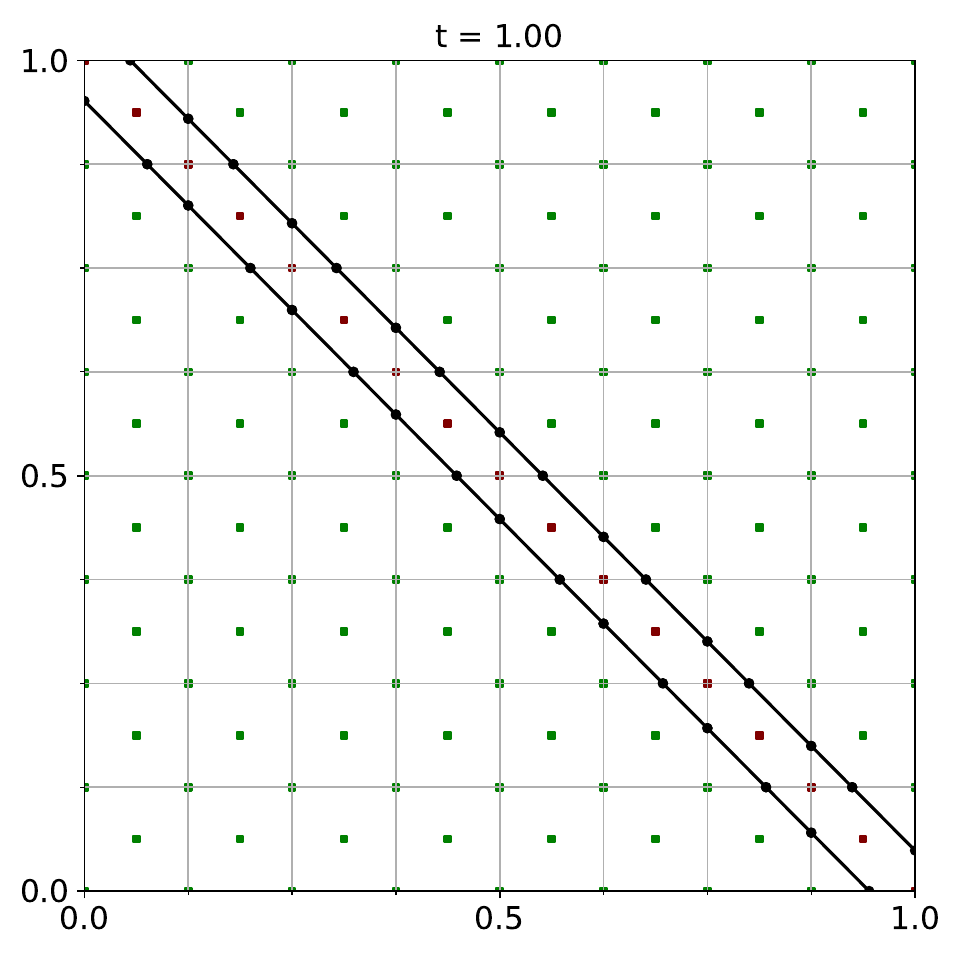}
\\
(a) & (b) & (c)
\end{tabular}
\end{center}
\caption{Stagnation flow: interface lines and Color Vertex (reference phase is denoted by brown squares) at times $t = 0, 0.5, 1$.}
\label{Fig_stagna_intfs}
\end{figure}

In this test two parallel linear interfaces oriented at $45$ degrees are placed inside the unit square domain. The initial thickness of the film is $D = 0.2$ (see Fig.~\ref{Fig_stagna_intfs}a). A stagnation flow $(u, v) = (0.5 - y, 0.5 - x)$ with a stagnation point at $(0.5, 0.5)$ is applied, so that the reference phase is drained out from the computational domain along a diagonal direction and the thickness of the film decreases. Eventually, a sub-cell film is formed inside the domain.
We consider this test only to show the capability of the EBIT method to preserve a sub-cell structure, therefore we use a rather coarse mesh with resolution $N_x \times N_x = 8 \times 8$ and a constant timestep $\Delta t = 1/64$ so that $CFL = u_{max} \Delta t / \Delta = 0.0625$.

The interface lines and the Color Vertex distribution at several times are shown in Fig.~\ref{Fig_stagna_intfs}. Initially, the film is thick enough so that there is no sub-cell structure (Fig.~\ref{Fig_stagna_intfs}a). As the drainage goes on, the film becomes thinner than the diagonal of the cell (Fig.~\ref{Fig_stagna_intfs}b) and it is correctly preserved by the EBIT method even at later times (Fig.~\ref{Fig_stagna_intfs}c).

\subsection{Translation with uniform velocity}

In this test a circular interface of radius $R=0.15$ and center at $(0.25, 0.75)$ is placed inside the unit square domain. The domain is meshed with $N_x \times N_x$ square cells of size $\Delta =1/N_x$, where $N_x = 32, 64, 128, 256, 512$. A uniform and constant velocity field $(u, v)$ with $u = -v$ is applied, so that the interface is advected along a diagonal direction. At halftime $t = 0.5\,T$ the center reaches the position $(0.75, 0.25)$, 
the velocity field is then reversed and the circular interface should return to its initial position at $t = T = 1$ with no distortion.

We use this case to test the new EBIT method, where the unaligned markers are computed by a circle fit. The accuracy and mass conservation of the method are measured by the area, shape, and symmetric difference errors. The area error $E_{area}$ is defined as the absolute value of the relative difference between the area occupied by the reference phase at the initial time $t=0$ and that at $t=T$
\begin{equation}
E_{area} = \frac{|A(T) - A(0)|}{A(0)}
\label{Eq_error_surface}
\end{equation}
The shape error, in a $L^\infty$ norm, is defined as the maximum distance between any marker $\boldsymbol{x}_i$ on the interface and the corresponding closest point on the analytical solution
\begin{equation}
E_{shape} = \max\limits_i |dist (\boldsymbol{x}_i)|, \quad
dist (\boldsymbol{x}_i) = \sqrt{(x_i - x_c)^2 + (y_i - y_c)^2} - R
\label{Eq_error_shape}
\end{equation}
where $(x_c, y_c)$ are the coordinates of the circle center and $R$ its radius. 
For all kinematic tests discussed in the following sections, the shape error is evaluated at the end of the advection.

For two domains, $A$ and $B$, the symmetric difference $A \bigtriangleup B$ is defined as follows
\begin{equation}
A \bigtriangleup B = (A \cup B) \,\backslash\, (A \cap B)
\label{Eq_sym_diff}
\end{equation}

In our test, the interface reconstructions at the beginning and at the end of the simulation are set to be $A$ and $B$, respectively. We use the area of symmetric difference
\begin{equation}
E_{sym} = |A \bigtriangleup B| = |A| + |B| - 2 |A \cap B|
\label{Eq_sym_diff_error}
\end{equation}
to measure the accuracy of an interface advection method.

\begin{figure}
\begin{center}
\begin{tabular}{cc}
\includegraphics[width=0.45\textwidth]{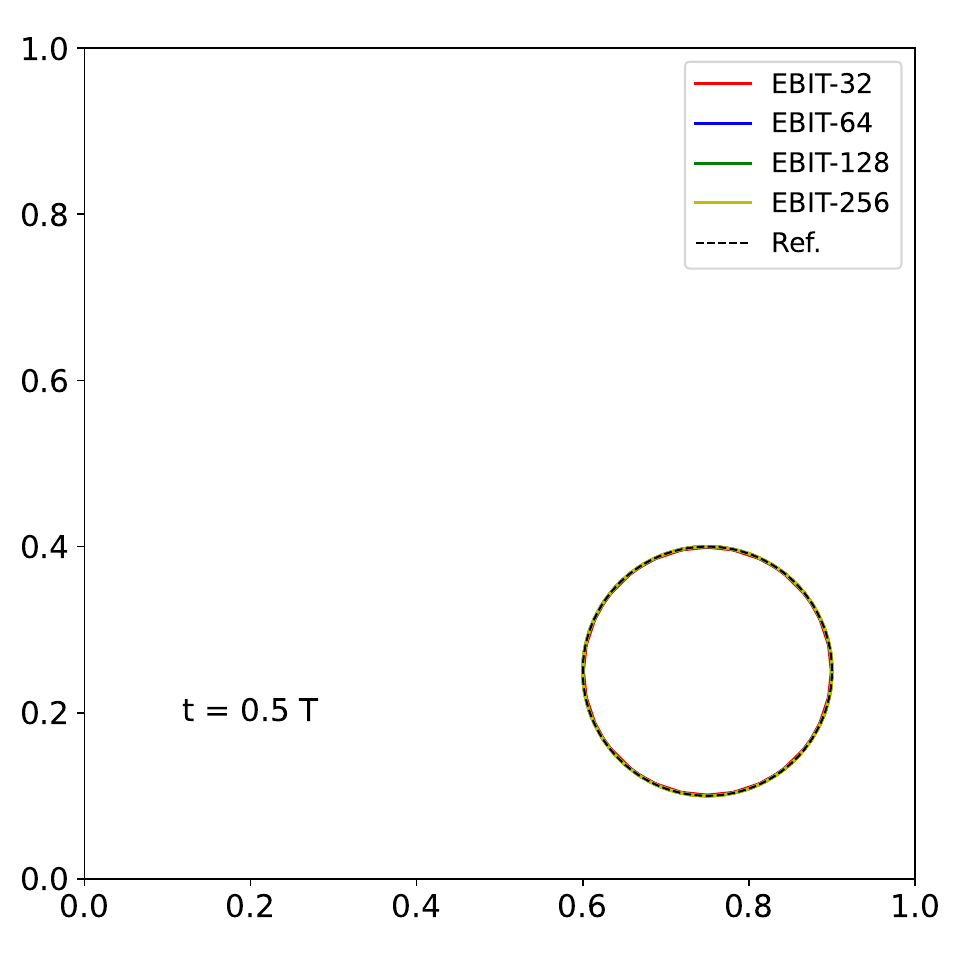} &
\includegraphics[width=0.45\textwidth]{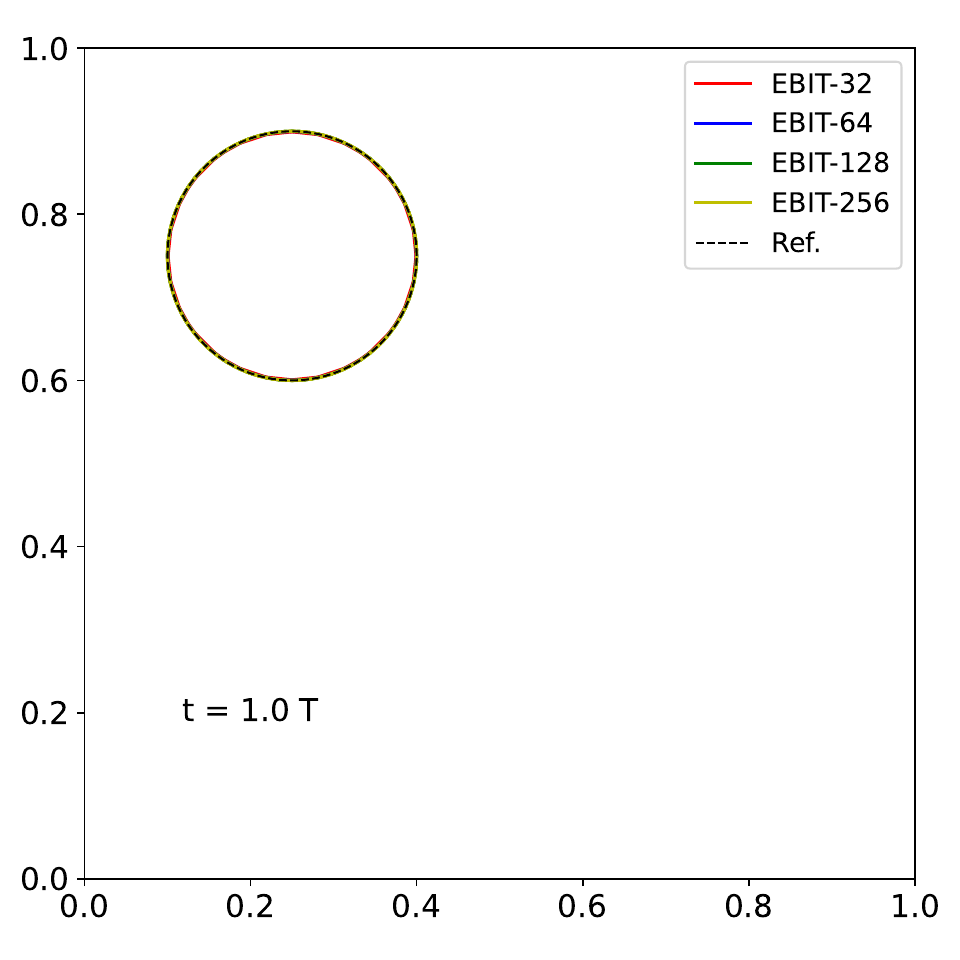}\\
(a) & (b) 
\end{tabular}
\end{center}
\caption{Translation test with the new EBIT method at different resolutions: (a) interface lines at halftime; (b) interface lines at the end of the simulation.}
\label{Fig_translation_intf}
\end{figure}

To initialize the markers on the grid lines we first compute the signed distance \eqref{Eq_error_shape} on the cell vertices, then we use a root-finding routine, when the sign of the distance is opposite on the two endpoints of a cell side, to calculate the position of a marker. Hence, there is a small numerical error in the initial data, that accumulates as the interface is translated. However, because of the circle fit in the EBIT method this error remains rather limited during the translation.
We employ a relatively small $CFL$ number $CFL = (u\,\Delta t)/\Delta = 0.125$. The interface lines at halftime and at the end of the simulation are shown in Fig.~\ref{Fig_translation_intf} for different mesh resolutions. 

\begin{figure}
\begin{center}
\begin{tabular}{cc}
\includegraphics[width=0.45\textwidth]{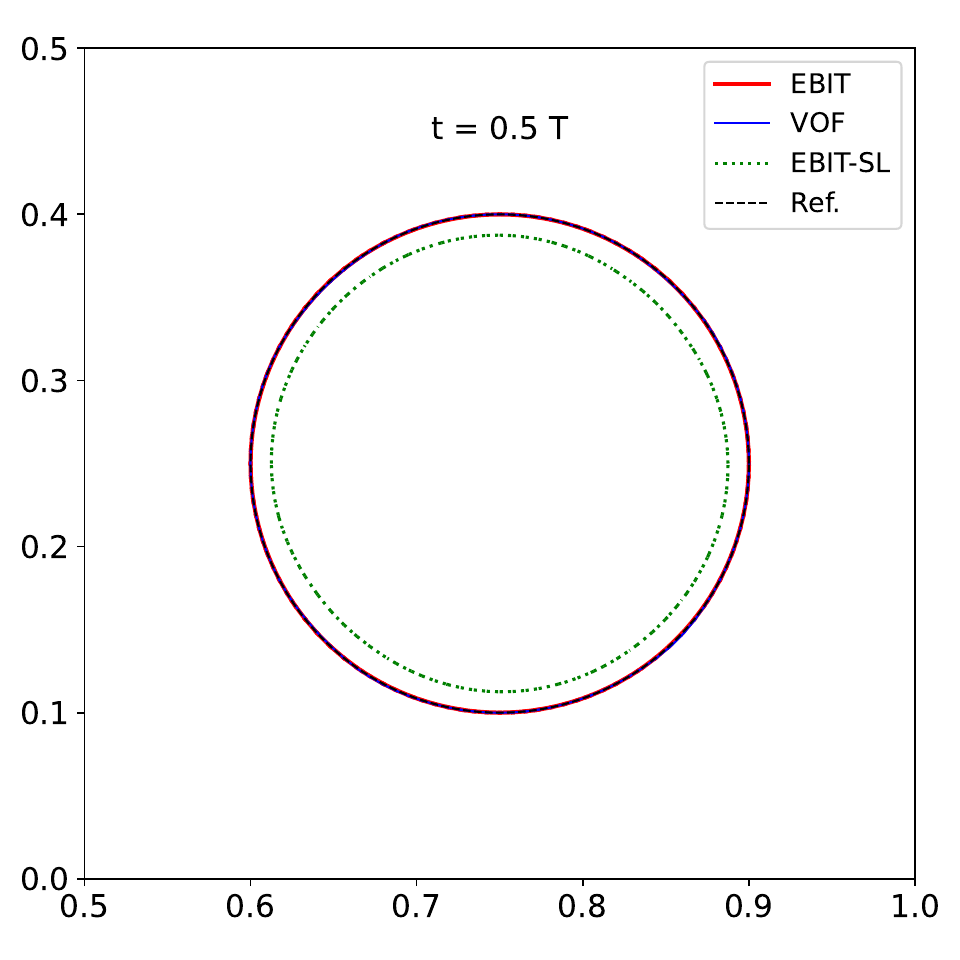} &
\includegraphics[width=0.45\textwidth]{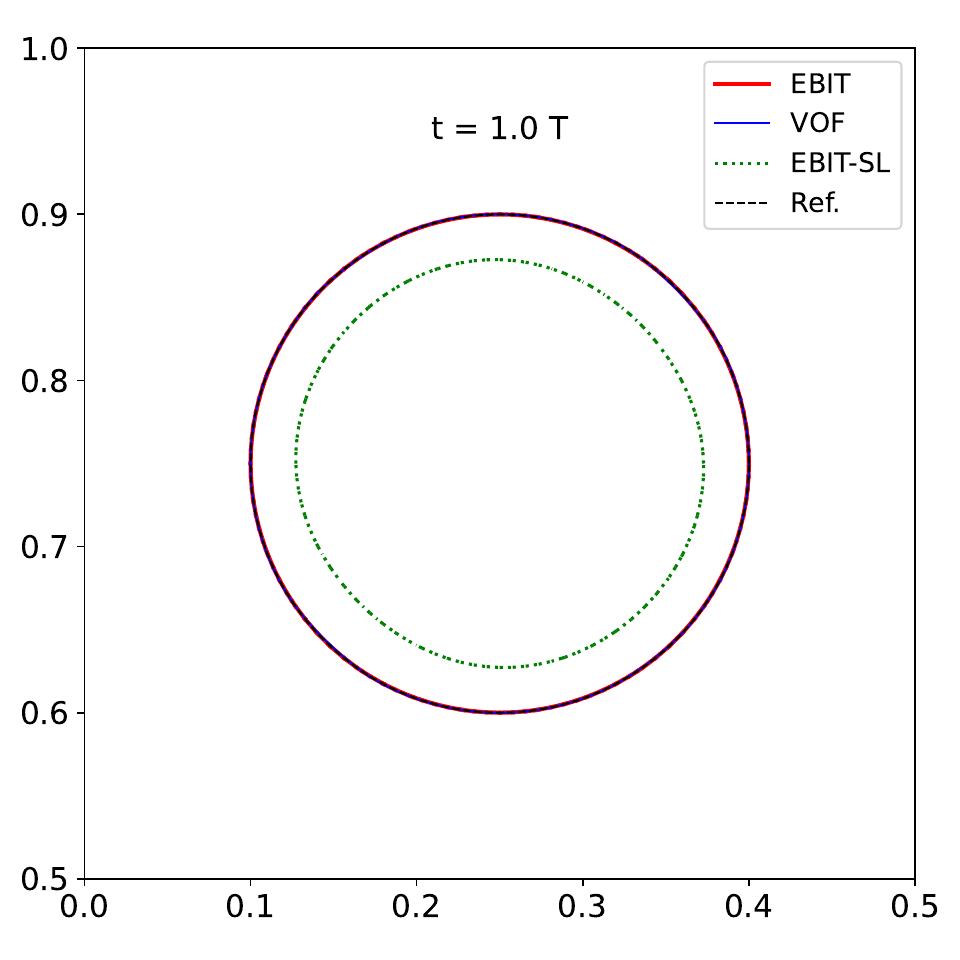}\\
(a) & (b) 
\end{tabular}
\end{center}
\caption{Translation test with different methods at resolution $N_x=128$: (a) interface lines at halftime; (b) interface lines at the end of the simulation.}
\label{Fig_translation_intf_cmp}
\end{figure}
\begin{figure}
\begin{center}
\begin{tabular}{ccc}
\includegraphics[width=0.33\textwidth]{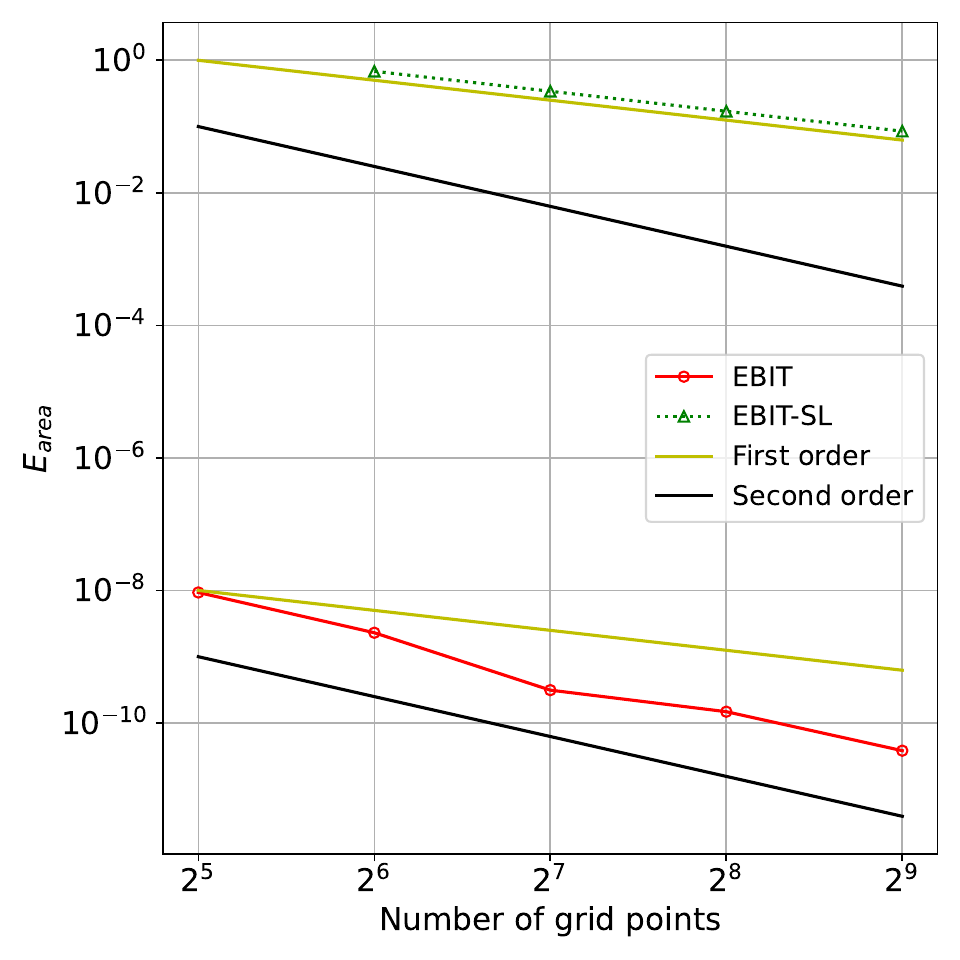} &
\includegraphics[width=0.33\textwidth]{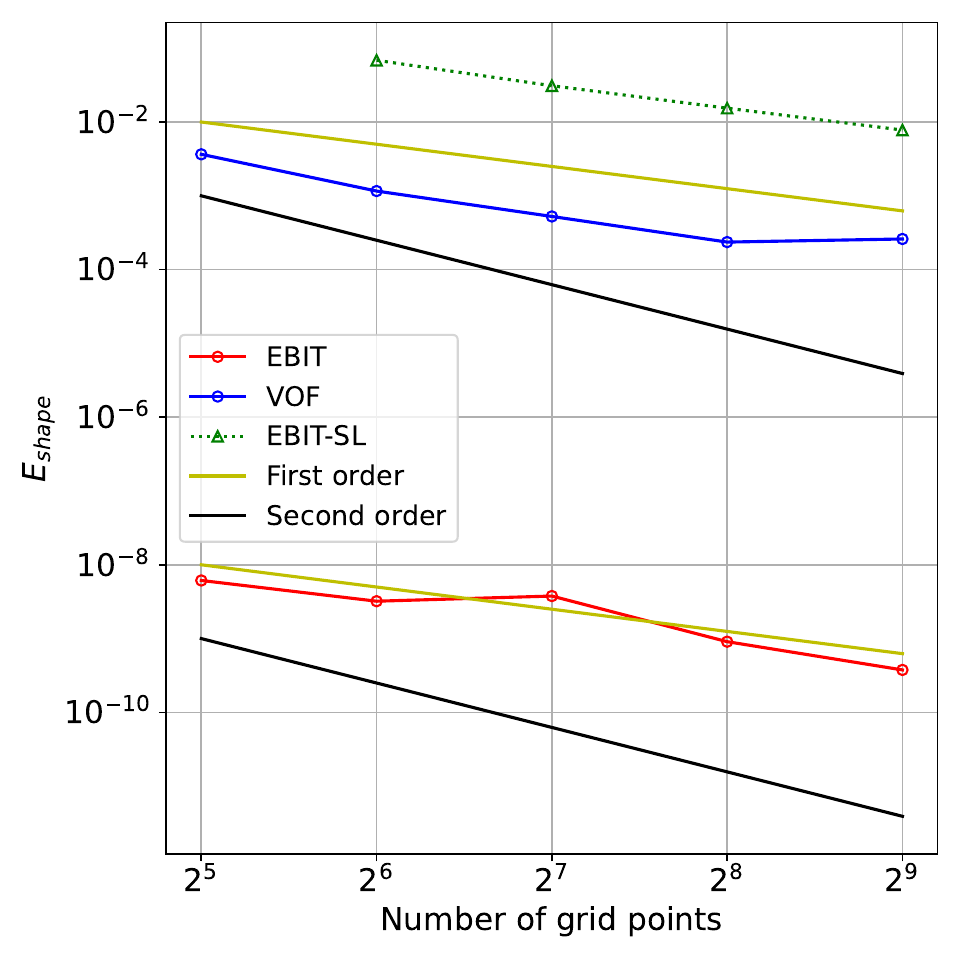} &
\includegraphics[width=0.33\textwidth]{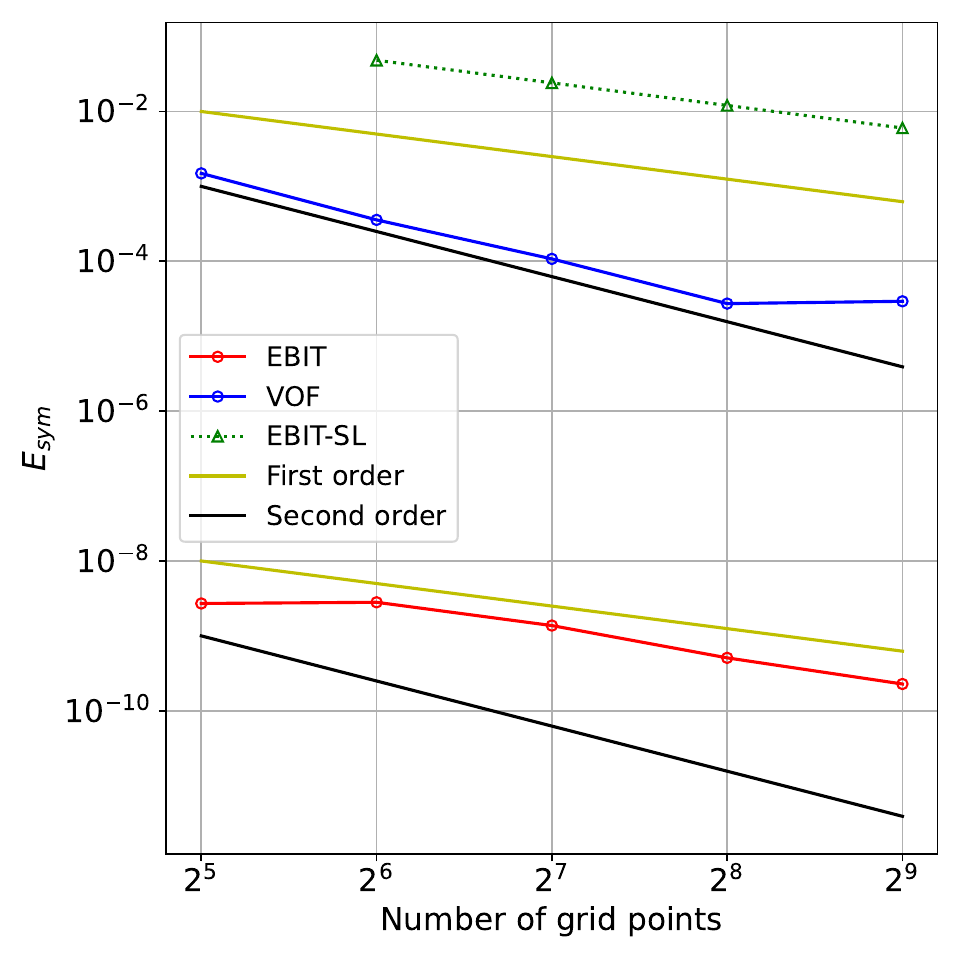}\\
(a) & (b) & (c)
\end{tabular}
\end{center}
\caption{Errors in the translation test for different methods as a function of grid resolution: (a) area error $E_{area}$; (b) shape error $E_{shape}$; (c) symmetric difference error $E_{sym}$.}
\label{Fig_translation_error}
\end{figure}

The interface lines are also calculated with different reconstruction and advection methods for a medium mesh resolution, $N_x = 128$, and are shown in Fig.~\ref{Fig_translation_intf_cmp}. Both the new EBIT method and the PLIC-VOF method maintain rather well the circular shape of the interface during the translation. On the other hand, the first version of the EBIT method, that considers a straight line approximation (SL) for the calculation of the position of unaligned markers, loses mass continuously during the simulation.

The area errors $E_{area}$ and the shape errors $E_{shape}$ are listed in Table~\ref{Tab_translation_error} and are shown in Fig.~\ref{Fig_translation_error} for the different methods. 
For the new EBIT method we observe a second-order convergence rate for the area error and approximately a first-order convergence rate for both the shape error and symmetric difference error, as shown in Fig.~\ref{Fig_translation_error}. For the first version of the EBIT method (SL), both the area, shape and symmetric difference errors are much larger. Furthermore, this method loses all the reference phase at the lowest grid resolution, $N_x = 32$ (this fact is denoted by ``NA'' in Table~\ref{Tab_translation_error}). 

\begin{table}[hbt!]
\footnotesize
\caption{Mesh convergence study for the translation test.}
\centering
\begin{tabular}{cc|ccccc}
\hline 
 &$N_x$ &32 & 64 & 128 & 256 & 512\\ 
\hline 
EBIT&$E_{area}$ & $9.32\times 10^{-9}$ & $2.30\times 10^{-9}$ & $3.13\times 10^{-10}$ & $1.48\times 10^{-10}$ & $3.82\times 10^{-11}$\\ 
&$E_{shape}$ & $6.13\times 10^{-9}$ & $3.21\times 10^{-9}$ & $3.77 \times 10^{-9}$ & $9.08 \times 10^{-10}$ & $3.76\times 10^{-10}$\\
&$E_{sym}$ & $2.75\times 10^{-9}$ & $2.81\times 10^{-9}$ & $1.37 \times 10^{-9}$ & $5.09 \times 10^{-10}$ & $2.28\times 10^{-10}$\\
\hline 
VOF&$E_{shape}$ & $3.88\times 10^{-3}$ & $1.06\times 10^{-3}$ & $4.72 \times 10^{-4}$ & $2.59 \times 10^{-4}$ & $2.68\times 10^{-4}$\\
&$E_{sym}$ & $1.49\times 10^{-3}$ & $3.57\times 10^{-4}$ & $1.08 \times 10^{-4}$ & $2.73 \times 10^{-5}$ & $2.93\times 10^{-5}$\\
\hline 
EBIT-SL&$E_{area}$ & NA & $6.83\times 10^{-1}$ & $3.41\times 10^{-1}$ & $1.70\times 10^{-1}$ & $8.52\times 10^{-2}$\\ 
&$E_{shape}$ & NA & $6.86\times 10^{-2}$ & $3.10 \times 10^{-2}$ & $1.54 \times 10^{-2}$ & $7.78\times 10^{-3}$\\
&$E_{sym}$ & NA & $4.82\times 10^{-2}$ & $2.41 \times 10^{-2}$ & $1.20 \times 10^{-2}$ & $6.02\times 10^{-3}$\\
\hline 
\end{tabular}
\label{Tab_translation_error}
\normalsize
\end{table}

For the PLIC-VOF method, the mass conservation is accurate to machine error and it is not shown in the figure, while the shape error is evaluated with the two endpoints of the PLIC-VOF reconstruction in each cut cell. For this error we observe in Fig.~\ref{Fig_translation_error} a first-order convergence rate. 
But for the symmetric difference error, we observe a second-order convergence rate. The lower order of convergence rate of the shape error may due to the discontinuous feature of the PLIC-VOF reconstruction on the cell edge and to the maximum norm we used for the evaluation of shape error.
Due to the fact that in the new EBIT method we are fitting a circle with a circle, the shape and symmetric difference errors obtained with the PLIC-VOF method are larger than those of the new EBIT method.

\begin{table}[hbt!]
\caption{Performance in weak scaling on IRENE.}
\centering
\begin{tabular}{ccccc}
\hline 
 Number of cells & $N_c$ & $T_s$ ($10^{-3}$ s/step) & $S$/Ideal  & $E$ \\ 
\hline 
$128^2$ & $4$ & $2.16$ & $1 / 1$ & $1$ \\ 
$256^2$ & $16$ & $2.62$ & $3.3 / 4$ & $0.82$ \\ 
$512^2$ & $64$ & $2.80$ & $12.3 / 16$ & $0.77$ \\ 
$1024^2$ & $256$ & $3.26$ & $42.4 / 64$ & $0.66$ \\ 
$2048^2$ & $1024$ & $3.69$ & $149.7 / 256$ & $0.58$ \\ 
\hline 
\end{tabular}
\label{Tab_weak_scaling}
\end{table}

We also use this simple case to test the scalability of the EBIT method. The simulations for the scalability test were run on ROME partition of the IRENE supercomputer. This partition contains 2286 nodes and each node contains 128 CPU cores. Only one thread is used for each CPU core, and each thread is mapped to an MPI process. 
For a Cartesian mesh, Basilisk employs a common domain decomposition strategy to subdivide the computational domain and distributes one subdomain to each MPI process. Since the markers are bound to a cell, as the interface moves across different subdomains, the markers are distributed to different processes automatically by the build-in subroutines of Basilisk.

For most of the 2D simulations, it is more reasonable to use a large number of cores when a very high resolution is needed. Hence we are more interested in the ``weak scaling'' performance of the EBIT method, where we vary the number of CPU cores and keep the number of cells distributed to each core fixed. A subdomain grid of $64 \times 64$ is used for test, and a minor difference of performance is observed when we further increase the resolution of the subdomain. The scalability is measured by the speedup $S(N_c)$ and efficiency $E(N_c)$:
\begin{equation}
S(N_c) = \frac{N_c T_s (N_{c,ref})}{N_{c, ref} T_s (N_c)}, \quad E(N_c) = \frac{N_{c,ref} }{N_c} S (N_c)
\label{Eq_speedup}
\end{equation}
where $N_c$ is the number of cores and $T_s$ the wall time per advection step. The number of cores used for reference ($N_{c, ref}$) is $4$. Table~\ref{Tab_weak_scaling} shows the overall performance of the EBIT method in weak scaling for five different runs.
For the ideal case without the overhead of communications and synchronizations among processes, the wall time should remain constant when the number of cores is increased. But due to the overheads in a real simulation, the wall time slightly increases as we increase the number of cores, as shown in Table~\ref{Tab_weak_scaling}. Consequently, the overheads incur a degradation of the speedup and efficiency and will eventually affect the overall scalability. The code is 149.7 times faster with 1024 cores, but the efficiency decreases to only $58\%$. Since this is our very first parallel study of the EBIT method, further optimization of the communications and synchronizations can still be performed to enhance its scalability.

\subsection{Single vortex}

The single vortex test was designed to test the ability of an interface tracking method to follow the evolution in time of an interface that is highly stretched and deformed \cite{Rider_1998_141}. A circular interface of radius $R=0.15$ and center at $(0.5, 0.75)$ is placed inside the unit square domain. 
A divergence-free velocity field $(u, v) = (\partial \phi \big/ \partial y, - \partial \phi \big/ \partial x)$ described by the stream function $\phi = \pi^{-1} \sin^2(\pi x) \sin^2(\pi y) \cos(\pi t/T)$ is imposed in the domain. The cosinusoidal time-dependence slows down and reverses the flow, so that the maximum deformation occurs at $t = 0.5\,T$, then the interface returns to its initial position without distortion at $t = T$. Furthermore, as the value of the period $T$ is increased, a thinner and thinner revolving ligament develops.

The timestep is kept constant in a simulation, and is computed from the maximum value of the velocity, $u_{max}$, at time $t=0$ so that $CFL = u_{max} \,\Delta t / \Delta = 0.125$. The error is again measured by the area, shape and symmetric difference errors.

To obtain the reference solution, we connect with segments an ordered list of $512$ Lagrangian markers, which are placed on the initial interface, and advect them numerically along the flow streamlines. A fourth-order Runge-Kutta method, with an adaptive timestep, in the Python library SciPy is used to solve the ordinary differential equations 
$d \X \big/ d t = \U(x(t), y(t), t)$, which describe the motion of the markers. A user-defined maximum timestep, $\Delta t_{max} = 0.01$, is used in the numerical integration.

\begin{figure}
\begin{center}
\begin{tabular}{cc}
\includegraphics[width=0.45\textwidth]{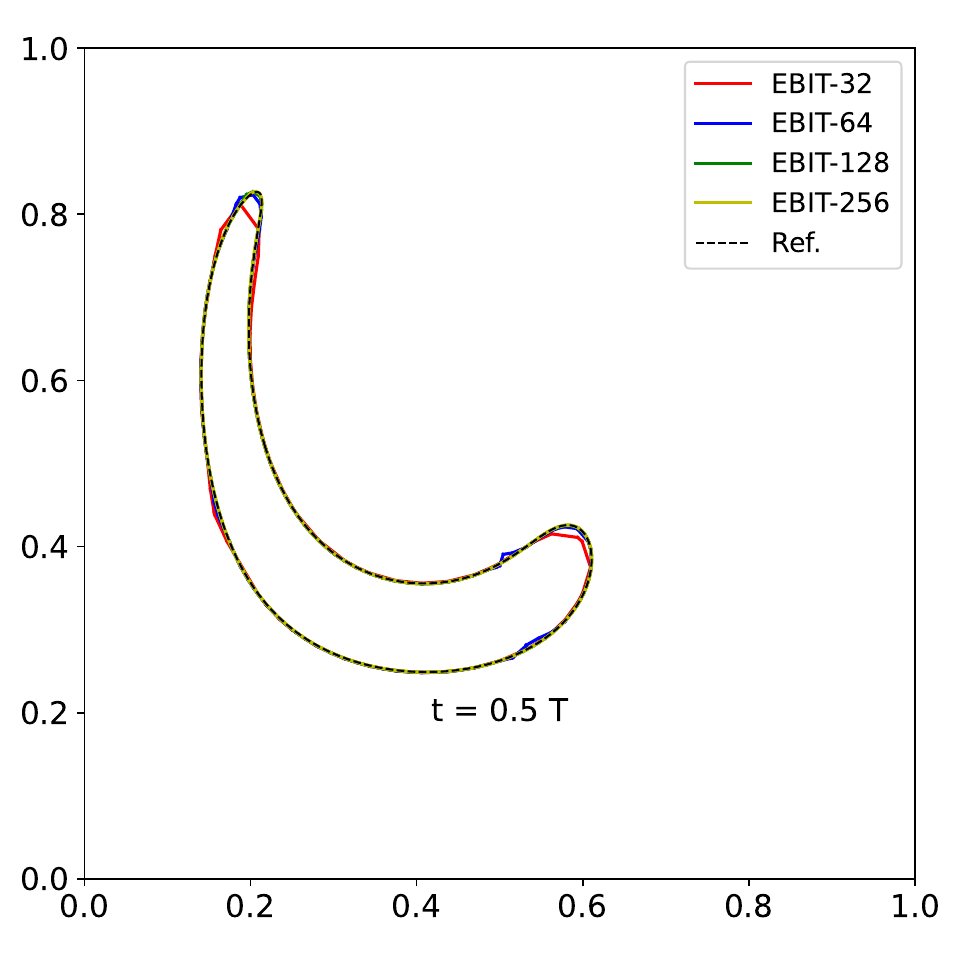} &
\includegraphics[width=0.45\textwidth]{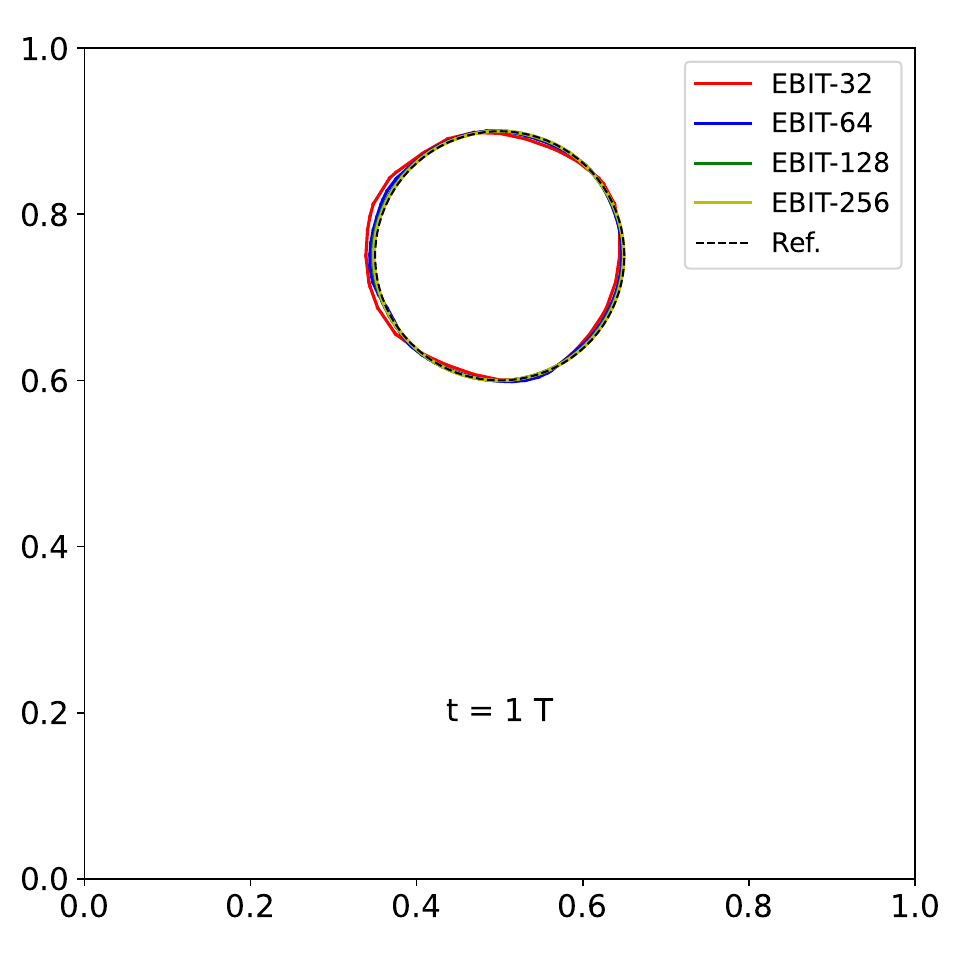}\\
(a) & (b) 
\end{tabular}
\end{center}
\caption{Single vortex test with period $T = 2$ at different resolutions: (a) interface lines at halftime; (b) interface lines at the end of the simulation.}
\label{Fig_vortex_intf_t2}
\end{figure}
\begin{figure}
\begin{center}
\begin{tabular}{cc}
\includegraphics[width=0.45\textwidth]{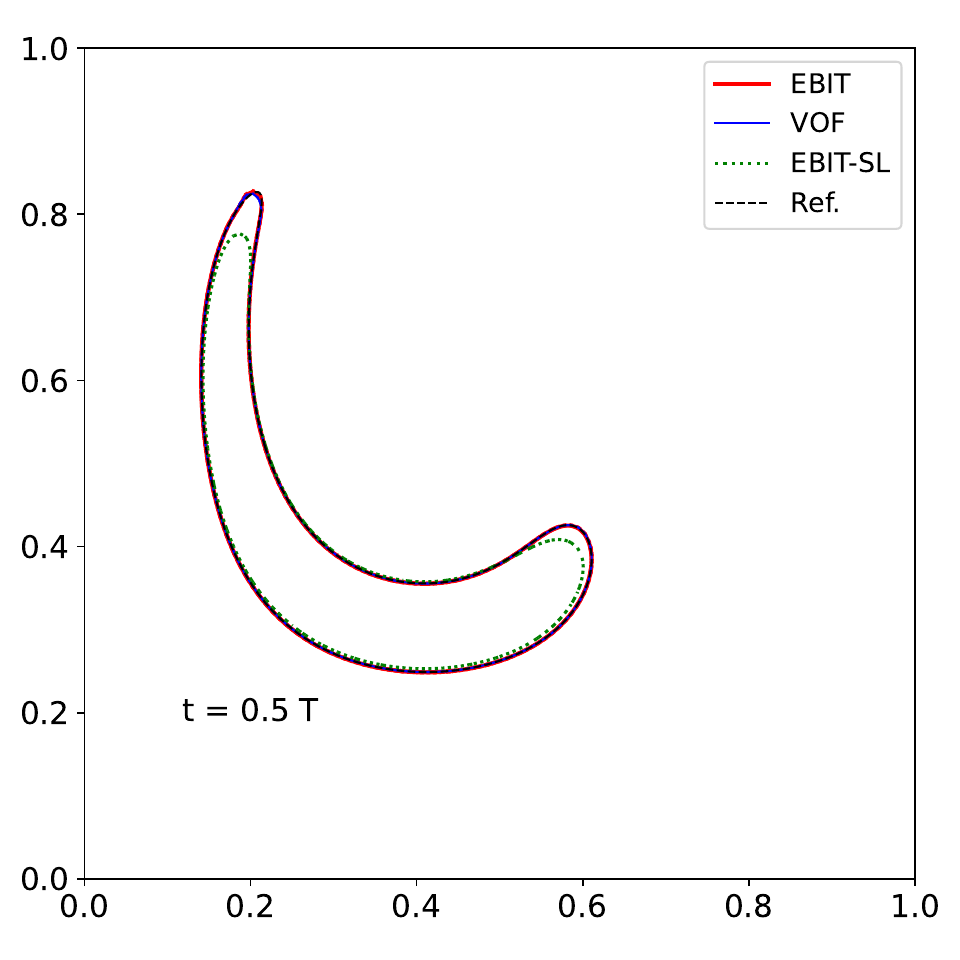} &
\includegraphics[width=0.45\textwidth]{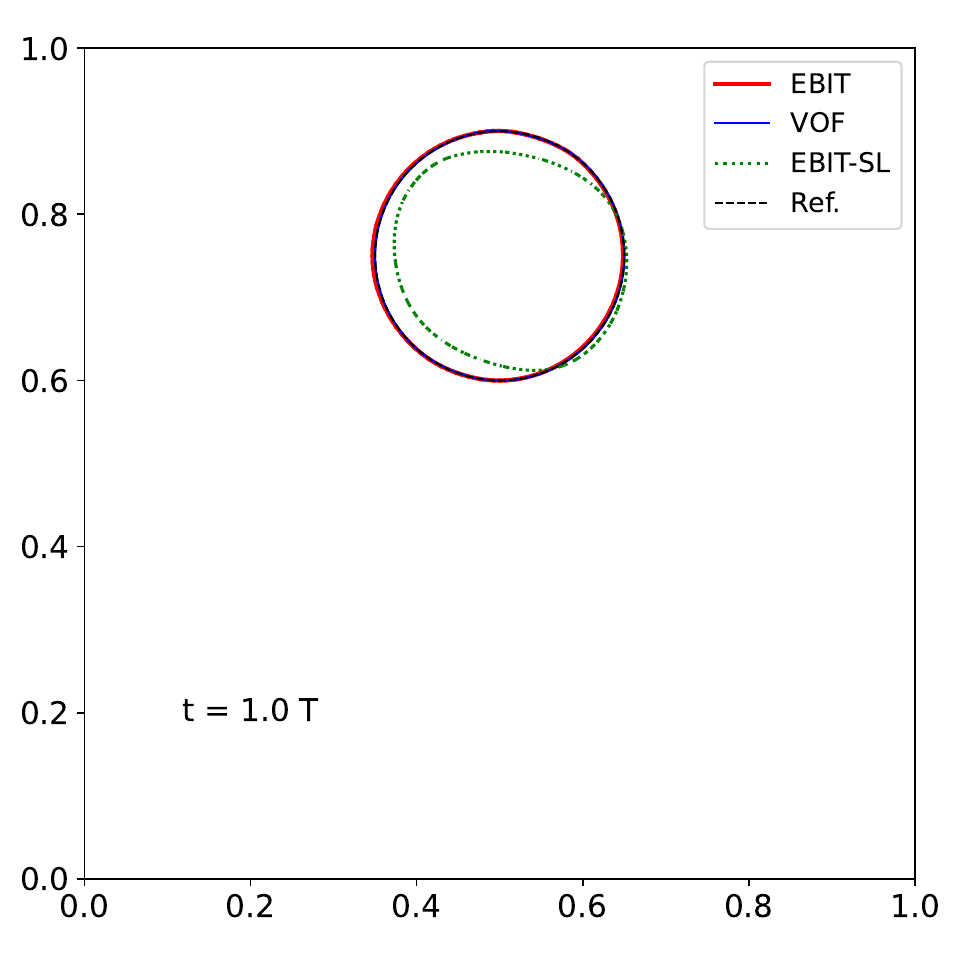}\\
(a) & (b) 
\end{tabular}
\end{center}
\caption{Single vortex test with period $T = 2$ at resolution $N_x = 128$ with different methods: (a) interface lines at halftime; (b) interface lines at the end of the simulation.}
\label{Fig_vortex_intf_t2_cmp}
\end{figure}
\begin{table}[hbt!]
\footnotesize
\caption{Mesh convergence study for the single vortex test with period $T = 2$.}
\centering
\begin{tabular}{cc|cccccc}
\hline 
 &$N_x$& 32 & 64 & 128 & 256 & 512 & 1024\\ 
\hline 
EBIT&$E_{area}$ & $1.69\times 10^{-2}$ & $7.45 \times 10^{-3}$ & $2.62 \times 10^{-3}$ & $1.25 \times 10^{-3}$ & $5.99 \times 10^{-4}$ & $2.87 \times 10^{-4}$\\ 
&$E_{shape}$ & $1.45 \times 10^{-2}$ & $6.74 \times 10^{-3}$ & $3.07 \times 10^{-3}$ & $1.54 \times 10^{-3}$ & $7.75 \times 10^{-4}$ & $3.85 \times 10^{-4}$\\
&$E_{sym}$ & $4.59 \times 10^{-3}$ & $2.16 \times 10^{-3}$ & $1.06 \times 10^{-3}$ & $5.28 \times 10^{-4}$ & $2.62 \times 10^{-4}$ & $1.31 \times 10^{-4}$\\
\hline
VOF &$E_{shape}$ & $8.79 \times 10^{-3}$ & $3.00 \times 10^{-3}$ & $1.17 \times 10^{-3}$ & $4.11 \times 10^{-4}$ & $1.21 \times 10^{-4}$ & $3.38 \times 10^{-5}$\\
&$E_{sym}$ & $3.16 \times 10^{-3}$ & $6.93 \times 10^{-4}$ & $1.43 \times 10^{-4}$ & $3.14 \times 10^{-5}$ & $7.48 \times 10^{-6}$ & $2.11 \times 10^{-6}$\\
\hline 
EBIT-SL &$E_{area}$ & $8.26\times 10^{-1}$ & $3.66 \times 10^{-1}$ & $1.71 \times 10^{-1}$ & $8.19 \times 10^{-2}$ & $4.02 \times 10^{-2}$ & $1.99 \times 10^{-2}$\\ 
&$E_{shape}$ & $1.20 \times 10^{-1}$ & $5.77 \times 10^{-2}$ & $2.68 \times 10^{-2}$& $1.33 \times 10^{-2}$ & $6.99 \times 10^{-3}$ & $3.77 \times 10^{-3}$\\
&$E_{sym}$ & $5.80 \times 10^{-2}$ & $2.80 \times 10^{-2}$ & $1.38 \times 10^{-2}$& $6.82 \times 10^{-3}$ & $3.38 \times 10^{-3}$ & $1.69 \times 10^{-3}$\\
\hline 
\end{tabular}
\label{Tab_vortex_error_t2}
\normalsize
\end{table}
\begin{figure}
\begin{center}
\begin{tabular}{ccc}
\includegraphics[width=0.33\textwidth]{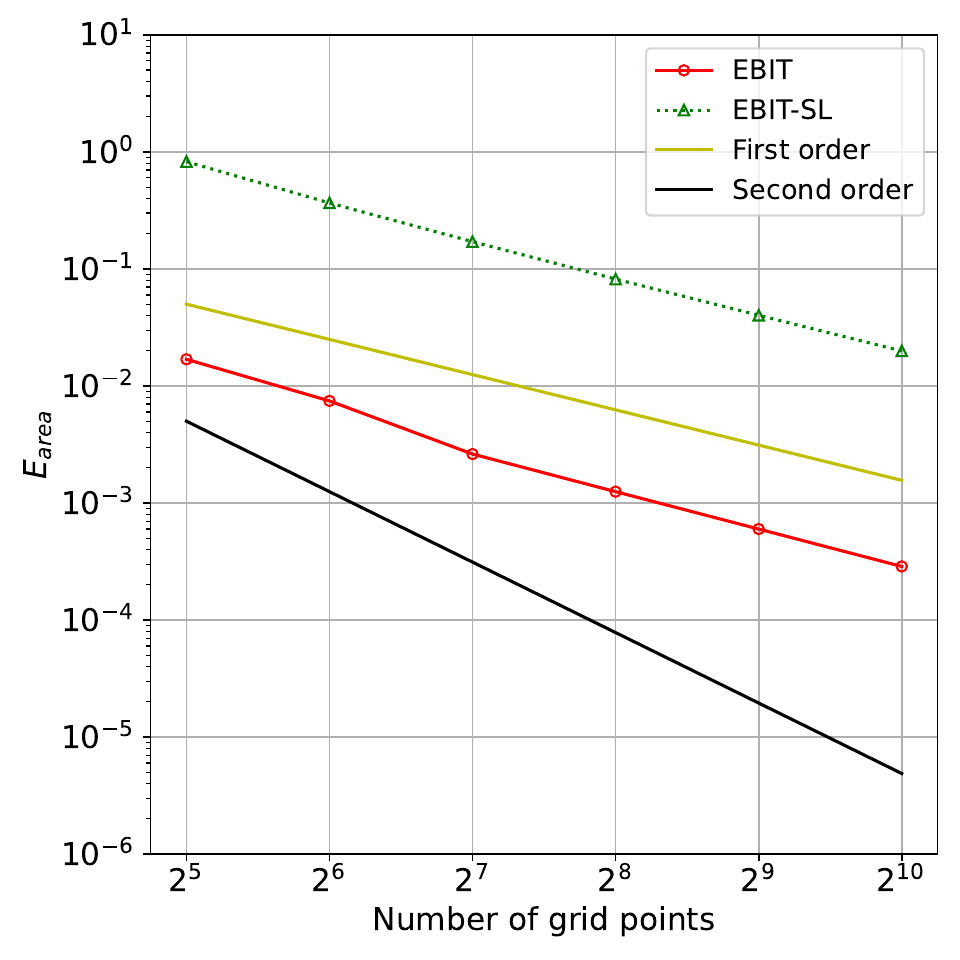} &
\includegraphics[width=0.33\textwidth]{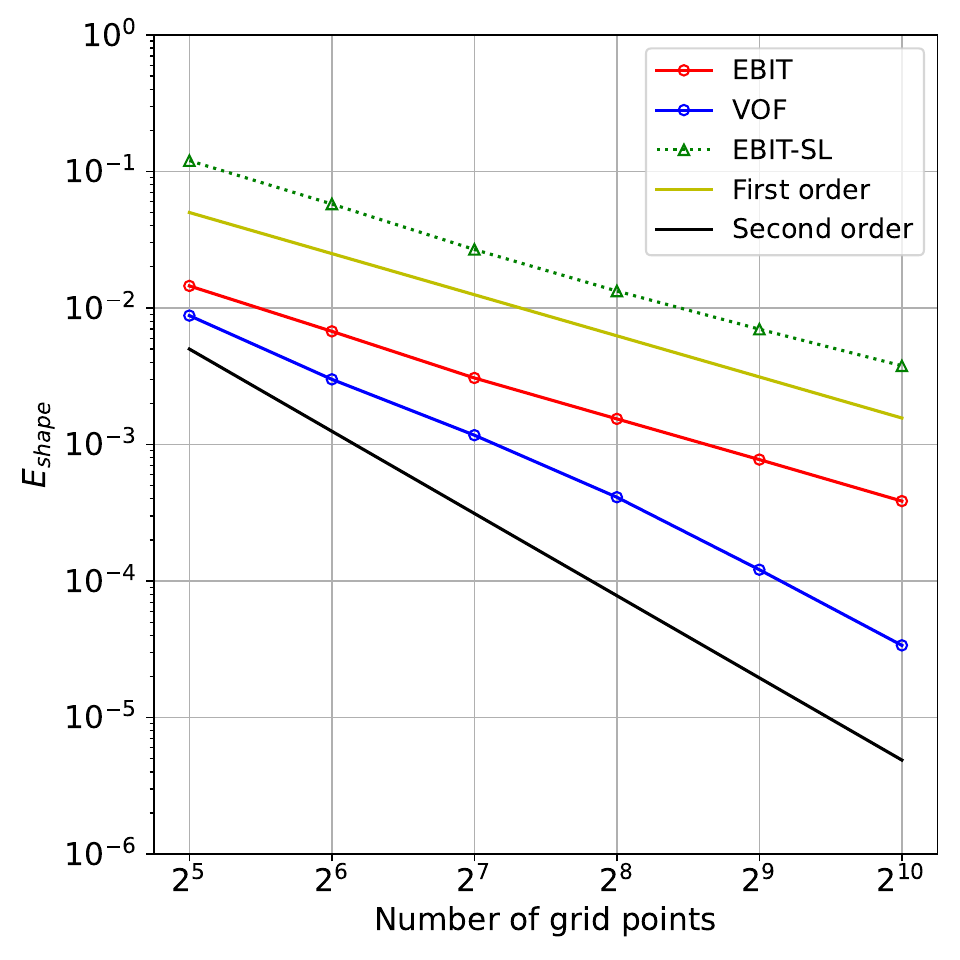} &
\includegraphics[width=0.33\textwidth]{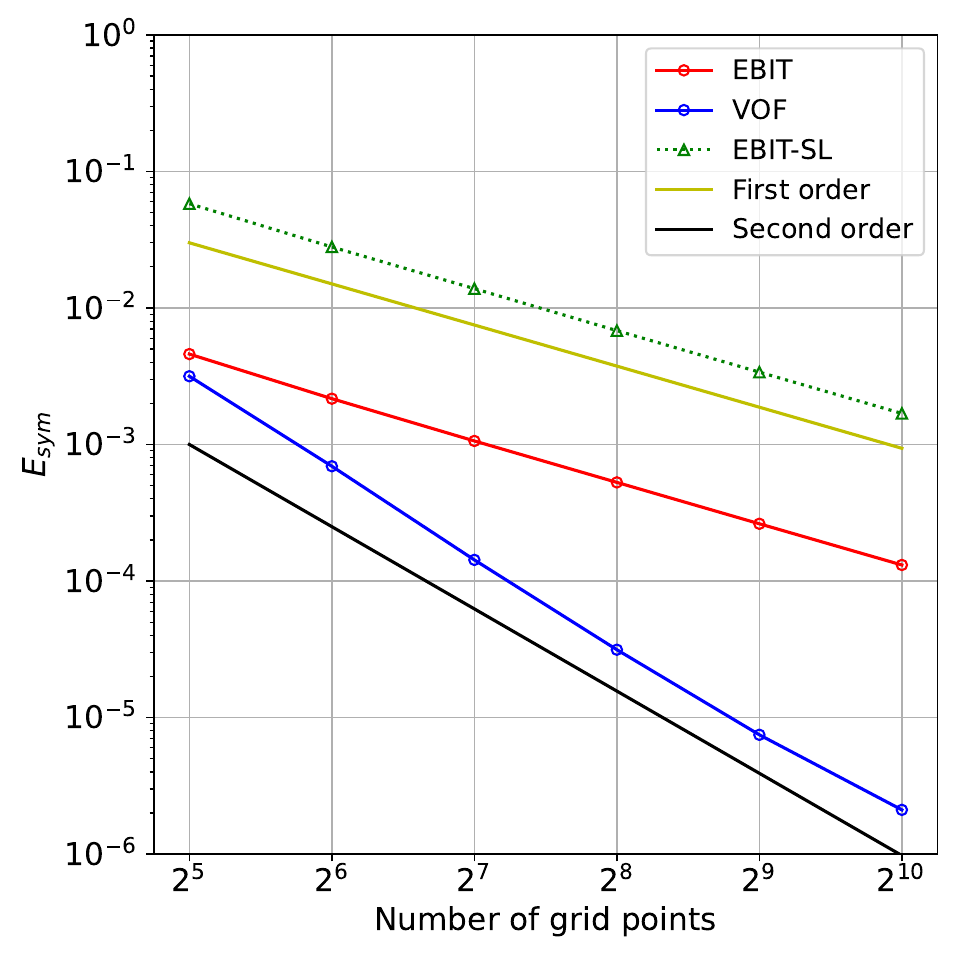}\\
(a) & (b) & (c)
\end{tabular}
\end{center}
\caption{Errors in the single vortex test with period $T=2$ for different methods as a function of grid resolution: (a) area error $E_{area}$; (b) shape error $E_{shape}$; (c) symmetric difference error $E_{sym}$.}
\label{Fig_vortex_error_t2}
\end{figure}

The interface line at maximum deformation and back to its initial position, for the test with period $T = 2$, is shown in Fig.~\ref{Fig_vortex_intf_t2} for different mesh resolutions. Even at the lowest resolution $N_x = 32$, with the new EBIT method we still recover the initial shape and lose little mass. 
The interface line obtained with different methods is shown in Fig.~\ref{Fig_vortex_intf_t2_cmp} for the resolution $N_x = 128$. The results obtained with the new EBIT method and the PLIC-VOF method agree rather well with each other. For the EBIT method with a straight line fit, we observe a considerable amount of area loss.

The area error $E_{area}$, the shape error $E_{shape}$ and the symmetric difference error $E_{sym}$ are listed in Table~\ref{Tab_vortex_error_t2} and are shown in Fig.~\ref{Fig_vortex_error_t2} for the different methods here considered. For the new EBIT method that implements a circle fit, a first-order convergence rate is observed for both errors. The shape errors calculated with the new EBIT method converge less fast than those with the PLIC-VOF method. For the PLIC-VOF method, a second-order convergence rate is observed for the symmetric difference errors, which is higher than that of the shape error.

\begin{figure}
\begin{center}
\begin{tabular}{cc}
\includegraphics[width=0.45\textwidth]{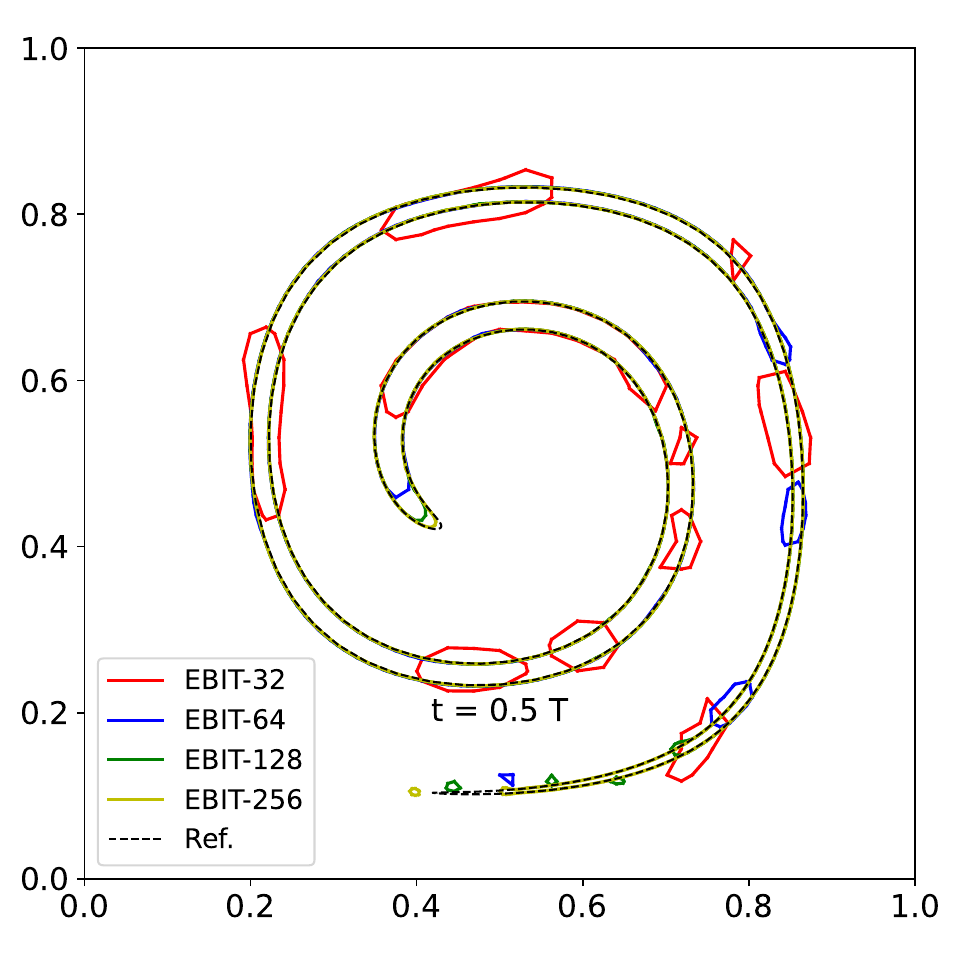} &
\includegraphics[width=0.45\textwidth]{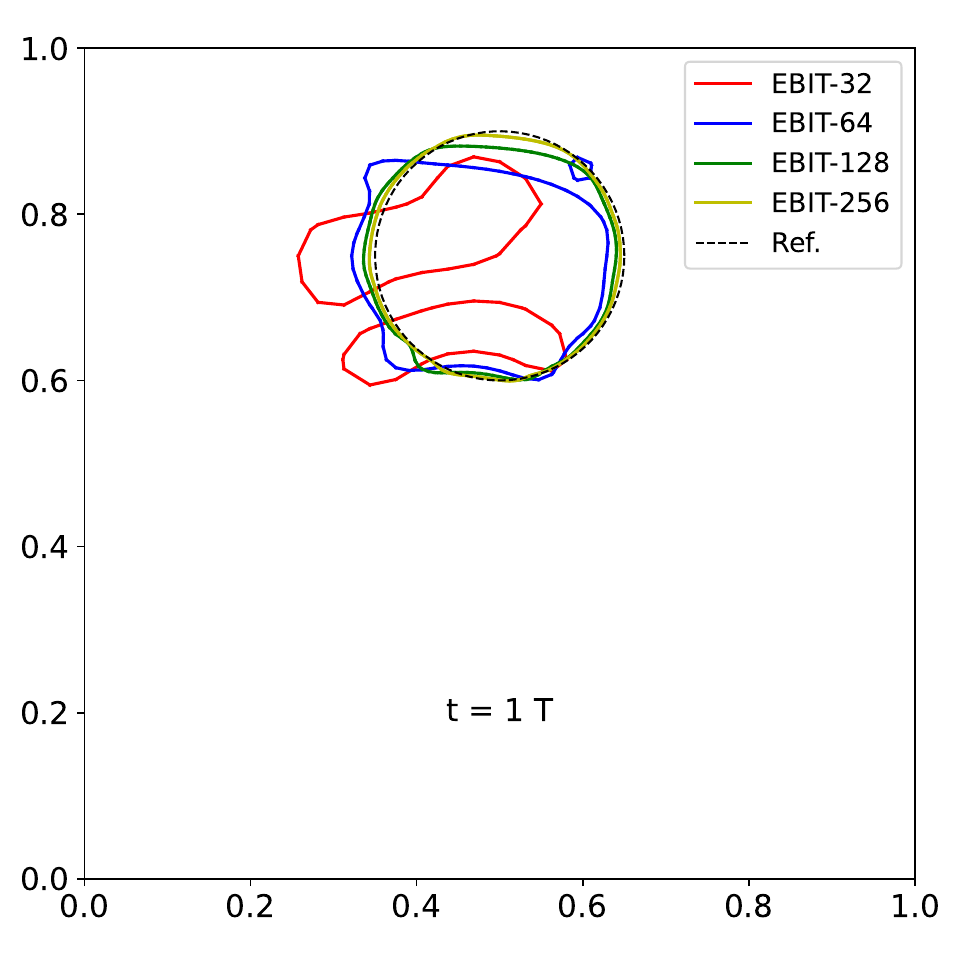}\\
(a) & (b)
\end{tabular}
\end{center}
\caption{Single vortex test with period $T = 8$ at different resolutions: (a) interface lines at halftime; (b) interface lines at the end of the simulation.}
\label{Fig_vortex_intf_t8}
\end{figure}
\begin{figure}
\begin{center}
\begin{tabular}{cc}
\includegraphics[width=0.45\textwidth]{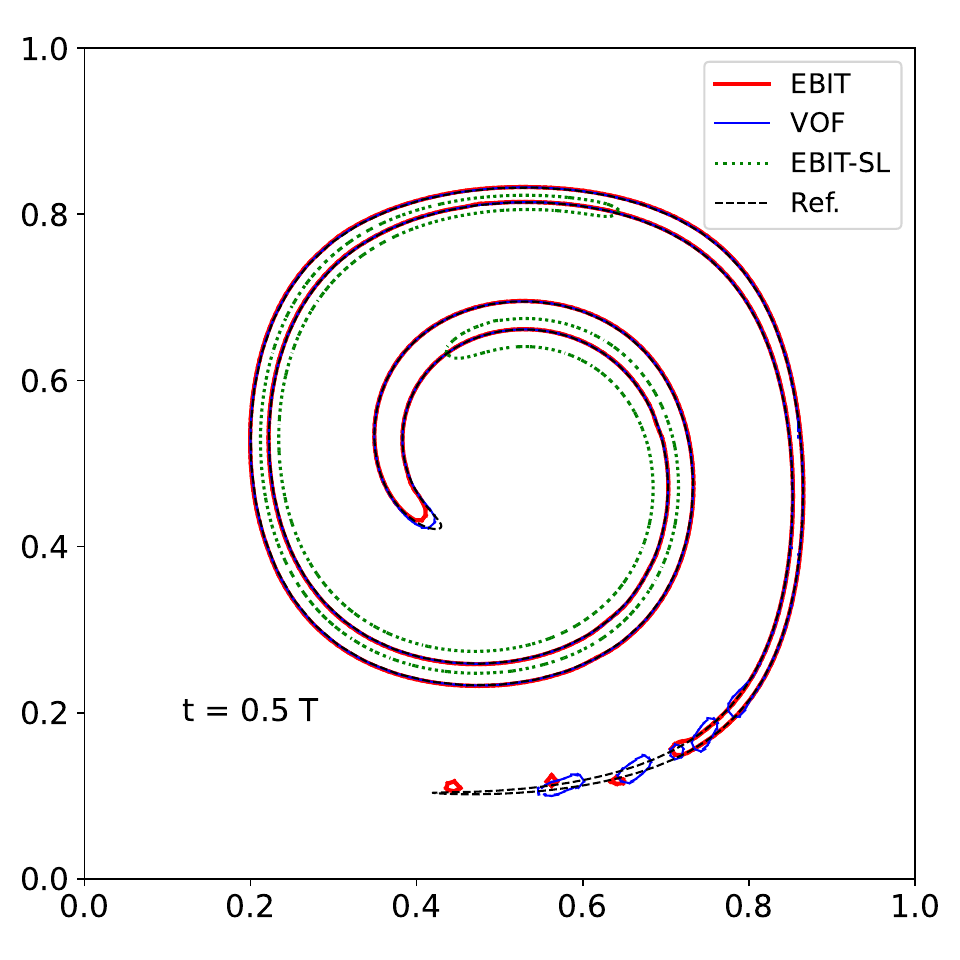} &
\includegraphics[width=0.45\textwidth]{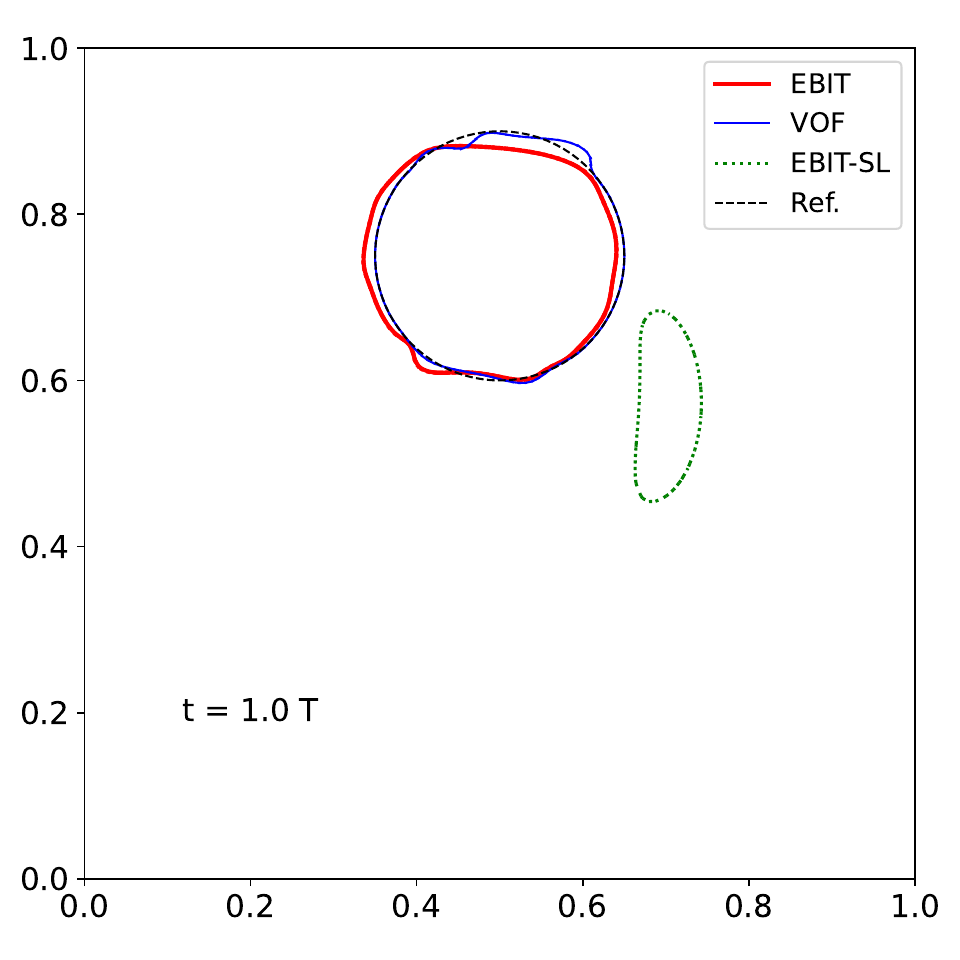}\\
(a) & (b) 
\end{tabular}
\end{center}
\caption{Single vortex test with period $T = 8$ at resolution $N_x = 128$ with different
methods: (a) interface lines at halftime; (b) interface lines at the end of the simulation.}
\label{Fig_vortex_intf_t8_cmp}
\end{figure}
\begin{table}[hbt!]
\footnotesize
\caption{Mesh convergence study for the single vortex test with period $T = 8$.}
\centering
\begin{tabular}{cc|ccccc}
\hline 
 &$N_x$& 32 & 64 & 128 & 256 & 512\\ 
\hline 
EBIT &$E_{area}$ & $3.80 \times 10^{-1}$ & $7.71 \times 10^{-2}$ & $2.43 \times 10^{-2}$ & $2.32 \times 10^{-3}$ & $8.23 \times 10^{-5}$\\ 
&$E_{shape}$ & $1.47 \times 10^{-1}$ & $4.99 \times 10^{-2}$ & $2.09 \times 10^{-2}$ & $7.07 \times 10^{-3}$ & $3.54 \times 10^{-3}$\\
&$E_{sym}$ & $4.13 \times 10^{-2}$ & $2.11 \times 10^{-2}$ & $8.09 \times 10^{-3}$ & $3.27 \times 10^{-3}$ & $1.56 \times 10^{-3}$\\
\hline 
VOF &$E_{shape}$ & $2.57 \times 10^{-1}$ & $6.58 \times 10^{-2}$ & $1.35 \times 10^{-2}$ & $7.67 \times 10^{-3}$ & $1.34 \times 10^{-3}$\\
&$E_{sym}$ & $7.01 \times 10^{-2}$ & $1.58 \times 10^{-2}$ & $1.95 \times 10^{-3}$ & $6.02 \times 10^{-4}$ & $1.03 \times 10^{-4}$\\
\hline 
EBIT-SL &$E_{area}$ & NA & NA & $7.98 \times 10^{-1}$ & $ 3.55 \times 10^{-1}$ & $ 1.61 \times 10^{-1}$\\ 
&$E_{shape}$ & NA & NA & $2.02 \times 10^{-1}$ & $1.29 \times 10^{-1}$ & $8.53 \times 10^{-2}$\\
&$E_{sym}$ & NA & NA & $8.49 \times 10^{-2}$ & $7.24 \times 10^{-2}$ & $4.07 \times 10^{-2}$\\
\hline 
\end{tabular}
\label{Tab_vortex_error_t8}
\normalsize
\end{table}

The interface line at maximum deformation and back to its initial position, for the test with period $T = 8$ is shown in Fig.~\ref{Fig_vortex_intf_t8} for different mesh resolutions. In this test, the interface line at maximum deformation at halftime
is stretched into a long thin ligament. When the mesh resolution is too coarse, i.e. $N_x = 32$, the new EBIT method loses some mass due to the artificial topological change mechanism. As the mesh resolution is increased, the mass loss progressively decreases and the method recovers better and better the initial circular shape.

The interface line obtained with different methods is shown in Fig.~\ref{Fig_vortex_intf_t8_cmp} for the resolution $N_x = 128$. At this intermediate mesh resolution, there is some discrepancy between the final shape obtained with the new EBIT method and that with the PLIC-VOF method. However, they show the same level of deviation from the reference solution. 

For the EBIT method with a straight line fit, there is an even more pronounced mass loss. Furthermore, the interface does not return to its initial position due to the lateral shift of the interface with respect to the reference solution, as shown in Fig.~\ref{Fig_vortex_intf_t8_cmp}a at maximum deformation.

\begin{figure}
\begin{center}
\begin{tabular}{ccc}
\includegraphics[width=0.33\textwidth]{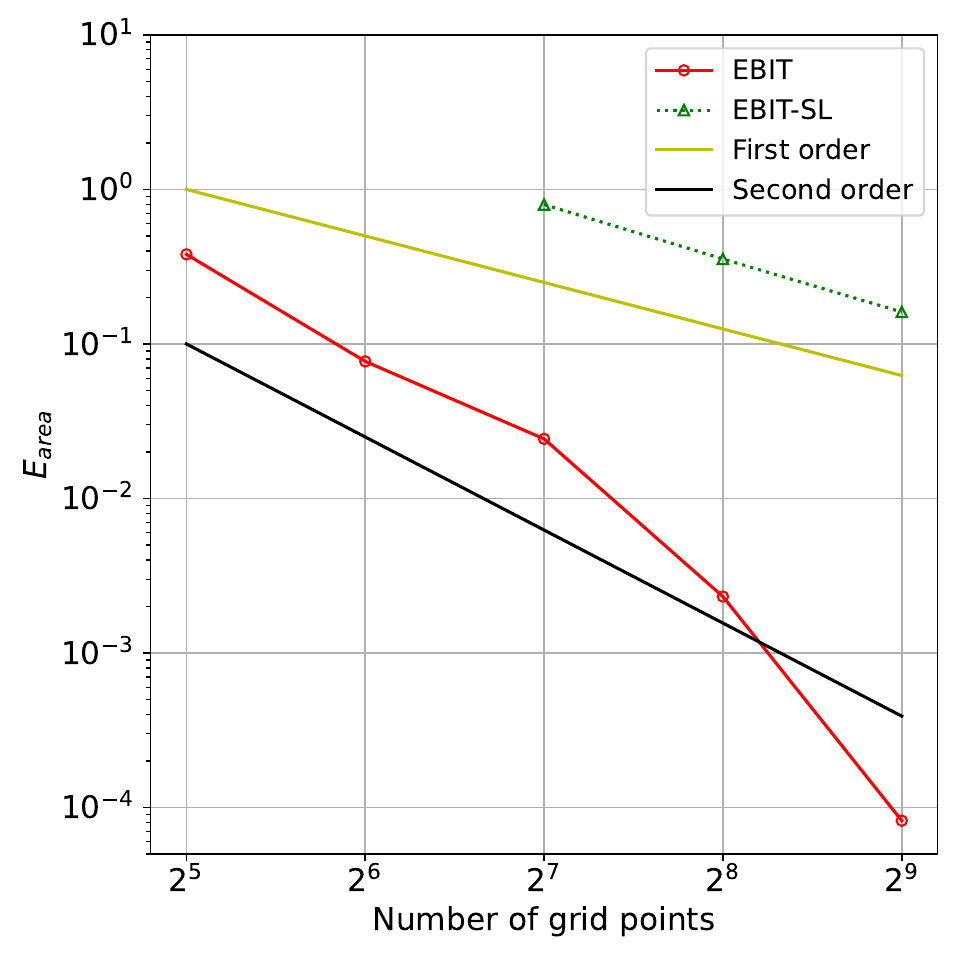} &
\includegraphics[width=0.33\textwidth]{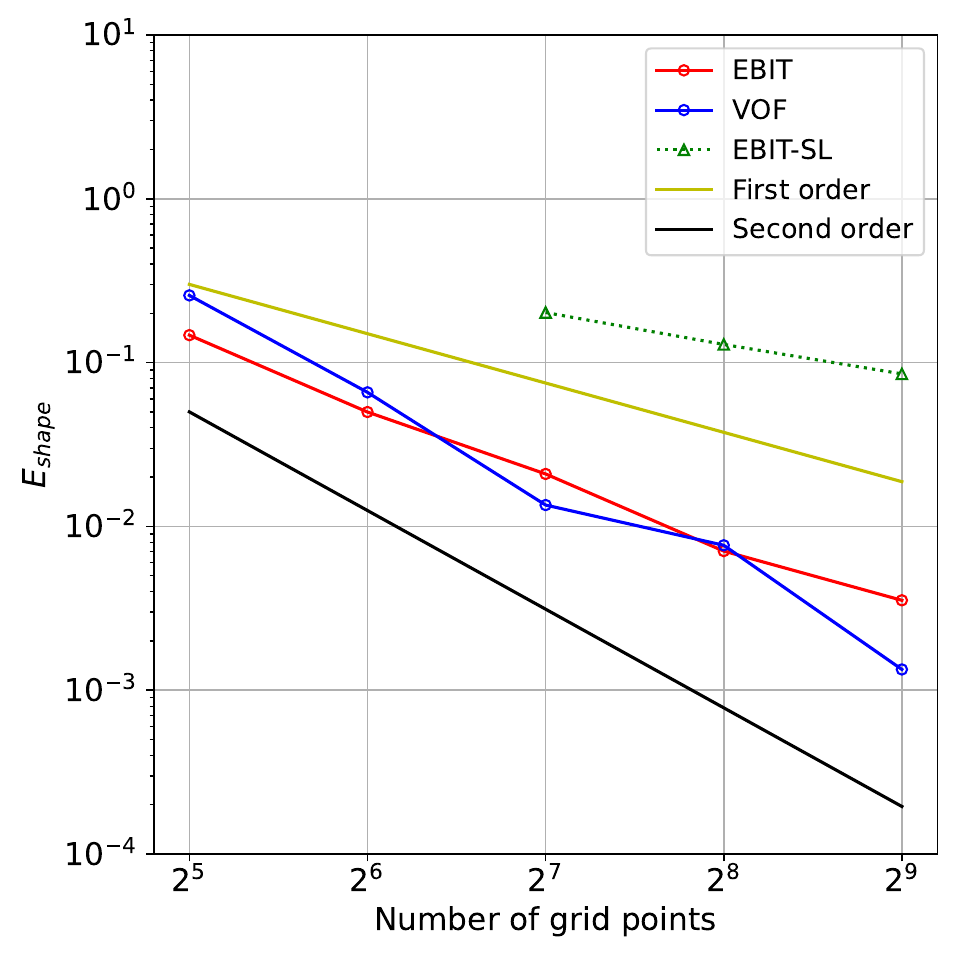}&
\includegraphics[width=0.33\textwidth]{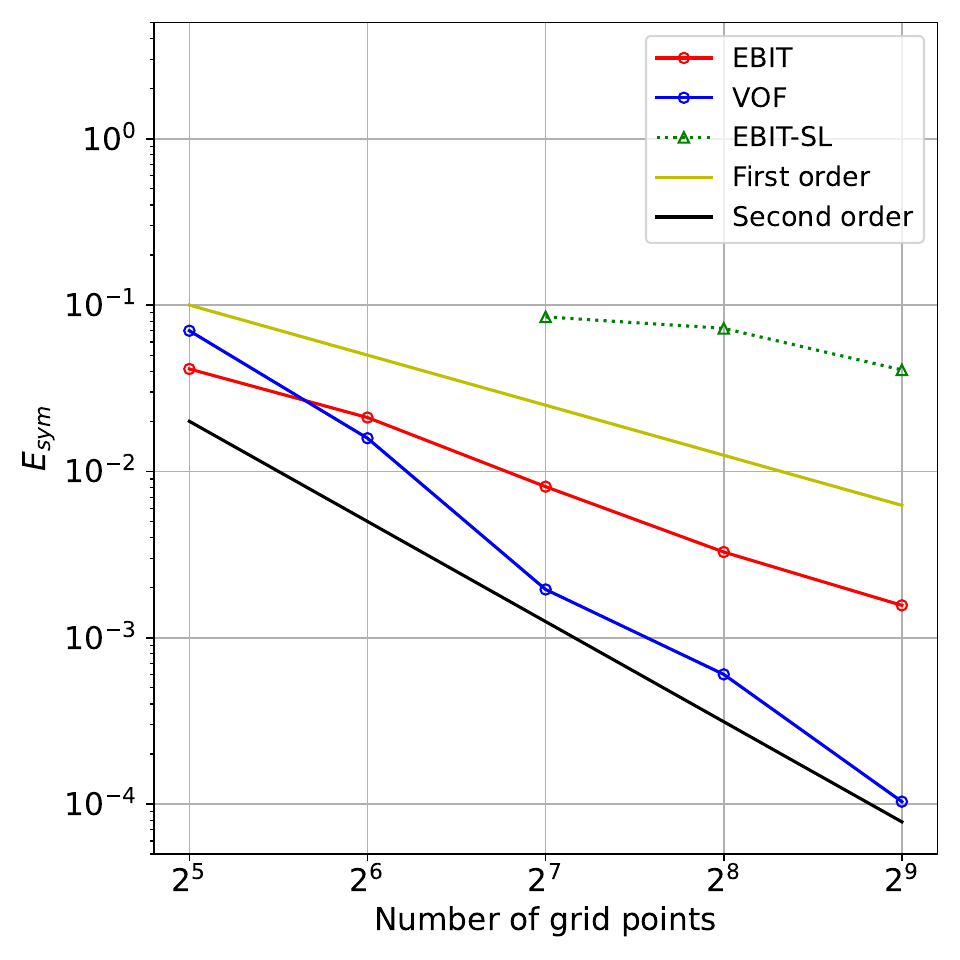}\\
(a) & (b) & (c)
\end{tabular}
\end{center}
\caption{Errors in the single vortex test with period $T = 8$ for different methods as a function of grid resolution: (a) area error $E_{area}$; (b) shape error $E_{shape}$; (c) symmetric difference error $E_{sym}$.}
\label{Fig_vortex_error_t8}
\end{figure}

The area error $E_{area}$, the shape error $E_{shape}$ and the symmetric difference error $E_{sym}$ are listed in Table~\ref{Tab_vortex_error_t8} and are shown in Fig.~\ref{Fig_vortex_error_t8}. For the new EBIT method, a convergence rate between first-order and second-order is observed for the area and shape errors, but only a  first-order convergence rate is observed for the symmetric difference error. This behavior is similar to that obtained in the previous test with period $T=2$. The shape errors obtained with the new EBIT method and the PLIC-VOF method become closer to each other as the mesh is refined. A second-order convergence rate is still observed for the symmetric difference error calculated with the PLIC-VOF method.
The EBIT method with a straight line fit loses all the reference phase even when an intermediate mesh resolution, $N_x = 64$, is used.
All the kinematic tests show that the new EBIT method with a circle fit does decrease the mass loss as the interface is reconstructed, thus increasing the accuracy of mass conservation.

\begin{figure}
\begin{center}
\begin{tabular}{cc}
\includegraphics[width=0.45\textwidth]{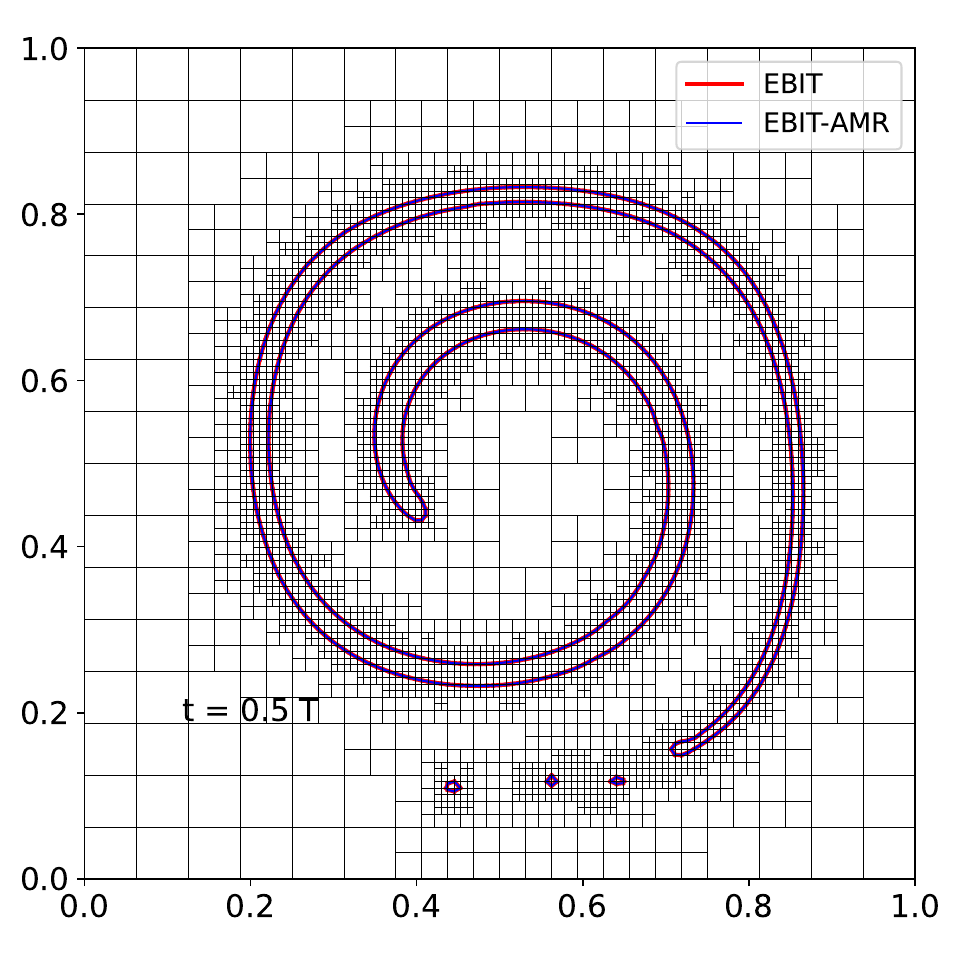} &
\includegraphics[width=0.45\textwidth]{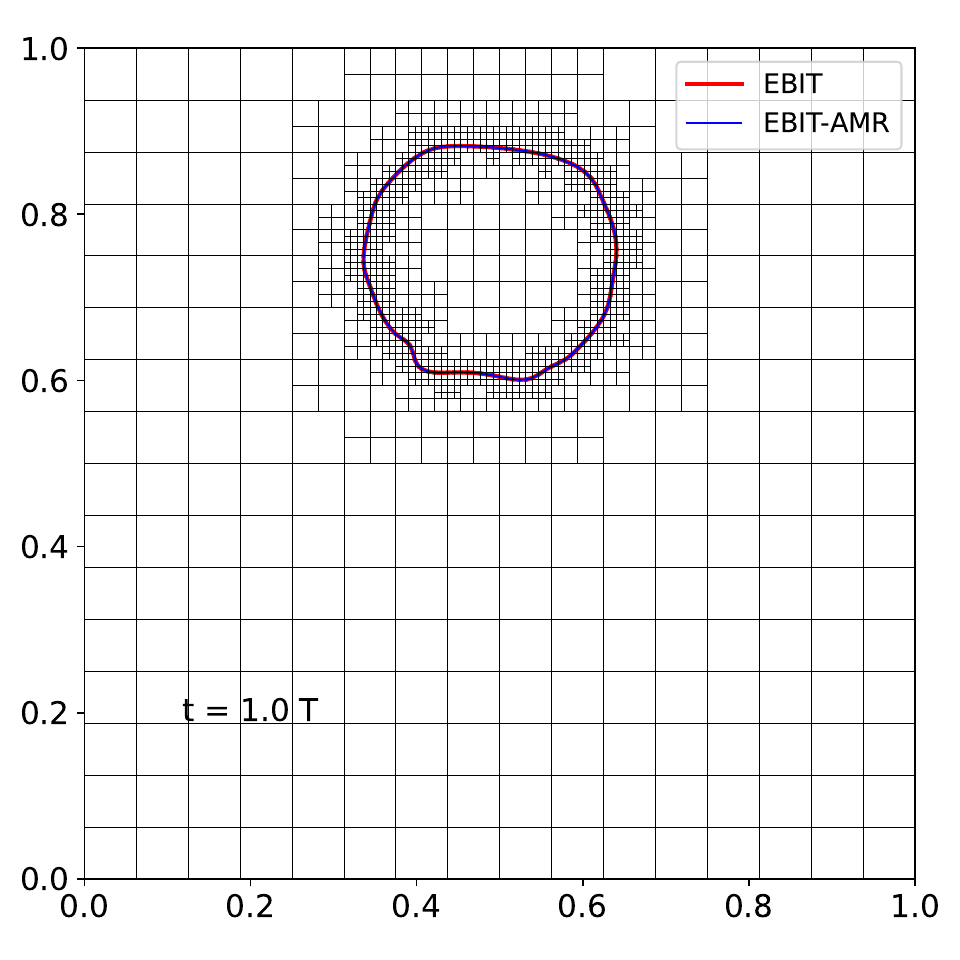}\\
(a) & (b) 
\end{tabular}
\end{center}
\caption{Single vortex test with period $T = 8$ at fixed resolution $N_x=128$ and AMR with $N_{l,max}=7$ and $N_{l,min}=4$: (a) interface lines at halftime; (b) interface lines at the end of the simulation.}
\label{Fig_vortex_intf_t8_amr}
\end{figure}

In order to demonstrate the feasibility of the integration of the new EBIT method with AMR, we run again the single-vortex test case on a quadtree grid in Basilisk. The interface lines at halftime and at the end of the simulation and the corresponding meshes are shown in Fig.~\ref{Fig_vortex_intf_t8_amr}. The maximum level of refinement is $N_{l,max}=7$, which corresponds to the mesh resolution $N_x =128$, while the minimum level is $N_{l,min}=4$. All the cells near the interface are refined to the maximum level, to avoid an inconsistency like that shown in Fig.~\ref{Fig_ebit_amr}. Since the applied velocity field is not affected by the mesh resolution and the same stencil is used for the velocity interpolation, the interface line obtained on the quadtree grid coincides with that on the fixed Cartesian mesh.

\subsection{Zalesak's disk}

The Zalesak's disk \cite{Zalesak_1979_31} test case is used to assess the ability of the EBIT method to deal with sharp edges and corners. A notched circular interface of radius $R = 0.15$ and center at $(0.5, 0.75)$ is placed inside the unit square domain. The notched width is $0.05$ and its length $0.25$. A constant velocity field, $(u, v) = (2\pi (0.5 - y), 2\pi (x - 0.5))$, is applied, so that the interface performs a rotation around the center of the computational domain and returns to its initial position at $t = T = 1.0$.

In the simulations we use a constant timestep that varies with the mesh resolution. Its value is determined by the maximum velocity at time $t=0$ so that $CFL = u_{max} \Delta t / \Delta = \pi / 16 \approx 0.2$. The error is again measured by the area and symmetric difference errors.

\begin{figure}
\begin{center}
\begin{tabular}{cc}
\includegraphics[width=0.45\textwidth]{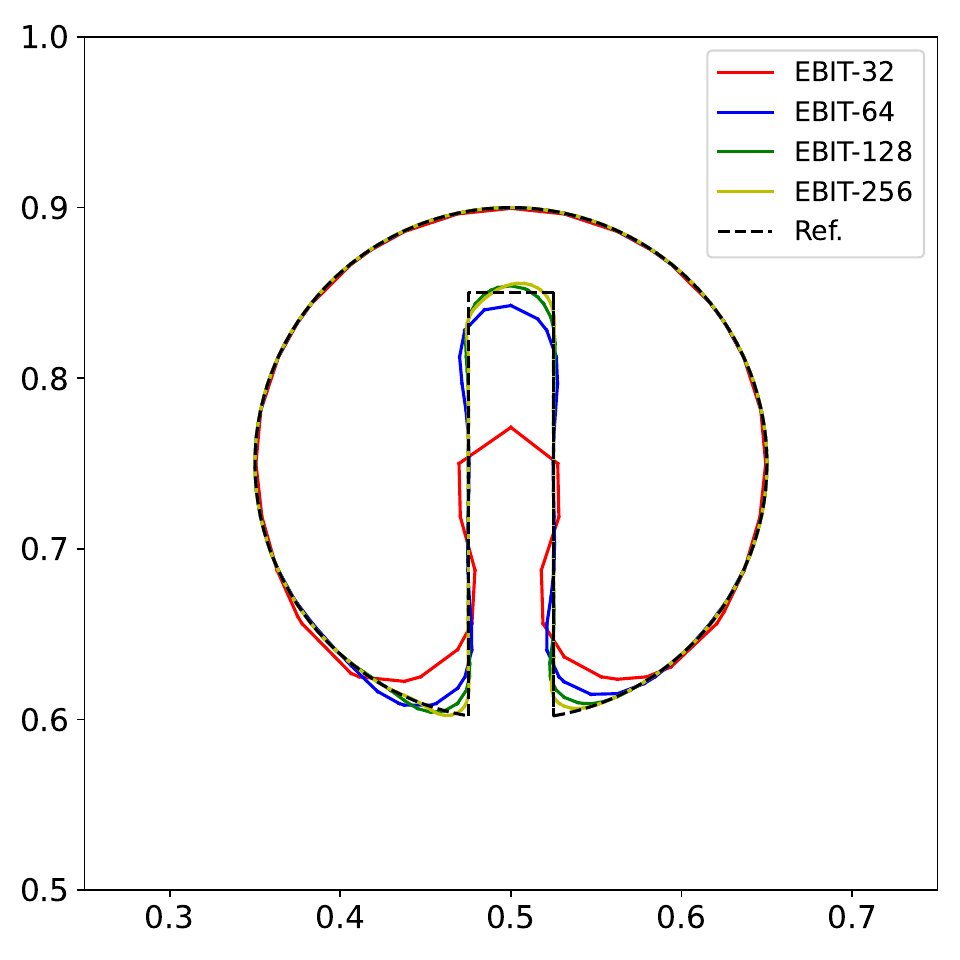} &
\includegraphics[width=0.45\textwidth]{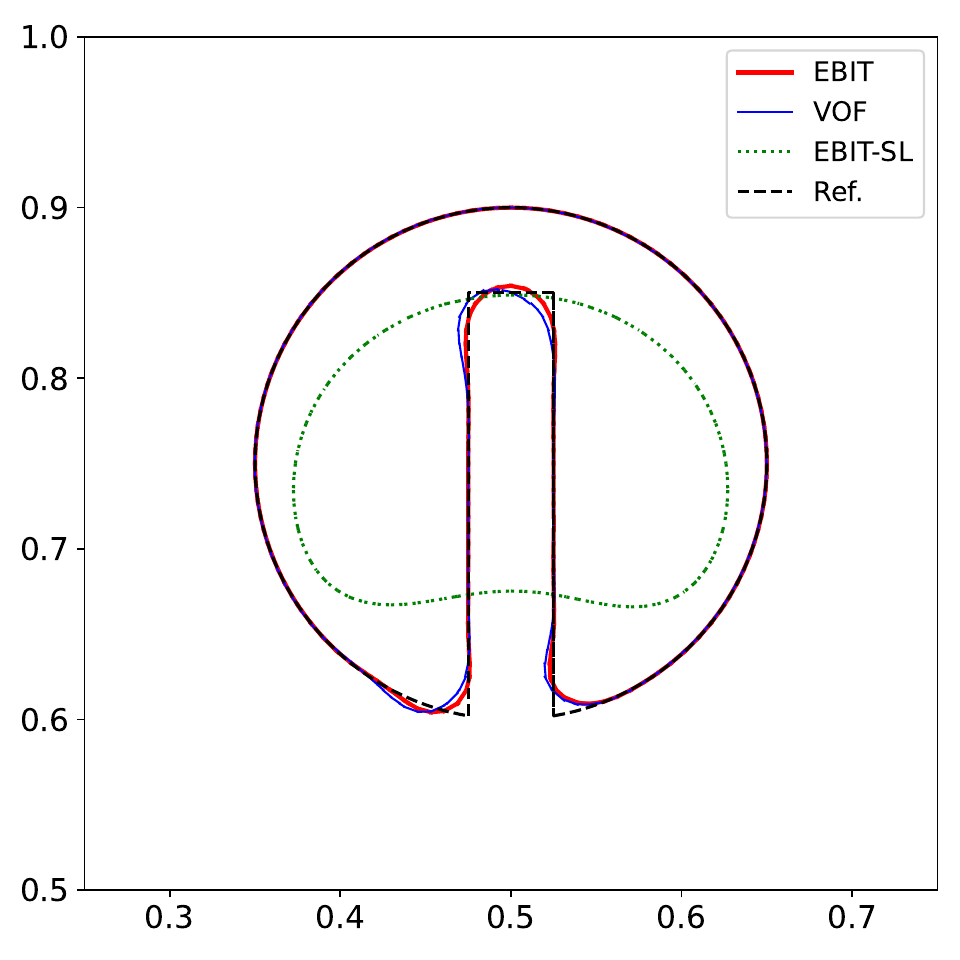}\\
(a) & (b) 
\end{tabular}
\end{center}
\caption{Zalesak's disk, interface lines at the end of the simulation: (a) interface lines for different mesh resolutions; (b) interface lines with different methods ($N_x = 128$).}
\label{Fig_zalesak_intf}
\end{figure}

The interface line at the end of the simulation is shown in Fig.~\ref{Fig_zalesak_intf}a for different mesh resolutions. The new EBIT method retains rather well the circular section of the interface for all mesh resolutions, except near the corners of the notch. As the mesh resolution is increased, the lateral sides of the notch are better approximated by the new EBIT method, while the interface smoothing near the corners progressively decreases.

The interface line obtained with different methods is shown in Fig.~\ref{Fig_zalesak_intf}b for the resolution $N_x = 128$. For the EBIT method with a straight line fit, the whole notched region has disappeared at the end of the simulation and there is also a relevant mass loss. The new EBIT method and the PLIC-VOF method present a similar level of smoothing near the four corner regions.

\begin{figure}
\begin{center}
\begin{tabular}{cc}
\includegraphics[width=0.45\textwidth]{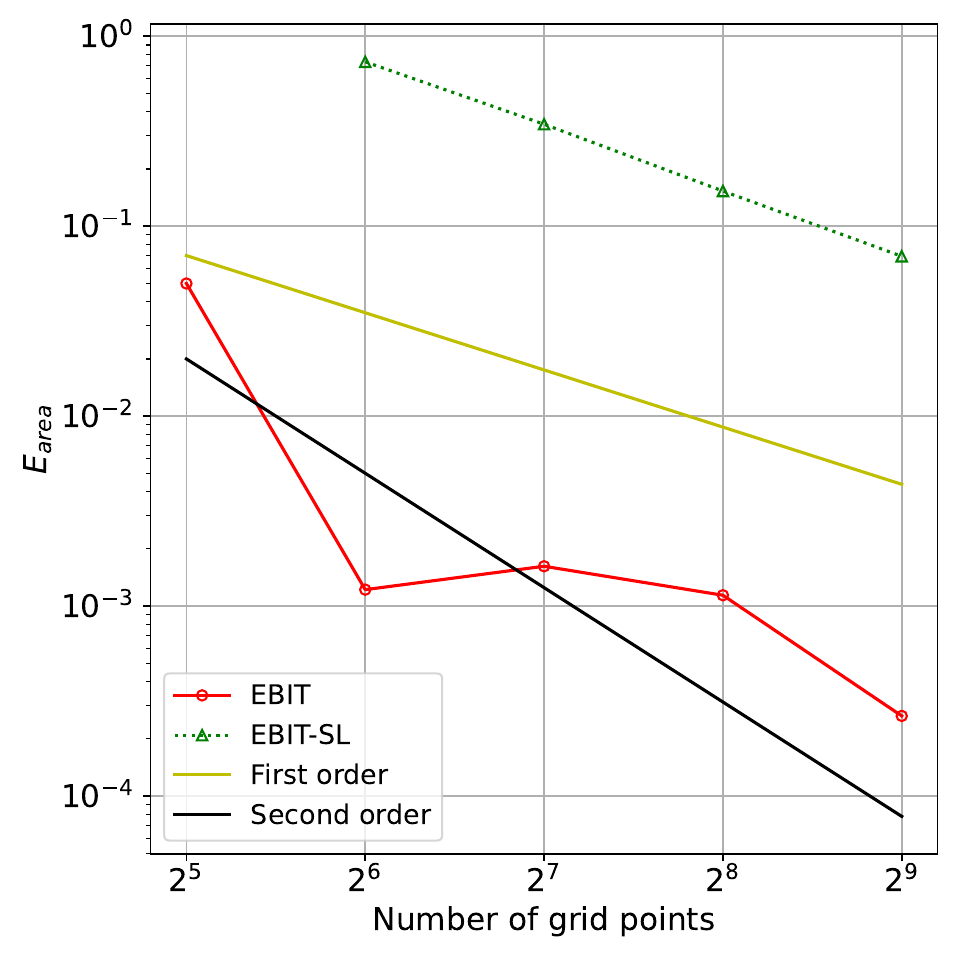} &
\includegraphics[width=0.45\textwidth]{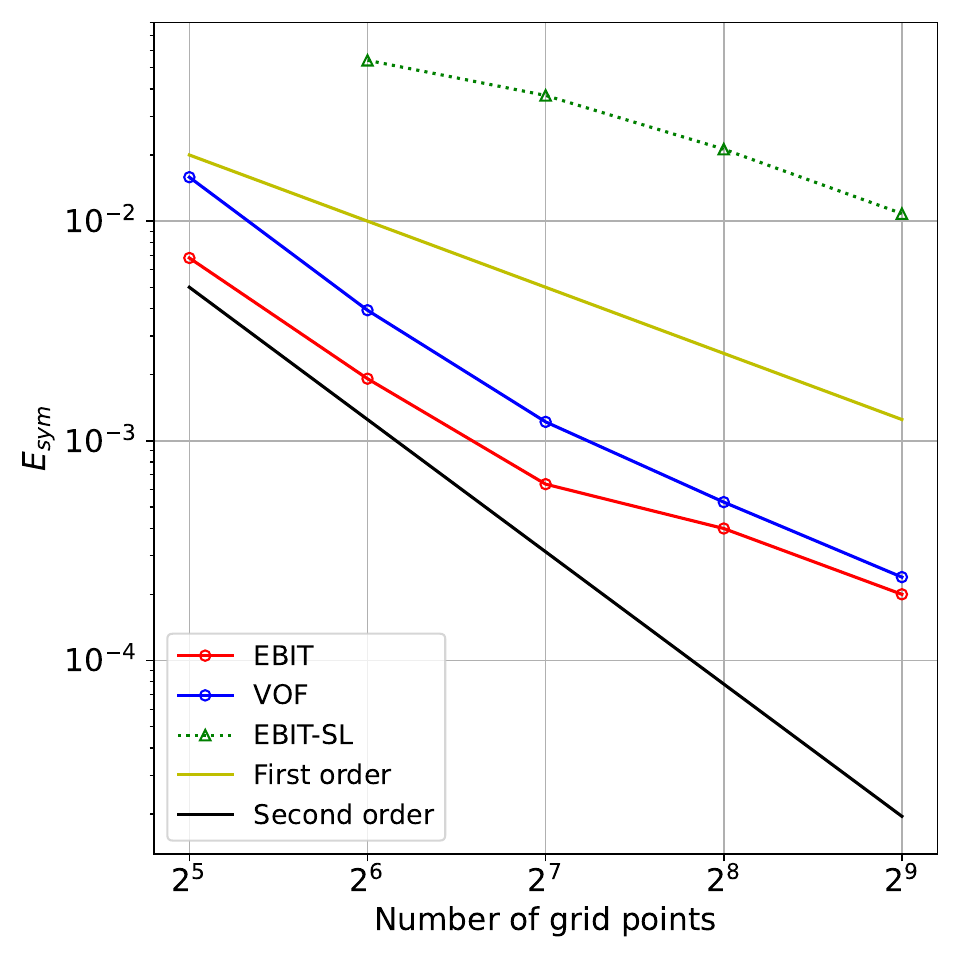}\\
(a) & (b)
\end{tabular}
\end{center}
\caption{Errors in the Zalesak's disk test for different methods as a function of grid resolution: (a) area error $E_{area}$; (b) symmetric difference error $E_{sym}$.}
\label{Fig_zalesak_error}
\end{figure}

\begin{table}[hbt!]
\footnotesize
\caption{Mesh convergence study for the Zalesak's disk test.}
\centering
\begin{tabular}{cc|ccccc}
\hline 
 &$N_x$& 32 & 64 & 128 & 256 & 512\\ 
\hline 
EBIT &$E_{area}$ & $4.99 \times 10^{-2}$ & $1.22 \times 10^{-3}$ & $1.62 \times 10^{-3}$ & $1.14 \times 10^{-3}$ & $2.64 \times 10^{-4}$\\ 
&$E_{sym}$ & $6.80 \times 10^{-3}$ & $1.92 \times 10^{-3}$ & $6.35 \times 10^{-4}$ & $3.99 \times 10^{-4}$ & $2.00 \times 10^{-4}$\\
\hline 
VOF & $E_{sym}$ & $1.58 \times 10^{-2}$ & $3.93 \times 10^{-3}$ & $1.22 \times 10^{-3}$ & $5.26 \times 10^{-4}$ & $2.40 \times 10^{-4}$\\
\hline 
EBIT-SL &$E_{area}$ & NA & $7.33 \times 10^{-1}$ & $3.44 \times 10^{-1}$ & $ 1.53 \times 10^{-1}$ & $ 6.94 \times 10^{-2}$\\ 
&$E_{sym}$ & NA & $5.39 \times 10^{-2}$ & $3.73 \times 10^{-2}$ & $2.13 \times 10^{-2}$ & $1.08 \times 10^{-2}$\\
\hline 
\end{tabular}
\label{Tab_zalesak_error}
\normalsize
\end{table}

The area error $E_{area}$ and the symmetric difference error $E_{sym}$ are listed in Table~\ref{Tab_zalesak_error} and are shown in Fig.~\ref{Fig_zalesak_error}. For the new EBIT method, the convergence rate of the area error fluctuates at low resolution but it seems to settle down at second-order at the highest resolution. For the PLIC-VOF method, the symmetric difference error presents a convergence rate close to first-order as the resolution is increased. The same error for the new EBIT method is approaching the same behavior but from below. This fact is clearly seen in the shape of the interface lines. Even if the two lines are close to each other, the interface with the new EBIT method is clearly more symmetric with respect to a vertical axis through the notch (see blue and red lines of Fig.~\ref{Fig_zalesak_intf}b).

\subsection{Capillary wave}

Capillary waves are a basic phenomenon of surface-tension-driven flows and their adequate numerical resolution is a prerequisite to more complex applications. The small-amplitude damped oscillations of a capillary wave are now a classical test case to check the accuracy of new numerical schemes that are developed to investigate the evolution in time of viscous, surface-tension-driven two-phase flows.

A sinusoidal perturbation is applied to a plane interface between two fluids initially at rest. Under the influence of surface tension, the interface begins to oscillate around its equilibrium position, while the amplitude of the oscillations decay in time due to viscous dissipation. The exact analytical solution was found by Prosperetti \cite{Prosperetti_1981_24} in the limit of very small amplitudes, and it is usually used as a reference.

In order to get a good agreement with the theory \cite{Popinet_2009_228}, it is necessary to move the top and bottom boundaries far away from the interface, in particular here we consider the rectangular computational domain $[0, \lambda] \times [0, 4 \lambda]$, where $\lambda$ is the wavelength of the perturbation. A symmetric boundary condition is applied on the four sides of the domain. The initial amplitude of the perturbation is 
$\lambda/100$, as in Popinet and Zaleski \cite{Popinet_1999_30}, Denner et al. \cite{Denner_2017_226} and Gerlach et al. \cite{Gerlach_2006_49}. The value of the other physical properties that are used in the simulation and of the Laplace number, 
$La = (\rho_1 \sigma \lambda) \big/ \mu_1^2$, is listed in Table~\ref{Tab_para_capwave}.
In this simulation, the maximum value $CFL = 0.5$ is used. Moreover, the timestep is also limited by the period of the shortest capillary wave:
\begin{equation}
\Delta t \leq \frac{(\rho_1 + \rho_2) \Delta^3}{2 \sigma}
\label{Eq_dt_cap}
\end{equation}

\begin{figure}
\centering
\includegraphics[width=\textwidth]{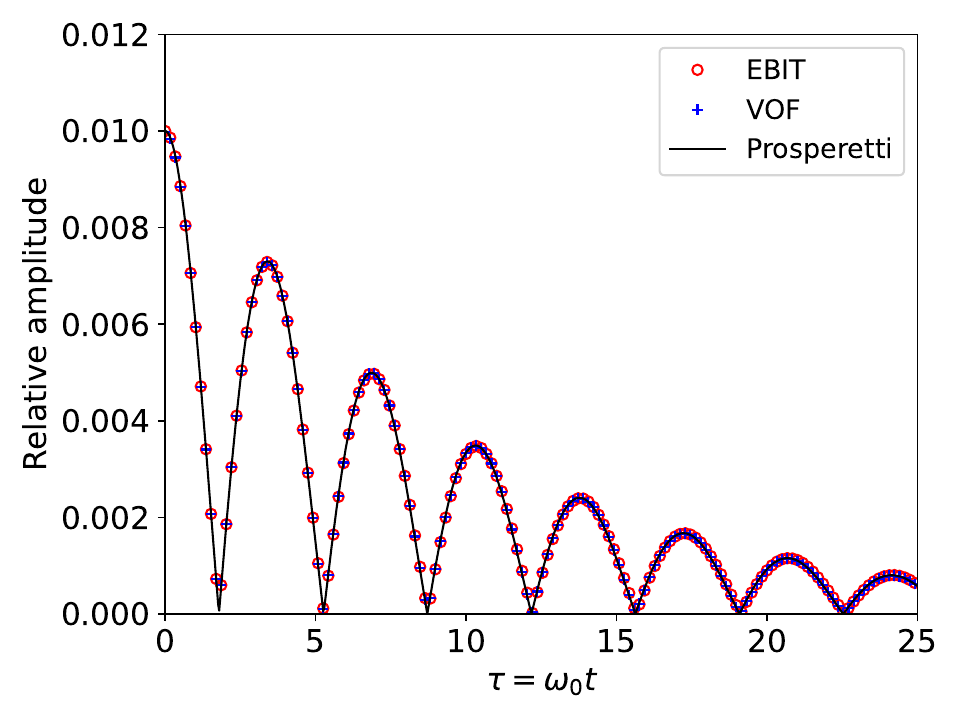}
\caption{Capillary wave test with different methods: time evolution of the maximum amplitude of the interface oscillation. The time $\tau = \omega_0\,t$ is non-dimensional and the grid resolution is $N_x \times N_y = 64 \times 256$.}
\label{Fig_capwave_amp}
\end{figure}

The time evolution of the maximum amplitude of the interface is shown in Fig.~\ref{Fig_capwave_amp}, together with those of the analytical solution and of the PLIC-VOF method. Time is made dimensionless by using the normal-mode oscillation frequency $\omega_0$, which is defined by the dispersion relation
\begin{equation}
\omega_0^2 = \frac{\sigma k^3}{2 \rho_1}
\end{equation}
where $k = 2\pi / \lambda$ is the wavenumber and $\lambda = 1$ in our simulation. The numerical results obtained with both the new EBIT and PLIC-VOF methods agree rather well with the analytical solution.

\begin{figure}
\centering
\includegraphics[scale=0.8]{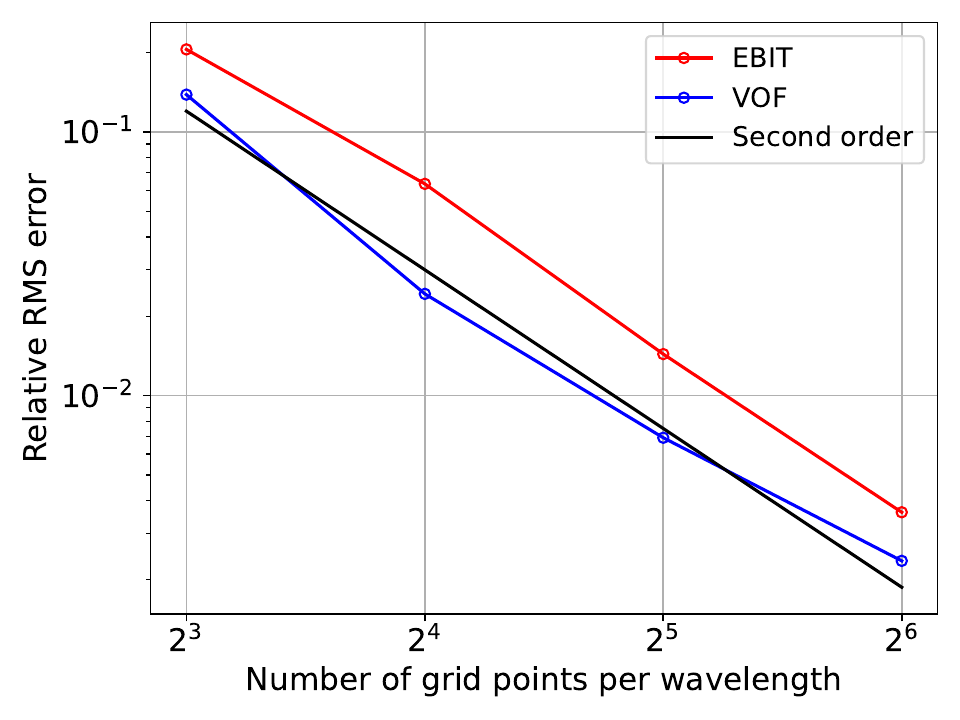}
\caption{Error $E_2$ in the capillary wave test for different methods as a function of the grid resolution.}
\label{Fig_capwave_acc}
\end{figure}

The error between the analytical solution \cite{Prosperetti_1981_24} and the two numerical solutions can be further analyzed with the $L_2$ norm
\begin{equation}
E_2 = \frac{1}{\lambda} \sqrt{\frac{\omega_0}{25} \int_{t=0}^{25 / \omega} (h - h_{exact})^2}
\label{Eq_E2}
\end{equation}
where $h$ is the maximum interface amplitude obtained with a numerical method
and $h_{exact}$ the reference value. The results are shown in Fig.~\ref{Fig_capwave_acc}
and a convergence rate close to second-order is observed for both methods,
the error of the PLIC-VOF method being always somewhat smaller.
In both simulations, the height function method has been used to calculate the curvature \cite{Popinet_2009_228}. More particularly, since we are considering a very small amplitude of the oscillations, hence of the interface curvature as well, the straight line approximation of the new EBIT method in each cut cell provides a fairly good approximation of the volume fraction and therefore of the calculation of the local height function.

The area error $E_{area}$ of Eq.~\eqref{Eq_error_surface} is used to measure the accuracy of mass conservation in the dynamics test cases. The area $A$ is that occupied by the reference phase. For the simulation at the highest resolution ($N_x \times N_y = 64 \times 256$), $E_{area}$ is $5.3 \times 10^{-6}$.
\begin{table}[hbt!]
\caption{Physical properties for the capillary wave test.}
\centering
\begin{tabular}{cccccc}
\hline 
 $\rho_1$ & $\rho_2$ &$\mu_1$ & $\mu_2$ & $\sigma$ & $La$ \\ 
\hline 
$1$ & $1$ &$ 0.01826$& $0.01826$ & $1$ & $3000$ \\ 
\hline 
\end{tabular}
\label{Tab_para_capwave}
\end{table}

\subsection{Rayleigh-Taylor instability}

In order to demonstrate the capability of the new EBIT method to deal with more complex flows, we investigate another classical test: the Rayleigh-Taylor instability at high Reynolds number, that involves a large deformation of the interface. 

The Rayleigh-Taylor instability occurs when a heavy fluid is on top of a lighter one, with the direction of gravity from top to bottom. The density difference between the two fluids plays an important role in the instability and is present in the definition of the dimensionless Atwood number $At$
\begin{equation}
At = \frac{\rho_1 - \rho_2}{\rho_1 + \rho_2}
\end{equation}
where $\rho_1$ and $\rho_2$ are the densities of the heavy and light fluids, respectively.

This instability has been investigated in several studies \cite{Tryggvason_1988_75, Guermond_2000_165, Ding_2007_226}, that consider an incompressible flow without surface tension effects, with $At = 0.5$ and different Reynolds numbers,
$Re = \big( \rho_1 g^{1/2} d^{3/2} \big) \big/ \mu_1$.

In this study, we consider the rectangular computational domain $[0, d] \times [0, 4d]$, partitioned with $N_x \times N_y = 128 \times 512$ grid cells. The plane interface $y_0(x) = 2\,d$ between the two fluids is perturbed by a sinusoidal wave $y_1(x) = 0.1\,d \cos (k x)$, so that the interface line at the beginning of the simulation is
\begin{equation}
y (x) = y_0 (x) + y_1 (x) = 2\,d + 0.1\,d \cos (k x), \quad k = \frac{2\pi}{\lambda}
\end{equation}
with $\lambda = d = 1$.
A no-slip boundary condition is enforced at the bottom and at the top of the computational domain, and a symmetric boundary condition on the two vertical sides. 
The value of the other physical properties that are used in the simulation and of the Reynolds number $Re$ is listed in Table~\ref{Tab_para_rti}. In this simulation, the timestep is limited by the $CFL$ number, $CFL=0.05$, and the maximum allowable timestep provided by the user, $\Delta t_{max} = 2 \times 10^{-4}$.

\begin{table}[hbt!]
\caption{Physical properties for the Rayleigh-Taylor instability test.}
\centering
\begin{tabular}{cccccc}
\hline 
 $\rho_1$& $ \rho_2$ &$\mu_1$ & $\mu_2$& $g$ & $Re$ \\ 
\hline 
$3 $ & $1$ & $ 0.00313$ & $ 0.00313$& $9.81$ &$3000$ \\ 
\hline 
\end{tabular}
\label{Tab_para_rti}
\end{table}
\begin{figure}
\begin{center}
\begin{tabular}{ccc}
\includegraphics[width=0.25\textwidth]{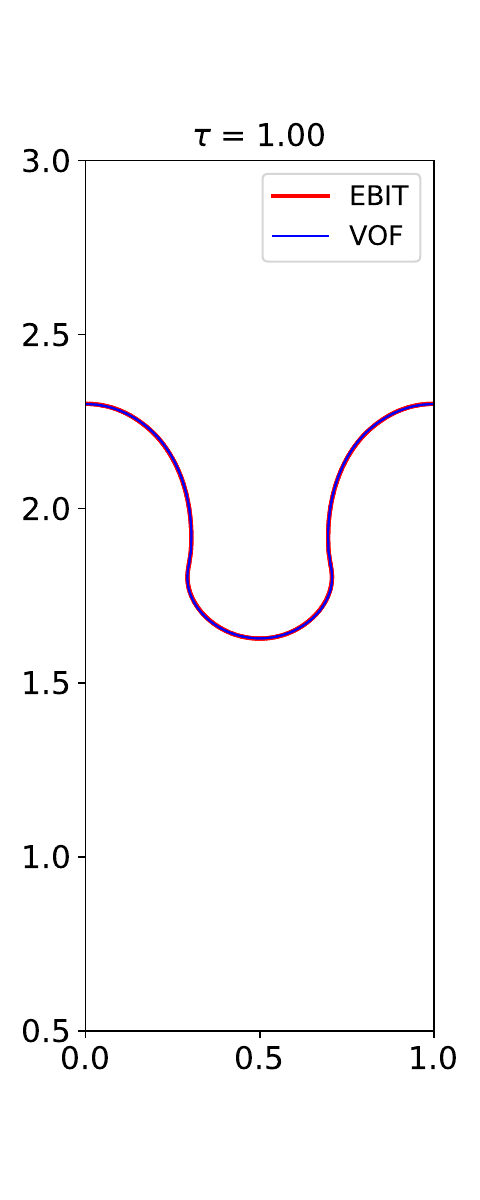} &
\includegraphics[width=0.25\textwidth]{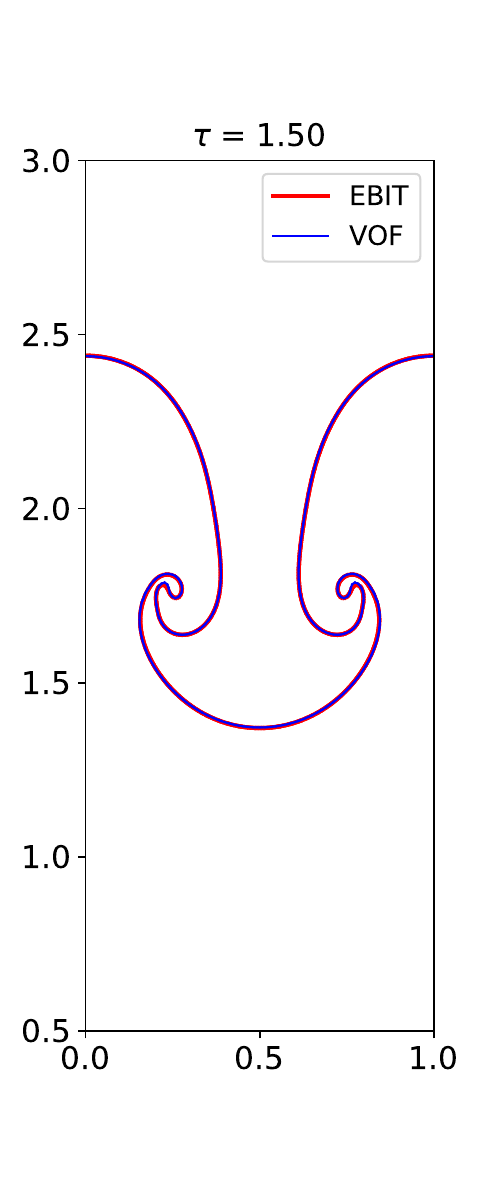} &
\includegraphics[width=0.25\textwidth]{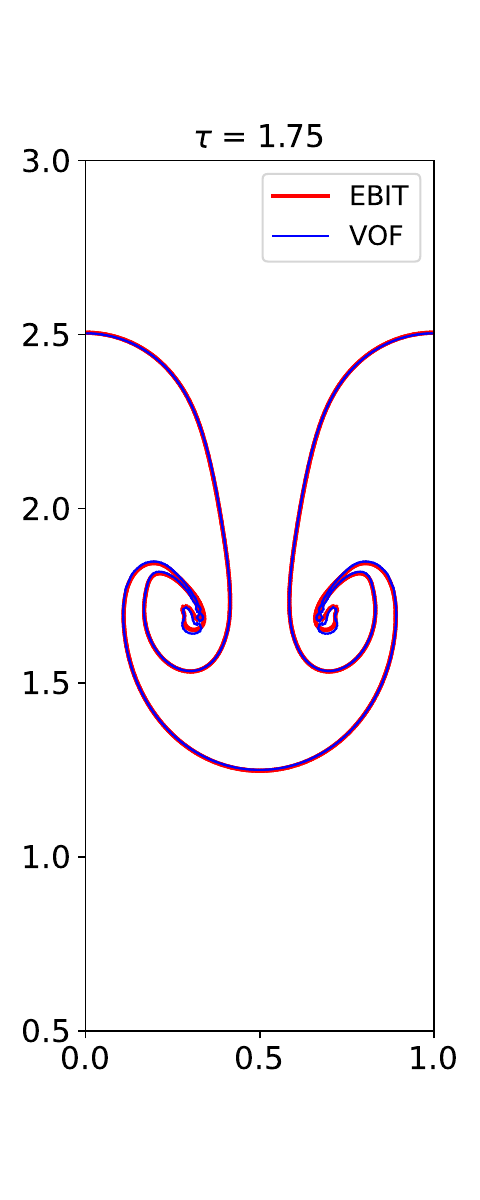} \\
\includegraphics[width=0.25\textwidth]{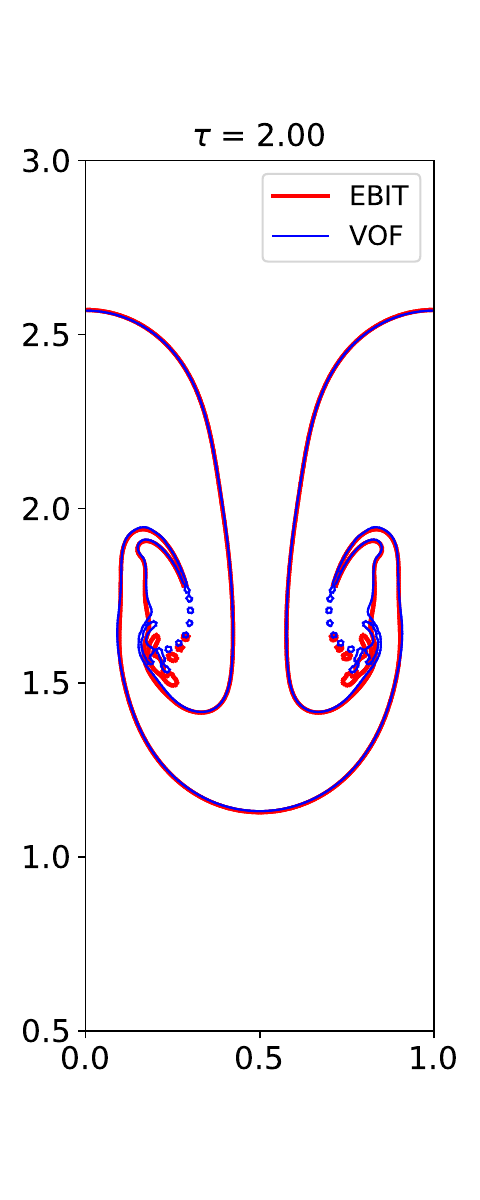} &
\includegraphics[width=0.25\textwidth]{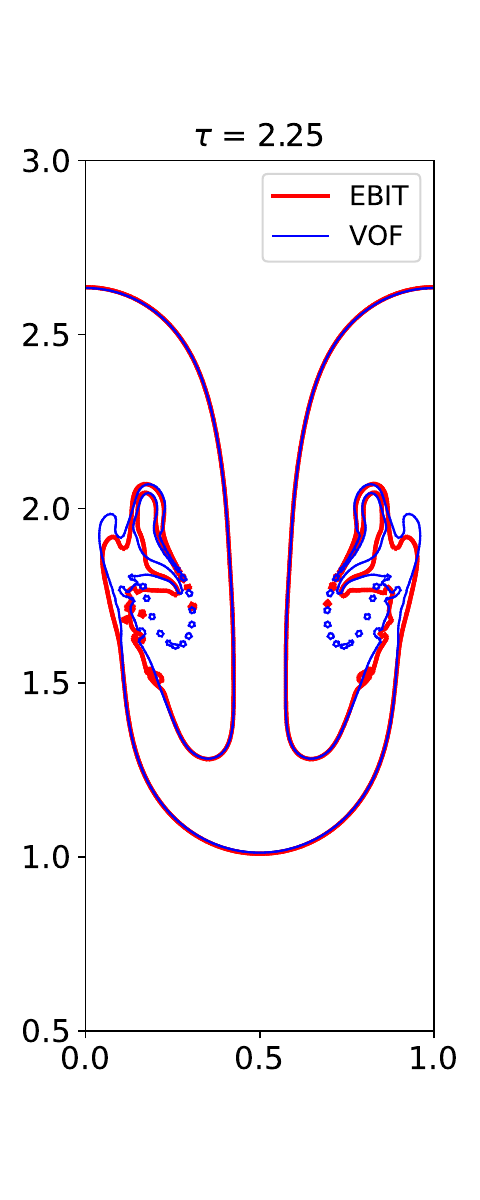} &
\includegraphics[width=0.25\textwidth]{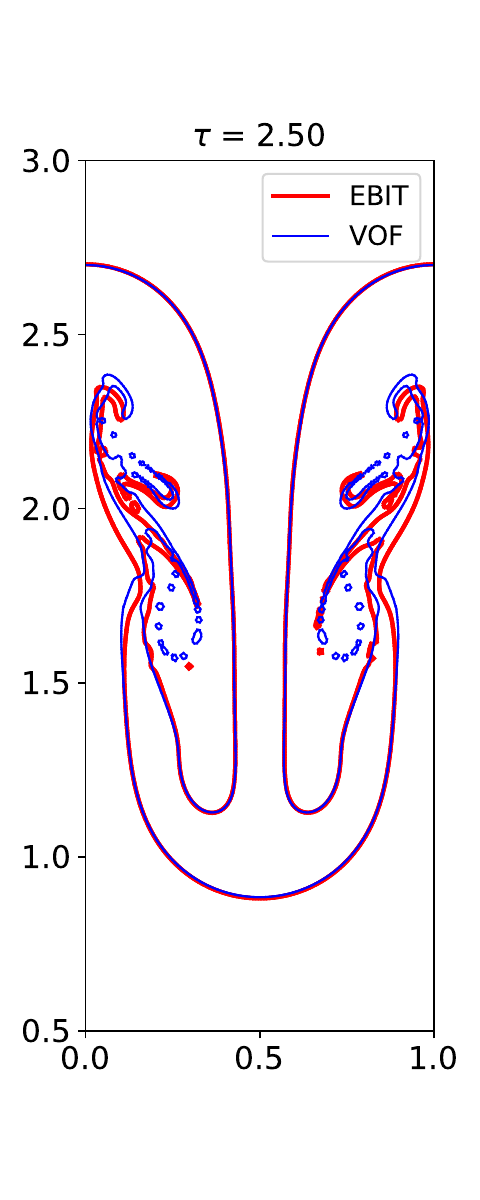} 
\end{tabular}
\end{center}
\caption{Rayleigh-Taylor instability test with different methods: interface lines at dimensionless times $\tau = 1.,1.5,1.75,2.,2.25,2.5$.}
\label{Fig_rti_intfs}
\end{figure}
\begin{figure}
\centering
\includegraphics[width=\textwidth]{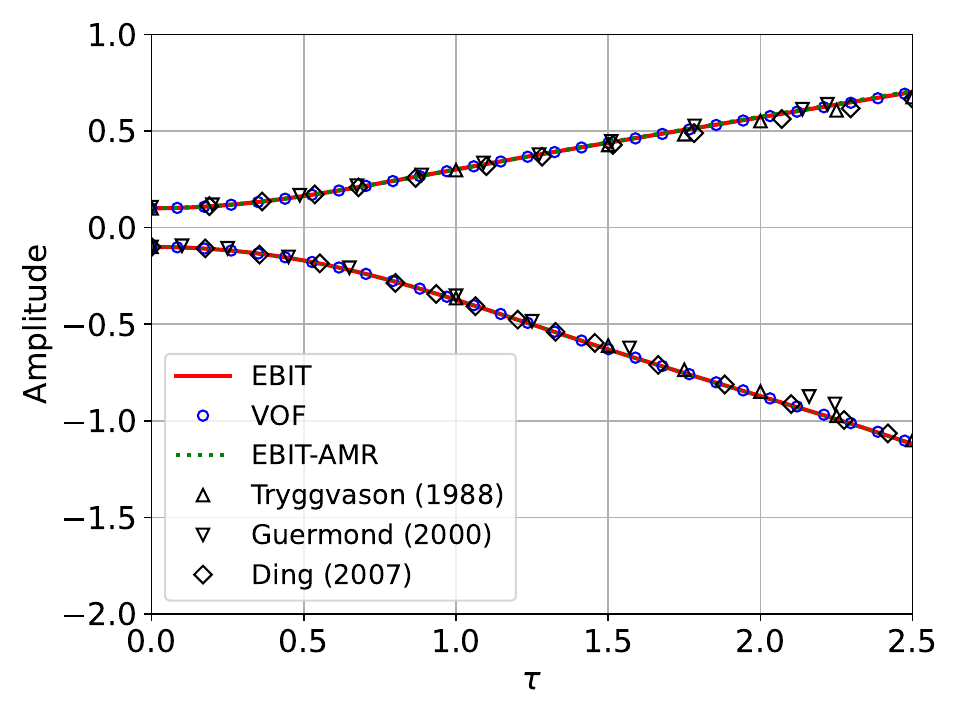}
\caption{Rayleigh-Taylor instability test with different methods: highest and lowest positions of the interface, with respect to the mean position at $y=2$, as a function 
of dimensionless time $\tau$.}
\label{Fig_rti_amp}
\end{figure}

The results of the simulation are first presented in Fig.~\ref{Fig_rti_intfs} with the representation of the interface line at several dimensionless times 
$\tau = t \sqrt{g\, At / d}$. Overall, these results compare rather well with those presented in \cite{Ding_2007_226}.

More specifically, in the early stages of the simulation, $\tau \leq 1.75$, the shape of interface calculated with the new EBIT method and that with the PLIC-VOF method are in very good agreement, small discrepancies are observed only in the roll-up region where complex structures with thin ligaments start to develop. 

Some remarkable differences occur at later times ($\tau \geq 2.00$) when the ligaments start to break up. Due to the mass conservation property of the PLIC-VOF method, many small droplets are formed when the ligaments tear apart. However, in the new EBIT method, these small droplets will soon disappear due to the topology change mechanism. Thus, in the roll-up region, the interface structure obtained with the new EBIT method agrees only qualitatively with that of the PLIC-VOF method.

In spite of this local but consistent difference, very good agreement is observed for the highest and lowest positions of the interface during the whole simulation. The highest
position of the rising fluid, near the two vertical boundaries at $x=0$ and $x=d$,
and the lowest position of the falling fluid, near the centerline at $x=d/2$, both computed
with respect to the mean position at $y = 2\,d$, are shown in Fig.~\ref{Fig_rti_amp}. They are in very good agreement with the results obtained with the PLIC-VOF method and those by Tryggvason \cite{Tryggvason_1988_75}, Guermond \cite{ Guermond_2000_165} and Ding \cite{Ding_2007_226}.

\begin{figure}
\begin{center}
\begin{tabular}{ccc}
\includegraphics[width=0.22\textwidth]{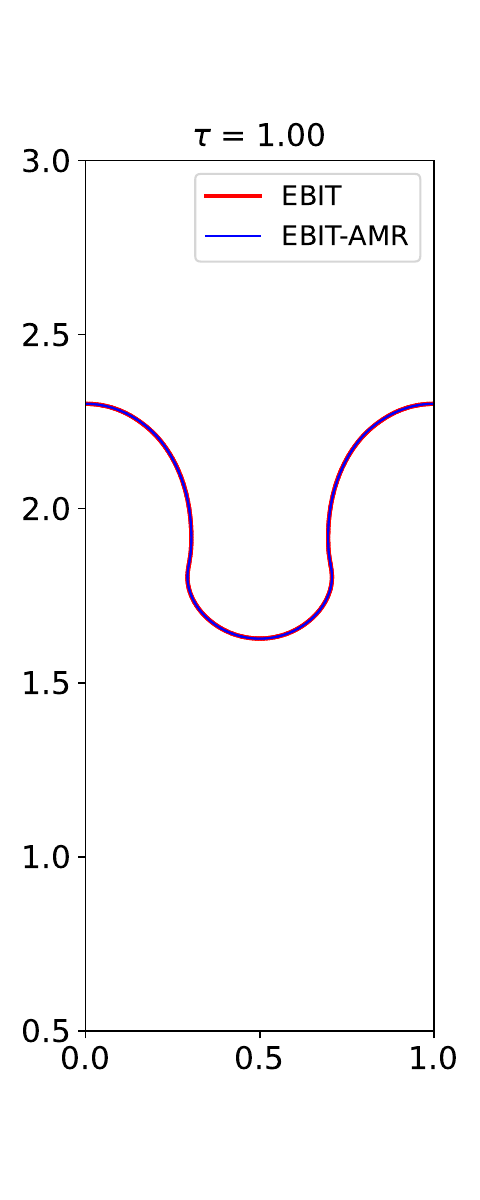} &
\includegraphics[width=0.22\textwidth]{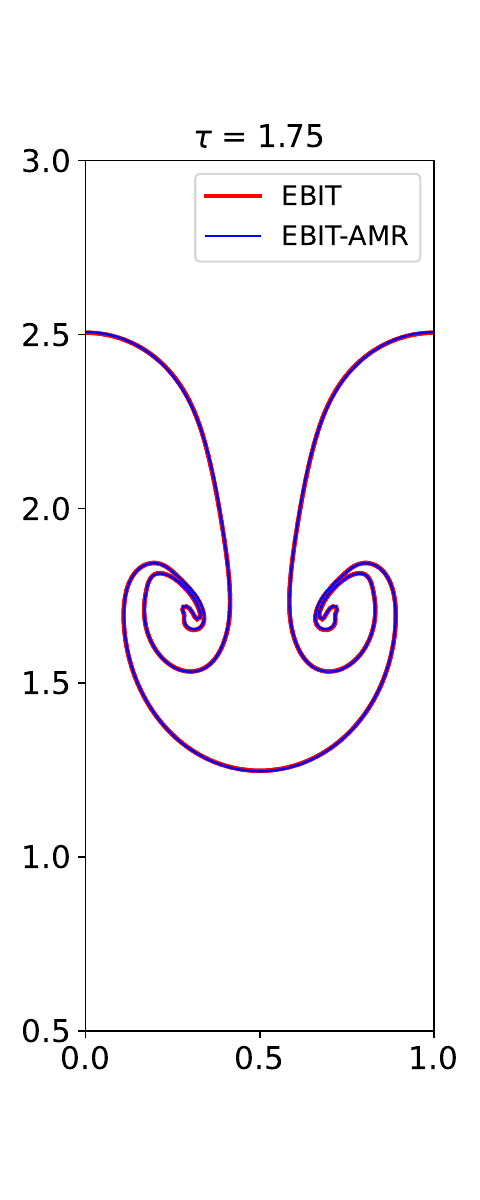} &
\includegraphics[width=0.22\textwidth]{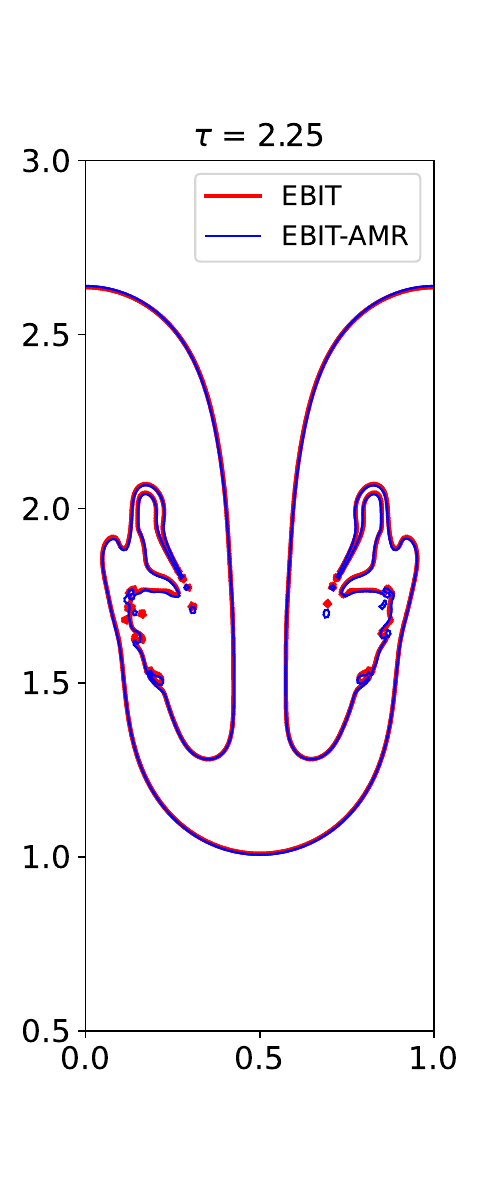} \\
 & (a) & \\[2pt]          
\includegraphics[width=0.22\textwidth]{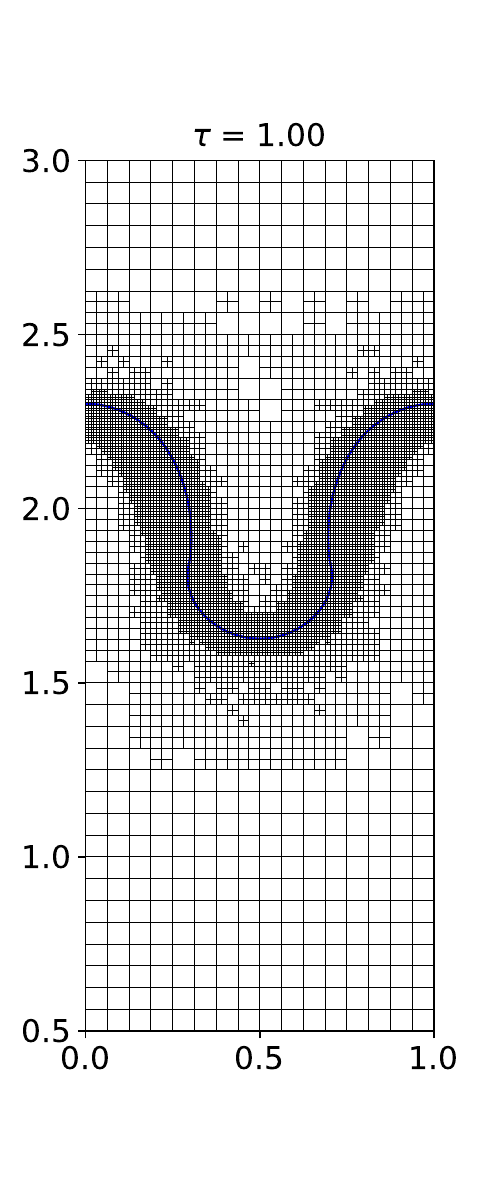} &
\includegraphics[width=0.22\textwidth]{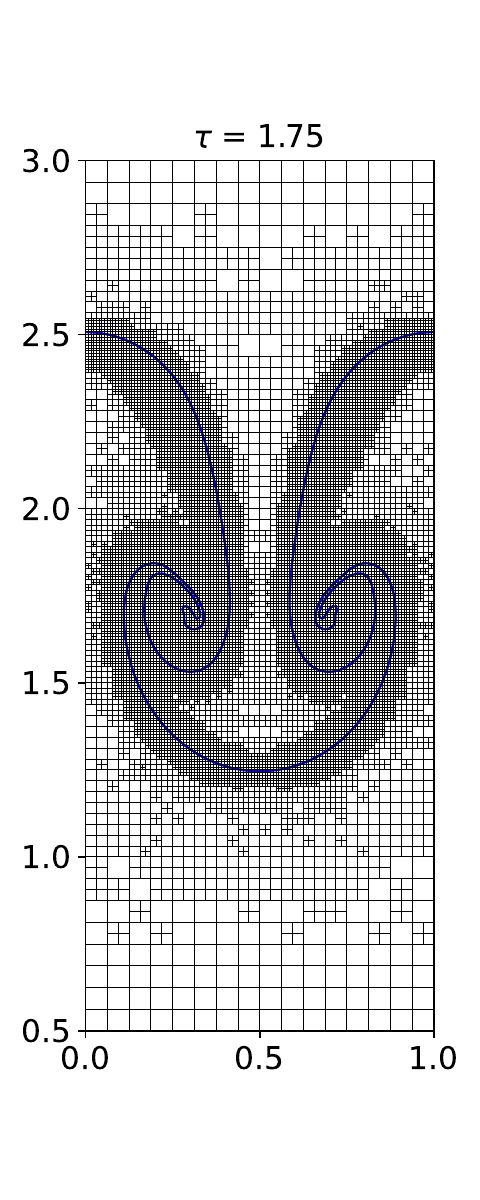} &
\includegraphics[width=0.22\textwidth]{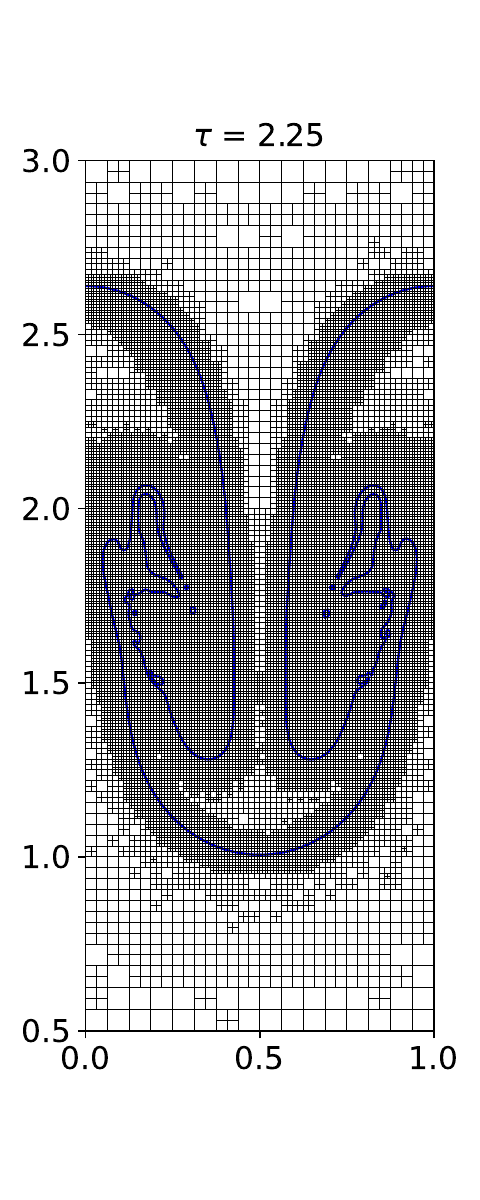} \\
 & (b) & 
\end{tabular}
\end{center}
\caption{Rayleigh-Taylor instability test at dimensionless times $\tau = 1.,1.75,2.25$: (a) interface lines with and without AMR; (b) computational grid with $N_{l,max}=9$ and $N_{l,min}=6$.}
\label{Fig_rti_intfs_amr}
\end{figure}

\begin{table}[hbt!]
\caption{Computational efficiency of the EBIT method with AMR, Rayleigh-Taylor instability test, $N_{l,max}=9$ and $N_{l,min}=6$.}
\centering
\begin{tabular}{cccc}
\hline 
 Method & Number of (leaf) cells & Time steps & Wall time (s) \\ 
\hline 
EBIT & $65536$ & $5656$ & $1741$ \\ 
EBIT-AMR & $27355$ & $5656$ & $207$ \\ 
\hline
VOF & $65536$ & $5656$ & $914$ \\ 
VOF-AMR & $27223$ & $5656$ & $151$ \\ 
\hline 
\end{tabular}
\label{Tab_rti_amr}
\end{table}

The interface lines obtained with the new EBIT method combined with AMR are finally shown in Fig.~\ref{Fig_rti_intfs_amr}. The maximum level of refinement is now $N_{l,max}=9$, while the minimum level is $N_{l,min}=6$. The refinement criteria are based not only on the position of the interface, but also on the velocity gradient, thus the cells within the roll-up region, characterized by a strong vorticity, are also refined to the maximum level, even if they are not very close to the interface (see Fig.~\ref{Fig_rti_intfs_amr}b). The interface lines calculated with the new EBIT method on the quadtree grid agree rather well with those on the fixed Cartesian grid (see Fig.~\ref{Fig_rti_intfs_amr}a), only minor differences are observed in the roll-up region.

The computational efficiency for this test of the new EBIT method with or without AMR is summarized in Table~\ref{Tab_rti_amr}. The wall time shown in the table includes both the interface advection and the Navier--Stokes solver. For the PLIC-VOF method with AMR (VOF-AMR), the mesh refinement criteria are based on both curvature and velocity gradient. Without AMR, the dynamics with the new EBIT method is about 2 times slower than with the PLIC-VOF method. 
When AMR is used, the wall times for the two methods are comparable. But the EBIT method still presents an about $33\%$ increase in the wall time. The lower efficiency of the EBIT method can be due to several reasons: i) The PLIC-VOF method updates the volume fractions directly. But for the EBIT method, both the position of markers and the value of Color Vertex need to be updated to calculate the volume fractions. Thus, more computation is involved in each advection step. ii) Since the connectivity of marker is represented by the Color Vertex implicitly, we need to access both the position of markers and the Color Vertex when we update each related field. Consequently, more memory access is involved in the advection step, which slows down the algorithm. iii) The PLIC-VOF method in Basilisk exists for more than decade and its efficiency has been highly optimized. In this work, we focus more on the accuracy of the EBIT method instead of its efficiency. Definitely, further improvement of the computational efficiency of the EBIT method is one of our goals.

For the simulation on a Cartesian grid, the area error is $E_{area} = 3.50 \times 10^{-3}$, which is much larger than that measured in the capillary wave test, $E_{area} = 5.3 \times 10^{-6}$. Here, topology changes take place at the later stages of the simulation and small droplets will be removed by the topology change mechanism.

\subsection{Rising bubble}

In this test we examine a single bubble rising under buoyancy inside a heavier fluid. This test case was first proposed by Hysing \cite{Hysing_2009_60} and it provides a standard benchmark for multiphase flow simulations, since this configuration is simple enough to be simulated accurately. Nevertheless, the bubble shows a strong deformation and even complex topology changes in some flow regimes \cite{Clift_1978, Legendre_2022_07}, thus giving an adequate challenge to interface tracking techniques. 

By taking into account the symmetry with respect to the vertical axis, we consider the rectangular computational domain $[0, D] \times [0, 4\,D]$, with $D=0.5$, partitioned with $N_x \times N_y = 128 \times 512$ grid cells. At the beginning of the simulation, a circular bubble of radius $R=D/2$ is positioned in the bottom part of the domain with center at $(0, D)$.
A no-slip boundary condition is enforced at the bottom and at the top of the computational domain, a free-slip boundary condition on the right vertical wall and a symmetric boundary condition on the left vertical boundary. The value of the relevant physical properties is that provided by Hysing \cite{Hysing_2009_60} and is listed in Table~\ref{Tab_para_rising},
where $Re = \big( \rho_1 g^{1/2} d^{3/2} \big) \big/ \mu_1$ is the Reynolds number and
$Bo = \big( \rho_1 g L^2 \big) \big/ \sigma$ the Bond number. In the following simulations, the timestep is limited by both the $CFL$ number, $CFL=0.5$, and the period of the shortest capillary, see Eq.~\eqref{Eq_dt_cap}.

\begin{table}[hbt!]
\caption{Physical properties for the rising bubble test.}
\centering
\begin{tabular}{ccccccccc}
\hline 
 Test case &$\rho_1$& $\rho_2$& $\mu_1$ & $\mu_2$ & $g$ & $\sigma$ & $Re$ & $Bo$ \\ 
\hline 
$1$ &$1000 $ & $100$ & $ 10$ & $1$ & $0.98$ &$24.5$ & $35$ & $10$ \\ 
$2$ &$1000 $ & $1$ & $ 10$ & $0.1$ & $0.98$ &$1.96$ & $35$ & $125$ \\ 
\hline 
\end{tabular}
\label{Tab_para_rising}
\end{table}
We consider the two different test cases of Table~\ref{Tab_para_rising}. In the first test, the bubble should end up in the ellipsoidal regime \cite{Hysing_2009_60}, since the surface tension forces are strong enough to hold the bubble together, hence no breakup is present in the simulation. For this first case, we solve both the axisymmetric problem and the two-dimensional Cartesian one.

\begin{figure}
\begin{center}
\begin{tabular}{cc}
\includegraphics[width=0.45\textwidth]{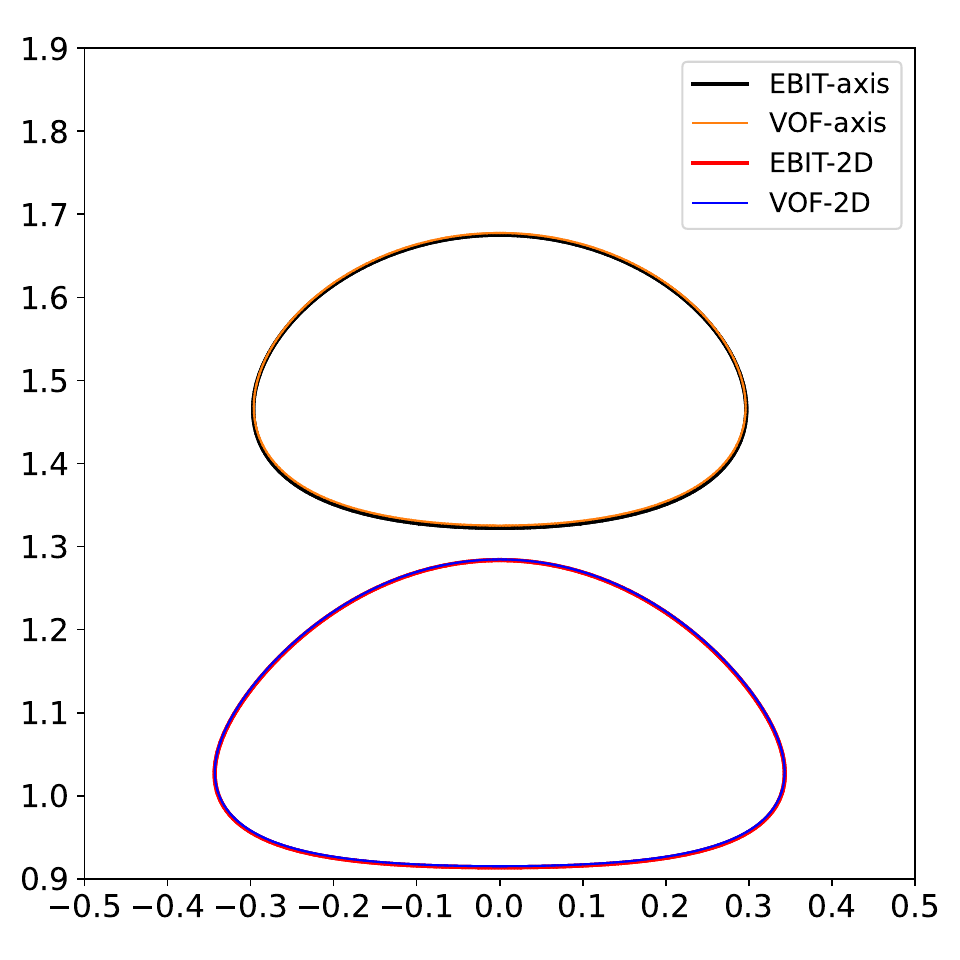} &
\includegraphics[width=0.45\textwidth]{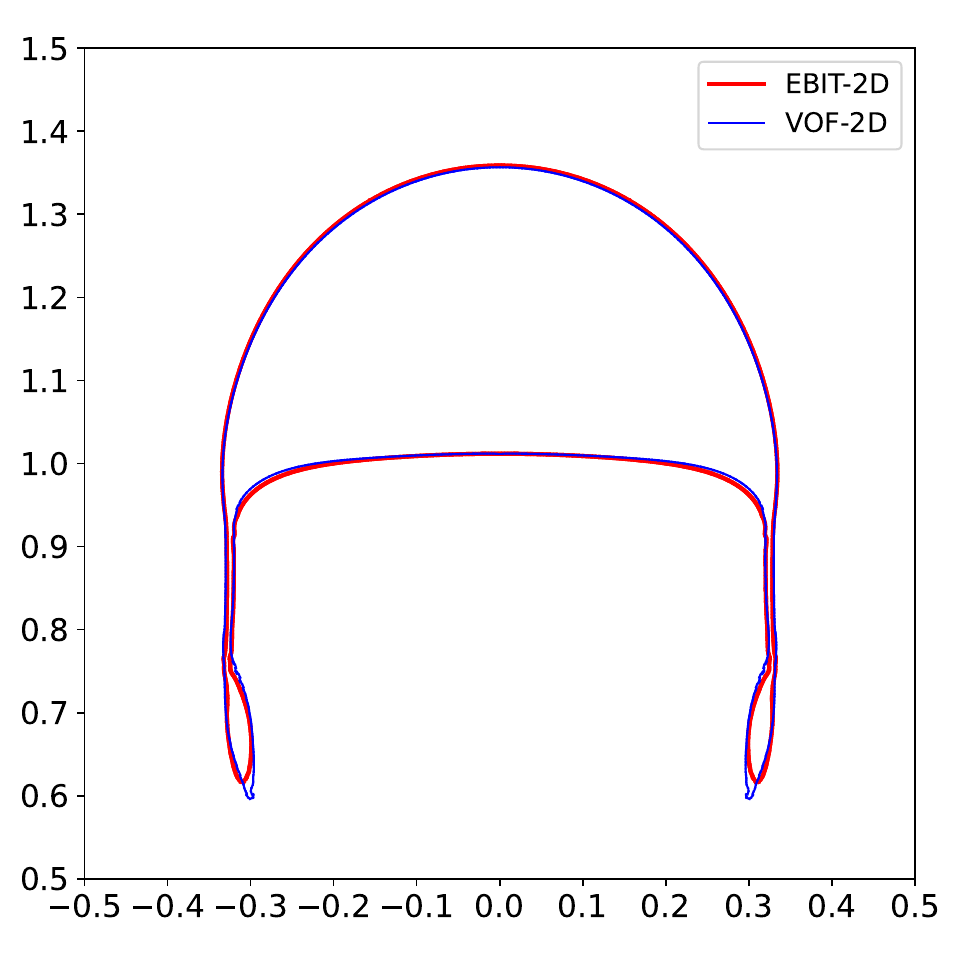}\\
(a) & (b) 
\end{tabular}
\end{center}
\caption{Rising bubble test. Interface lines at the end of the simulation $t = 3$ with different methods: (a) axisymmetric and Cartesian solutions for test case 1; (b) Cartesian solution for test case 2.}
\label{Fig_rising_intfs}
\end{figure}

The interface lines at the end of the simulation at time $t = 3$ are shown in Fig.~\ref{Fig_rising_intfs}a, for both the axisymmetric and Cartesian problems and with the new EBIT and PLIC-VOF methods. In general, good agreement is found between the two methods in both problems. More particularly, the bubble computed with the PLIC-VOF method
is always a very little ahead of the bubble with the new EBIT method.

\begin{figure}
\begin{center}
\begin{tabular}{cc}
\includegraphics[width=0.45\textwidth]{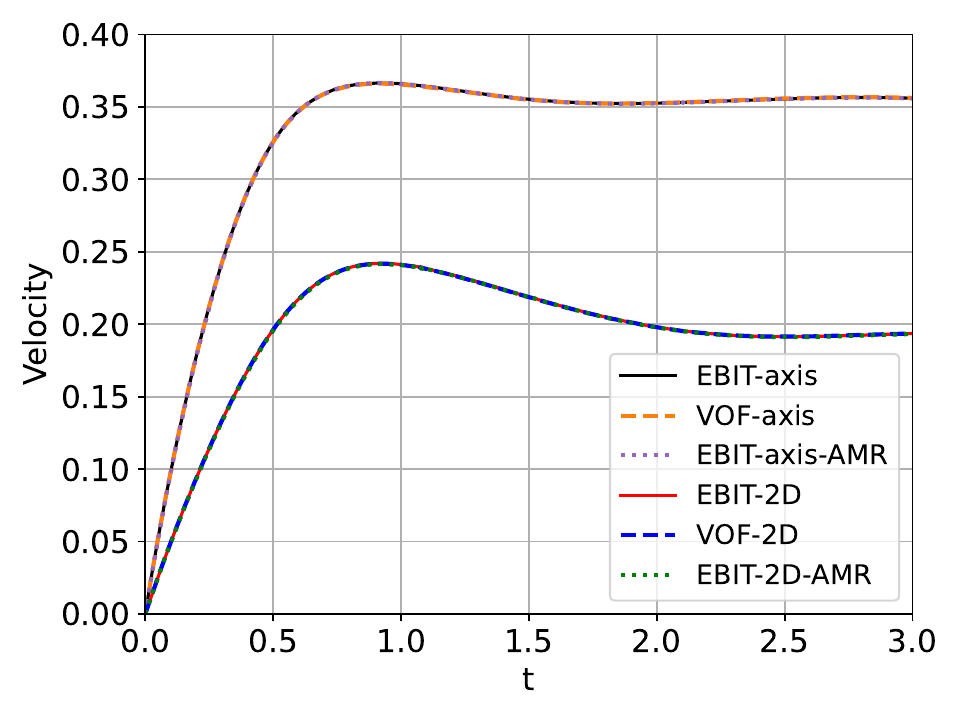} &
\includegraphics[width=0.45\textwidth]{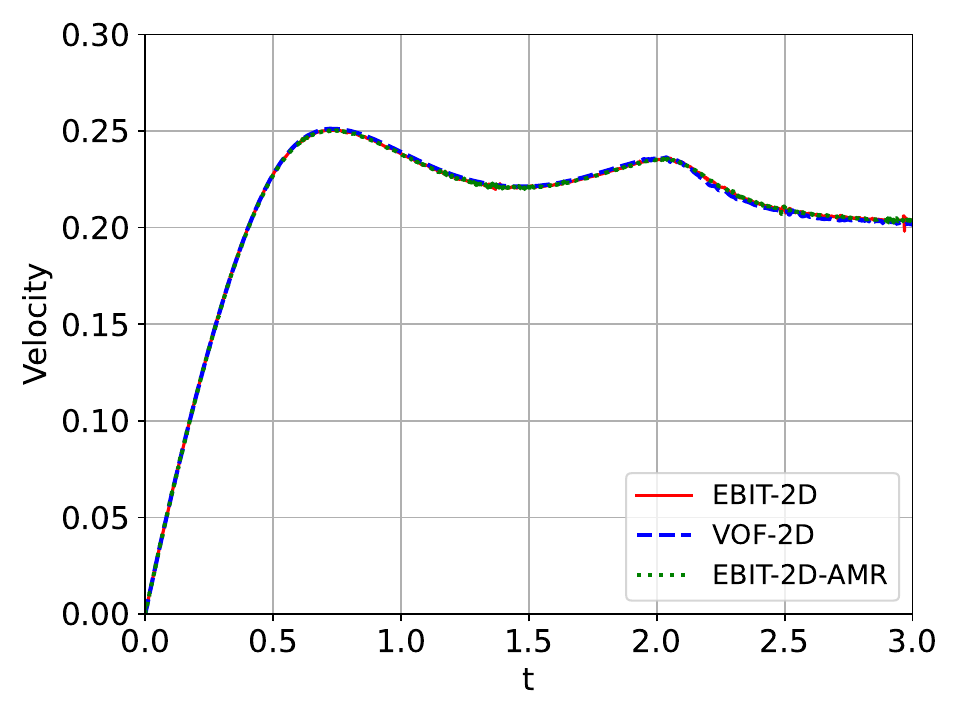}\\
(a) & (b) 
\end{tabular}
\end{center}
\caption{Rising bubble test. Bubble velocities as a function of time with different methods: (a) profiles for test case 1, including EBIT with AMR; (b) profiles for test case 2, including EBIT with AMR.}
\label{Fig_rising_vel}
\end{figure}

The rising velocities for the two problems and the two methods as a function of time are shown in Fig.~\ref{Fig_rising_vel}a. The figure includes also the results obtained with the new EBIT method in conjunction with AMR. The profiles of the rising velocities are very close to each other and this justify the fact that at the end of the simulation there is very little difference between the interface lines. For the Cartesian problem, the value of the maximum rising velocity is $0.2419$ for the new EBIT method and $0.2418$ for the PLIC-VOF method, which are basically the same value, $0.2419 \pm 0.0002$, reported by Hysing \cite{Hysing_2009_60}.

In the second test, the bubble lies somewhere between the skirted and the dimpled ellipsoidal-cap regimes, indicating that breakup can eventually take place. The simulation is carried out with both the new EBIT and PLIC-VOF methods, and the interface lines at the end of the simulation at time $t = 3$ are shown in Fig.~\ref{Fig_rising_intfs}b. At the given mesh resolution, the bubble skirt is observed with both methods, with no interface breakup.
Good agreement is observed between the two interface lines. More in detail, the new EBIT method predicts a slightly larger central part of the bubble and a smoother and shorter tail in the skirt region.

The rising velocities for the two methods as a function of time are shown in Fig.~\ref{Fig_rising_vel}b. In this case as well, the figure includes the results
obtained with the new EBIT method in conjunction with AMR. The presence of two peaks is well predicted in our simulation. The value of the first one is $0.2507$ for the new EBIT method and $0.2512$ for the PLIC-VOF method, in good agreement with the value $0.25 \pm 0.01$
indicated by Hysing \cite{Hysing_2009_60}.

\begin{figure}
\begin{center}
\begin{tabular}{cc}
\includegraphics[width=0.45\textwidth]{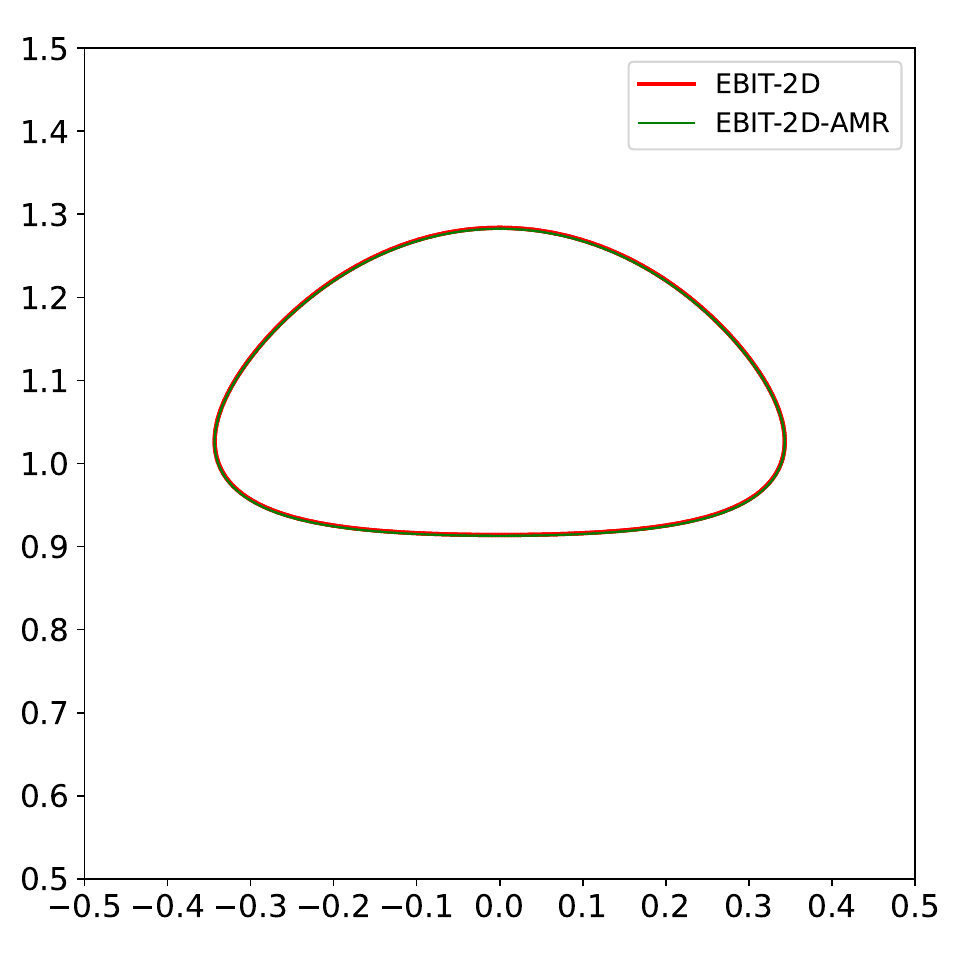} &
\includegraphics[width=0.45\textwidth]{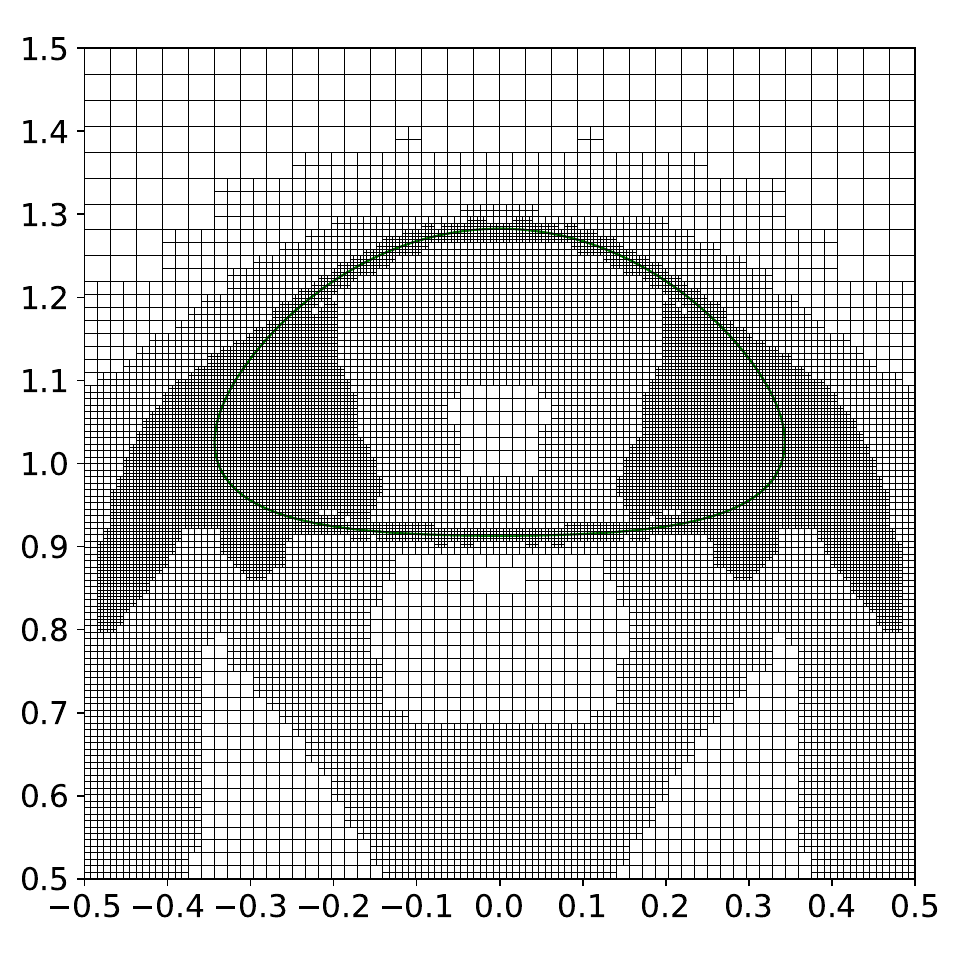}\\
(a) & (b) 
\end{tabular}
\end{center}
\caption{Rising bubble test case 1-2D: (a) interface lines at the end of the simulation $t = 3$ with and without AMR; (b) computational grid with $N_{l,max}=9$ and $N_{l,min}=6$.}
\label{Fig_rising_intfs_c1_amr}
\end{figure}

\begin{figure}
\begin{center}
\begin{tabular}{cc}
\includegraphics[width=0.45\textwidth]{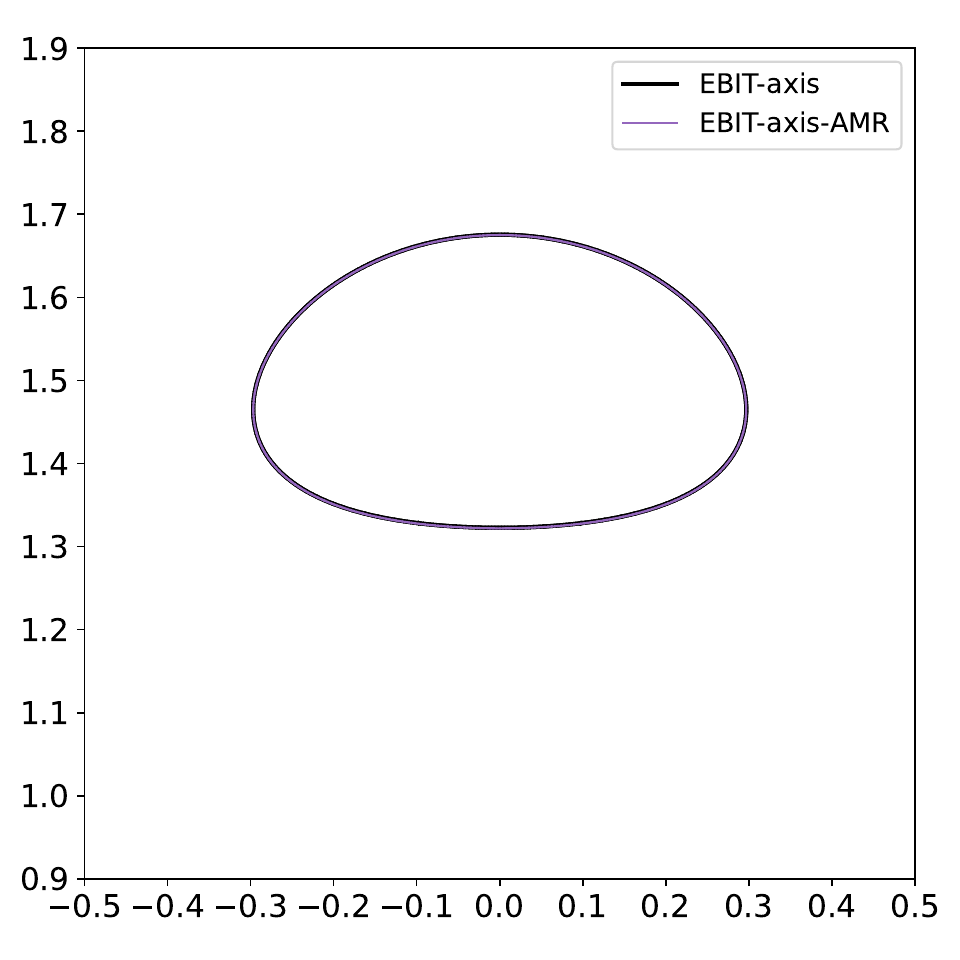} &
\includegraphics[width=0.45\textwidth]{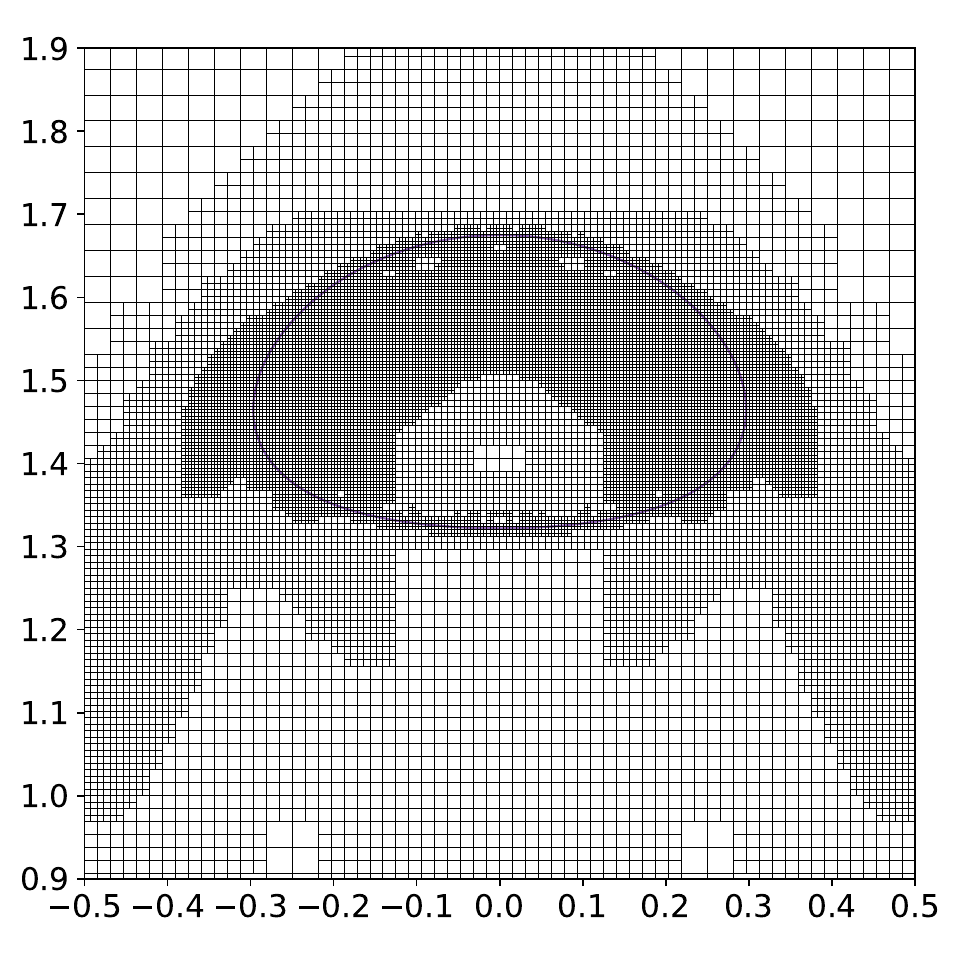}\\
(a) & (b) 
\end{tabular}
\end{center}
\caption{Rising bubble test case 1-Axisymmetric: (a) interface lines at the end of the simulation $t = 3$ with and without AMR; (b) computational grid with $N_{l,max}=9$ and $N_{l,min}=6$.}
\label{Fig_rising_intfs_c1_axi_amr}
\end{figure}

\begin{figure}
\begin{center}
\begin{tabular}{cc}
\includegraphics[width=0.45\textwidth]{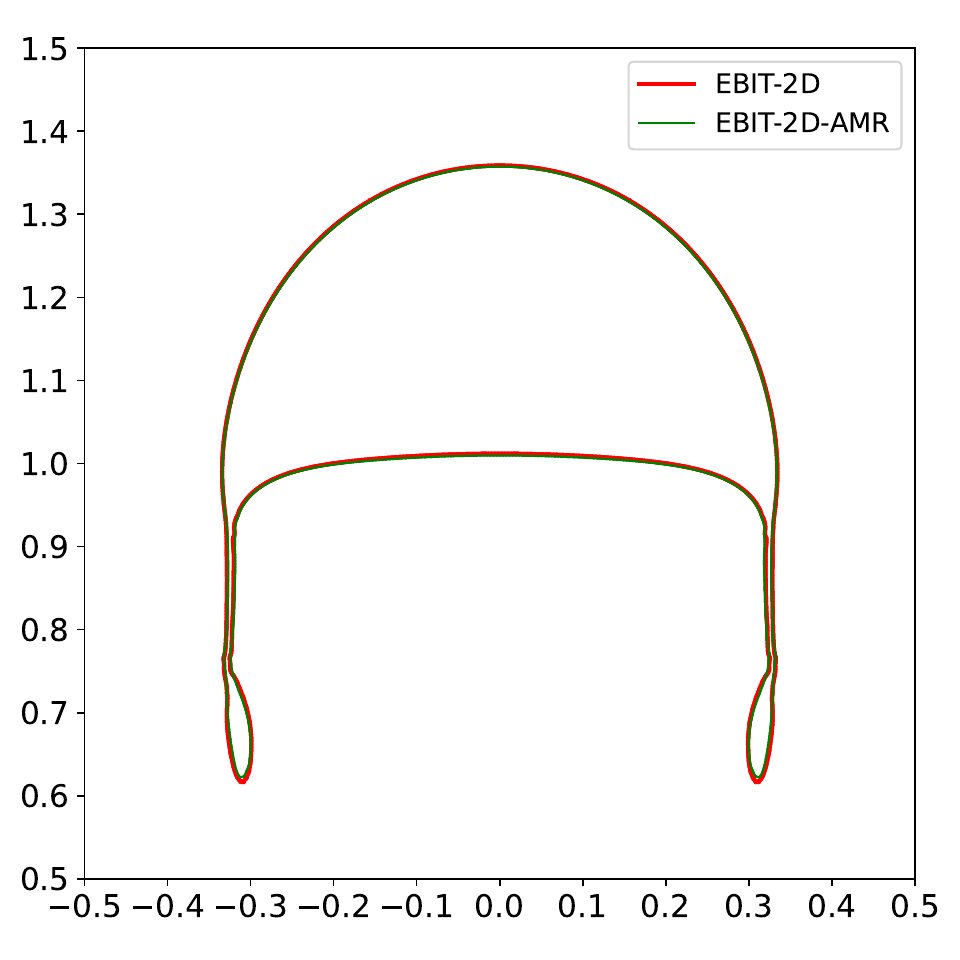} &
\includegraphics[width=0.45\textwidth]{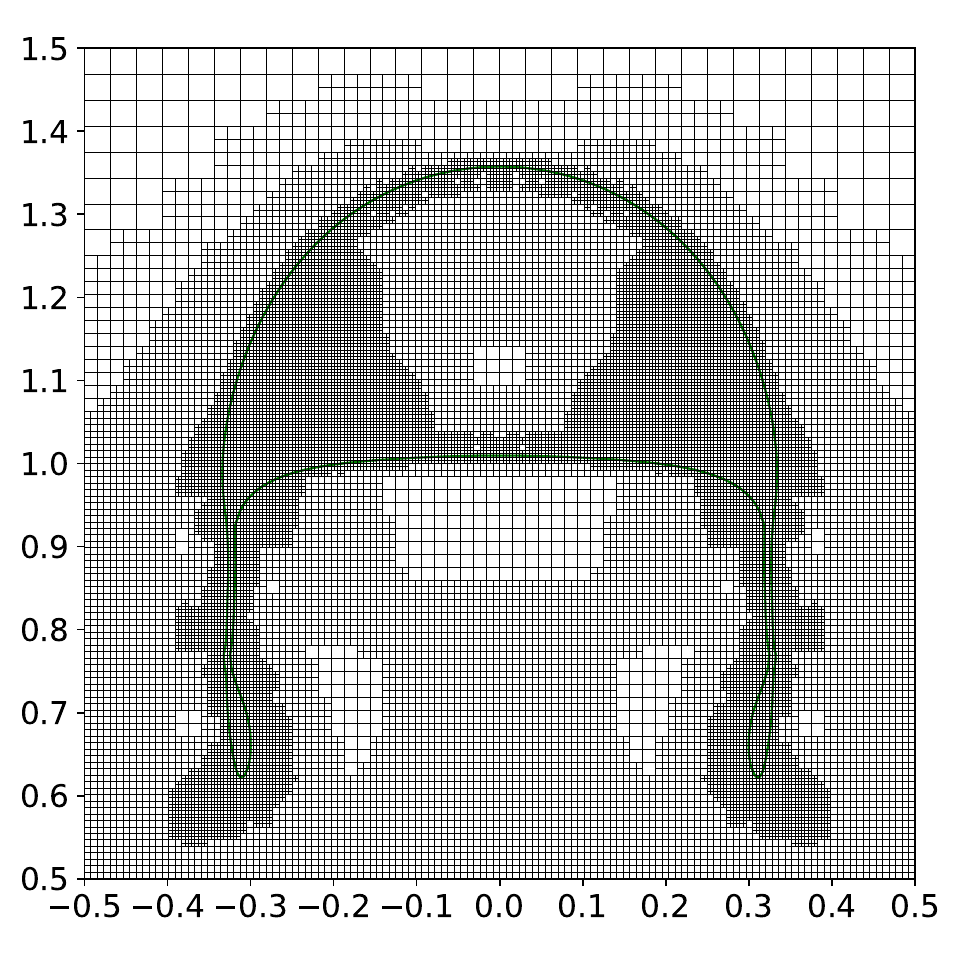}\\
(a) & (b) 
\end{tabular}
\end{center}
\caption{Rising bubble test case 2-2D: (a) interface lines at the end of the simulation $t = 3$ with and without AMR; (b) computational grid with $N_{l,max}=9$ and $N_{l,min}=6$.}
\label{Fig_rising_intfs_c2_amr}
\end{figure}

\begin{table}[hbt!]
\caption{Computational efficiency of the EBIT method with AMR, rising bubble test, $N_{l,max}=9$ and $N_{l,min}=6$.}
\centering
\begin{tabular}{ccccc}
\hline 
 &Method & Number of (leaf) cells & Time steps & Wall time (s) \\ 
\hline 
Case1-2D& EBIT & $65536$ & $4606$ & $1726$ \\ 
&EBIT-AMR & $11077$ & $4606$ & $172$ \\ 
&VOF & $65536$ & $4607$ & $935$ \\ 
&VOF-AMR & $10852$ & $4607$ & $147$ \\ 
\hline 
Case1-Axi & EBIT & $65536$ & $4606$ & $1824$ \\ 
&EBIT-AMR & $10141$ & $4606$ & $166$ \\ 
&VOF & $65536$ & $4607$ & $1136$ \\ 
&VOF-AMR & $9937$ & $4607$ & $145$ \\ 
\hline 
Case2-2D& EBIT & $65536$ & $1599$ & $4653$ \\ 
&EBIT-AMR & $14392$ & $1595$ & $729$ \\ 
&VOF & $65536$ & $1636$ & $2822$ \\ 
&VOF-AMR & $14095$ & $1634$ & $610$ \\ 
\hline 
\end{tabular}
\label{Tab_rising_amr}
\end{table}

When AMR is considered, the refinement criteria are those that have been used in the Rayleigh-Taylor instability test, i.e. proximity to the interface and velocity gradient. 
The maximum level of refinement is $N_{l,max}=9$, while the minimum level is $N_{l,min}=6$, for both test cases.
The interface lines at the end of the simulation at time $t = 3$ are shown in Figs.~\ref{Fig_rising_intfs_c1_amr} and \ref{Fig_rising_intfs_c1_axi_amr} for test case 1, and
in Fig.~\ref{Fig_rising_intfs_c2_amr} for test case 2.

For the axisymmetric problem, the interface line calculated by the new EBIT method on the quadtree grid is on top of the line on the fixed Cartesian grid. For the two Cartesian problems, the interface line on the quadtree grid is just a little bit behind that on the fixed Cartesian grid.

The computational efficiency for these two cases of the new EBIT method with
or without AMR is summarized in Table~\ref{Tab_rising_amr}. Without AMR, the new EBIT method represents a $60\%$ to $85\%$ increase in wall time compared to the PLIC-VOF method, which is smaller than that shown in the Rayleigh-Taylor instability test. When AMR is used, this increase becomes less than $20\%$ for all cases.

For the simulations on a Cartesian grid, the area errors $E_{area}$ for case1-2D, case1-Axi and case2-2D are $1.11 \times 10^{-4}$, $2.00 \times 10^{-4}$ and $4.56 \times 10^{-4}$, respectively, which are one order of magnitude smaller than those in the Rayleigh-Taylor instability test with topology changes. Furthermore, the case2-2D with a thin skirt structure shows a larger area loss.

\section{Conclusions}
We present a novel Front-Tracking method, the Edge-Based Interface Tracking (EBIT) method, which is suitable for almost automatic parallelization due to the lack of explicit connectivity.
Several new features have been introduced to improve the very first version of the EBIT method \cite{Chirco_2022_95}. First, a circle fit has been implemented to improve the accuracy of mass conservation in the reconstruction phase after the interface advection.
Second, a Color Vertex feature has been introduced to distinguish between ambiguous topological configurations and to represent the connectivity implicitly. Third, an automatic topological change mechanism has been discussed.

The new EBIT method has been implemented inside the free Basilisk platform in order to solve the Navier--Stokes equations for multiphase flow simulations with surface tension. Volume fractions are calculated based on the position of the markers and the Color Vertex, and are used to compute the physical properties and the surface tension force.

Numerical results for various cases, including both kinematic and dynamical tests, have been considered and compared with those obtained with the VOF method. Good agreement is observed for all test cases.

The extensions we foresee for EBIT are of two kinds: kinematic and dynamic. To give an example of a kinematic extension, if two markers per edge are allowed instead of one in the current method, dealing with thin strips of fluid will be possible. Two markers per edge would avoid reconnection as in VOF or evaporation as in Level-Sets, and would bring EBIT on a par with Front-Tracking in this respect. A dynamic extension would involve for example the direct computation of capillary forces or tensions from EBIT data instead of performing the detour through the VOF color fraction. There are other multiscale features, such as triple points, small droplets and bubbles and contact lines that could benefit from extensions of EBIT. To accurately represent the force balance near a triple point the direct computation of capillary forces would also be necessary. This improved treatment of surface tension could be done by implementing the momentum-conserving well-balanced method of Al-Saud et al. \cite{abu2018conservative}.

There is thus hope that EBIT, provided it is extended in certain ways, will be able to bring progress on multiscale issues. One often-used multiscale feature in both VOF and level set codes is the conversion of small connected patches to Lagrangian Particles. This is made easier if the exact location of the interface is known and its topology is decided. In VOF and Level-Set methods one often remarks instead some mass diffusion or ``evaporation'' near small drops. We finally note that the use of quad/octree AMR is particularly beneficial when studying flows with a wide range of scales, especially when combined with high performance computing. One originality of the new EBIT method is that it is merged with the quad/octree structure of a code with proven scalability on high performance computers.

\section{CRediT authorship contribution statement}

\textbf{J. Pan}: Conceptualization, Formal analysis, Code development, Simulations, Writing

\textbf{T. Long}: Formal analysis, Code development

\textbf{L. Chirco}: Conceptualization, Code development

\textbf{R. Scardovelli}: Formal analysis, Writing

\textbf{S. Popinet}: Basilisk code development

\textbf{S. Zaleski}: Conceptualization, Formal analysis, Supervision, Writing, Funding acquisition

\section{Declaration of competing interest}
The authors declare that they have no known competing financial interests or personal relationships that could have appeared to influence the work reported in this paper.

\section{Acknowledgements}
St\'{e}phane Zaleski and St\'{e}phane Popinet recall meeting Sergei Semushin in March 1995 and learning about his method. They thank him for the explanation of the method.
This project has received funding from the European Research Council (ERC) under the European Union's Horizon 2020 research and innovation programme (grant agreement number 883849). We thank the European PRACE group, the Swiss supercomputing agency CSCS, the French national GENCI supercomputing agency and the relevant supercomputer centers for their grants of CPU time on massively parallel machines, and their teams for assistance and the use of Irene-Rome at TGCC.






 \bibliographystyle{model1-num-names}
\bibliography{multiphase-jieyun,multiphase-stephane}
\end{document}